\documentclass[12pt]{JHEP3}

\usepackage{epsfig,amsfonts,amssymb,amscd,amsmath,mathrsfs,xspace,mathrsfs,amsthm,dsfont,bbm}

\usepackage{comment}
\usepackage{color}
\let\normalcolor\relax
\usepackage{bbold}

\newcommand{\sectiono}[1]{\section{#1}\setcounter{equation}{0}}

\newcommand{\be}{\begin{equation}}
\newcommand{\ee}{\end{equation}}
\newcommand{\ben}{\begin{eqnarray}\displaystyle}
\newcommand{\een}{\end{eqnarray}}
\newcommand{\bea}{\begin{eqnarray}}
\newcommand{\eea}{\end{eqnarray}}

\newcommand{\refb}[1]{(\ref{#1})}

\def\ZZZ{{\hbox{ Z\kern-1.6mm Z}}}
\def\RRR{{\hbox{ R\kern-2.4mm R}}}
\def\CCC{{\hbox{ C\kern-2.0mm C}}}
\def\zzz{{\hbox{z\kern-1mm z}}}

\def\ZZZ{{\mathbb Z} }
\def\RRR{{\mathbb R} }
\def\CCC{{\mathbb C} }

\newcommand{\IR}{\mathds{R}}
\newcommand{\IC}{\mathds{C}}
\newcommand{\IZ}{\mathds{Z}}

\newcommand{\IP}{\mathds{P}}

\newcommand{\p}{\partial}
\newcommand{\I}{\mathrm{i}}
\def\de{d}
\def\Tr{\,{\rm Tr}\, }

\def\Im{\,{\rm Im}\,}
\def\Re{\,{\rm Re}\,}

\def\({\left(}
\def\){\right)}
\def\[{\left[}
\def\]{\right]}
\def\hf{{1\over 2}}

\newcommand{\CP}{\IC P^1}

\newcommand{\non}{\nonumber}

\newcommand{\qeq}{{\hbox{=\kern-2.3mm ? \kern.5mm }}}
\renewcommand{\qeq}{=}

\newcommand{\bF}{{\bar F}}

\newcommand{\eps}{\epsilon}

\newcommand{\vp}{\varphi}
\newcommand{\ve}{\varepsilon}
\newcommand{\vrh}{\varrho}

\newcommand{\BBB}{{\cal B}}

\newcommand{\AAA}{{\cal A}}

\newcommand{\OO}{{\cal O}}

\newcommand{\XX}{{\cal X}}
\newcommand{\YY}{{\cal Y}}

\newcommand{\cV}{\mathcal{V}}
\newcommand{\cC}{\mathcal{C}}

\newcommand{\cK}{\mathcal{K}}
\newcommand{\cA}{\mathcal{A}}

\newcommand{\cI}{\mathcal{I}}
\newcommand{\cM}{\mathcal{M}}
\newcommand{\cW}{\mathcal{W}}
\newcommand{\cN}{\mathcal{N}}

\newcommand{\cR}{\mathcal{R}}
\newcommand{\cT}{\mathcal{T}}
\newcommand{\cY}{\mathcal{Y}}
\newcommand{\cJ}{\mathcal{J}}

\newcommand{\wt}{\widetilde}
\newcommand{\wh}{\widehat}

\newcommand{\NN}{{\cal N}}
\newcommand{\TT}{{\cal T}}

\def\bz{\bar z}

\def\bt{\bar t}

\def\bF{\bar F}

\def\bV{ \bar V }

\def\bZ{\bar Z}
\def\bchi{\bar \chi}

\newcommand{\tzeta}{\tilde\zeta}

\def\cl0{\tilde c_0}

\def\hGam{\hat\Gamma}

\def\one{{\hbox{ 1\kern-.8mm l}}}
\def\zero{{\hbox{ 0\kern-1.5mm 0}}}

\newcommand{\dalpha}{\beta}

\def\Om#1{\Omega_{#1}}
\def\bOm#1{\bar\Omega_{#1}}

\def\Min{M}

\def\Uin{\mathbf{U}}

\def\vv{w}

\def\vl{u}
\def\bvl{\bar \vl}
\def\CY{\mathfrak{Y}}

\def\ellg#1{\ell_{#1}}
\def\Xigi{\Xi_{\gamma}}
\def\Zg{Z_{\gamma}}
\def\bZg{\bar Z_{\gamma}}

\def\Thkl{\Theta_{\gamma}}

\def\qr{\sigma_\gamma}

\def\Ig{\cJ_{\gamma}}
\def\Ilg{\cJ^{(1)}}
\def\Ilog#1{\cJ^{(1,#1)}}
\def\Irt{\cJ^{(2)}}
\def\Irat#1{\cJ^{(2,#1)}}

\def\Igp{\Ilog{+}_{\gamma}}
\def\Igm{\Ilog{-}_{\gamma}}
\def\Igpm{\Ilog{\pm}_{\gamma}}

\def\Igg{\Ilg_{\gamma}}

\def\rIgamp#1{\Irat{+}_{#1}}
\def\rIgamm#1{\Irat{-}_{#1}}
\def\rIg{\Irt_{\gamma}}
\def\rIgp{\Irat{+}_{\gamma}}
\def\rIgm{\Irat{-}_{\gamma}}
\def\rIgpm{\Irat{\pm}_{\gamma}}

\def\Sg#1{S_\gamma^{(#1)}}

\def\CG{L_\gamma}

\def\NP{k}

\def\cTR{\cT^{\rm R}}
\def\ts{T}

\def\dalpha{{\dot\alpha}}

\def\dbeta{{\dot\beta}}

\def\dgamma{{\dot\gamma}}

\def\ddelta{{\dot\delta}}

\def\sp{\not  \hskip -.02in p}

\def\I{i}

\title{D-instantons in Type IIA string theory on  Calabi-Yau threefolds}

\author{Sergei Alexandrov$^1$, Ashoke Sen$^2$, Bogdan Stefa\'nski, jr.$^3$
\\
$^1$ {\it
Laboratoire Charles Coulomb (L2C), Universit\'e de Montpellier,
CNRS, F-34095, Montpellier, France}\\

$^2$ {\it Harish-Chandra Research Institute, HBNI \\ Chhatnag Road, Jhusi, Allahabad 211019, India}\\

$^3$ {\it Centre for Mathematical Science, City, University of London
\\ Northampton Square, EC1V 0HB London, UK}\\

\vspace*{2mm} {\tt e-mail:
\email{sergey.alexandrov@umontpellier.fr},
\email{sen@hri.res.in},
\email{Bogdan.Stefanski.1@city.ac.uk}
}

\vspace*{-3mm}

}

\abstract{Type IIA string theory compactified on a Calabi-Yau threefold has a hypermultiplet moduli
space whose metric is known to receive non-perturbative corrections from Euclidean D2-branes
wrapped on 3-cycles. These corrections have been
computed earlier by making use of mirror symmetry, S-duality and twistorial description of quaternionic geometries.
In this paper we compute the leading corrections in each  homology class using a direct
world-sheet approach without relying on any duality symmetry or supersymmetry. Our results are
in perfect agreement with the earlier predictions.
}

\begin{document}

\baselineskip 16.pt
\setlength{\parskip}{0.15cm}

\sectiono{Introduction}

D-instantons
describe classical saddle points in type II and type I string theories and give non-perturbative
contributions to string amplitudes. While the action $-\TT$ of a D-instanton is
easy to compute \cite{Becker:1995kb} and gives the prefactor
$e^{-\TT}$ multiplying all D-instanton amplitudes,\footnote{We call D-instanton
any Euclidean D-brane wrapped on a compact cycle. We shall use the sign convention
for the action $S$ such that
the Euclidean path integral is weighted by $e^S$. \label{fo1}}
the actual computation of the amplitude suffers from infrared divergences associated
with open string zero modes.
For this reason most of the initial study of D-instanton amplitudes, carried out in
ten-dimensional type IIB string theory,
used an indirect approach based on S-duality of this theory
\cite{Green:1997tv,Green:1998by}.
Perturbative amplitudes
in this theory are not usually S-duality invariant, but one finds that for certain amplitudes
containing limited powers of momenta, the constraint of supersymmetry together with the
requirement of S-duality fixes the amplitude completely. The weak coupling expansion of
the amplitude then generates the usual perturbative terms, but also contains terms
proportional to $e^{-k\, \TT}$ for integer $k$. These can be regarded as the contribution to the
amplitude from $k$ D-instantons. This procedure has been extended to the study of toroidal
compactifications of type II string theories, producing remarkable results on the effect of
D-instanton corrections to certain amplitudes in these theories
\cite{Green:1997di,Kiritsis:1997em,Pioline:1997pu,Green:1999pu,Basu:2007ru,Basu:2007ck,Green:2010wi,Green:2014yxa,Pioline:2015yea,Bossard:2020xod}.

In recent years a similar approach has been successful in computing the effect of
Euclidean D-branes in type II string theory compactified on Calabi-Yau (CY) 3-folds
(see \cite{Alexandrov:2011va,Alexandrov:2013yva} for a review).
The quantity of interest here is the metric on the moduli space of the theory. Supersymmetry
implies that the moduli space locally has a product structure $\cM_V\times \cM_H$ \cite{deWit:1984px}.
Here the first factor contains the vector multiplet moduli and is a special K\"ahler manifold, while the second
factor describes the hypermultiplet moduli and is restricted to be a quaternion-K\"ahler manifold \cite{Bagger:1983tt}.
Since the actions of  Euclidean branes depend on the four-dimensional string
coupling that is given by the vacuum expectation value of the dilaton field, and the latter is a part of the hypermultiplet moduli,
it follows that the instanton corrections can only affect the metric on $\cM_H$.
Remarkably, there exists a chain of dualities (see \cite[Fig.1]{RoblesLlana:2007ae})
involving S-duality of type IIB, electro-magnetic duality of type IIA and
mirror symmetry relating the two formulations, which allows us at least in principle to find all these non-perturbative corrections.
In practice, this program has been realized for D-instantons in \cite{Alexandrov:2008gh,Alexandrov:2009zh}
(based on earlier works \cite{RoblesLlana:2006is,RoblesLlana:2007ae})
and for NS5-brane instantons\footnote{NS5-branes generate
instanton corrections in CY compactifications which scale as $e^{-1/g_s^2}$ and thus are exponentially suppressed compared to D-instantons
which behave as $e^{-1/g_s}$. Nevertheless, they are expected to be indispensable for making sense of the metric on $\cM_H$
at the non-perturbative level (see, e.g., \cite{Pioline:2009ia}) and represent the last unknown
contribution to have the complete two-derivative effective action in this class of compactifications.
See \cite{Alexandrov:2014mfa,Alexandrov:2014rca} for an attempt to go beyond of the one-instanton approximation.}
in the one-instanton approximation in \cite{Alexandrov:2010ca}
where the instanton effects have been encoded into a holomorphic contact structure on the twistor space associated to $\cM_H$.
The explicit D-instanton corrected metric has been derived subsequently in \cite{Alexandrov:2014sya,Alexandrov:2017mgi}.

Despite these successes, it is important to develop techniques for direct computation of
D-instanton contributions to amplitudes. There are a variety of reasons for this; we shall
quote a few. First of all, S-duality is only a conjectured symmetry of the theory and it is
important to test its predictions whenever possible. Second, S-duality and supersymmetry allow
one to compute only a few of the terms in the effective action. In the particular example
of ten-dimensional type IIB string theory, S-duality and supersymmetry can only fix terms
in the four graviton amplitude that contain 14 or fewer
powers of momenta but terms with 16 or more
powers of momenta are not fixed. Therefore, direct D-instanton computation will be the
only way to fix these terms.
Finally, in theories with lower supersymmetry, there may be contributions that are protected
by supersymmetry but are either not fixed by S-duality or the consequences of S-duality
may be difficult to implement. For example, in certain flux compactifications of type II string theory
the superpotential and coefficients of gauge kinetic terms are
holomorphic functions of the moduli but receive contributions from
Euclidean D-branes \cite{Witten:1996bn}. Often such terms are important for studying various aspects of the
theory like moduli stabilization \cite{Kachru:2003aw,Demirtas:2021ote,Demirtas:2021nlu}
and  it is important to compute them.
At present there is no known way to fix them using S-duality.

For these reasons one would like to develop a systematic procedure for computing
D-instanton contribution to an amplitude without any help from supersymmetry or
duality symmetries.
During the last few years string field theory techniques have been used to develop
a systematic procedure for dealing with the infrared divergences that plague such analysis.
The initial study was
carried out for two-dimensional string
theory \cite{Balthazar:2019rnh,Sen:2019qqg,Balthazar:2019ypi,Sen:2021qdk},
but the procedure was later extended to
type IIB string theory in ten dimensions \cite{Sen:2021tpp,Sen:2021jbr}.
In particular, in both theories the overall normalization of the
D-instanton amplitude was determined by computing the exponential of the annulus
amplitude. This suffers from infrared divergences but string field theory gives an unambiguous
procedure for getting a finite result. In both theories the result is known with the help of duality
symmetries --- matrix model in the case of two-dimensional string theory and S-duality in the case
of type IIB string theory, and in both cases the result of direct D-instanton computation agrees
with the result predicted by duality.

Given this success, one could use this procedure for computing D-instanton contributions to
amplitudes where the result is not known otherwise, {\it e.g.} amplitudes in type IIB string theory
with large number of momentum factors \cite{Wang:2015jna}
or the superpotential and gauge kinetic
terms in $\NN=1$ supersymmetric string flux compactifications. On the other hand, it is also important
to test this procedure by computing other amplitudes where the result is known, {\it e.g.},
D-instanton corrections to the hypermultiplet moduli space in type II string theories on CY
3-folds. This is the problem that we shall address in this paper.

Since mirror symmetry relates type IIA and type IIB string compactifications on mirror CY
3-folds, we concentrate on type IIA string compactifications. In this theory the Kahler
moduli of the CY 3-fold belong to the vector multiplet and therefore cannot affect
the metric on the hypermultiplet moduli space. Since $\alpha'$ corrections are controlled
by the Kahler moduli, this means that the hypermultiplet moduli space metric cannot
receive any $\alpha'$ corrections. This allows us to work in the large volume limit and
use various geometric properties of the CY 3-fold without having to invoke
the exact conformal field theory description of the world-sheet theory.

Type IIA string theory has D$p$-branes for even $p$, spanning $(p+1)$-dimensional world-volume. Since the only
odd-dimensional cycles inside a CY 3-fold are 3-cycles, the instanton contributions
come from Euclidean D2-branes wrapped on these 3-cycles. We compute a specific class
of contributions which can be described as follows.
Let $\gamma$
label the homology class of a particular 3-cycle and $-\TT_\gamma$ denote the action of
a D2-brane wrapped on the 3-cycle in the sign convention described in footnote \ref{fo1}.
The contributions from the instantons in this sector
take the form of $e^{-\TT_\gamma}$ multiplied by a power series expansion in string coupling $g_s$.
We analyze only the leading
terms in the expansion in powers of $g_s$ (which are proportional to $g_s^{1/2}$).
We then compare our results with the earlier predictions based on S-duality, mirror symmetry
and supersymmetry and find perfect agreement. It is
certainly possible in principle to compute higher order terms in the power
series expansion, but this will involve more work.

Note that since $\TT_\gamma$ is proportional to the inverse string coupling,
terms of order $e^{-\TT_\gamma}$ for different $\gamma$ have exponential
suppression relative to each other. Therefore one could question the significance of
including the leading order term proportional to $e^{-\TT_\gamma}$ before
performing all order resummation of the contribution from a more dominant instanton.
Indeed, a common criticism of the analysis of instanton contributions to the amplitude
is that unless we can resum the whole series of perturbative contributions, instanton contributions cannot be defined
meaningfully. However, D-instantons for different $\gamma$'s
belong to different topological classes of saddle points compared to each other and
the perturbative saddle point. This is reflected in the fact that $\TT_\gamma$ contains an
imaginary part involving the component of the Ramond-Ramond (RR) 3-form field
along the 3-cycle represented by $\gamma$. These are
different for different $\gamma$'s, and perturbative corrections do not generate any
additional dependence on these components of the RR field.
This is a consequence of the fact that the usual string amplitudes involve the field strength
of the RR field instead of the potential, and the only contributions that depend on the
potential arise from the disk one-point function of the RR fields, whose exponentiation
generates the $e^{- \I \Im(\TT_\gamma)}$ factor in the D-instanton amplitude.
Therefore, in each non-trivial topological
sector, labelled by a specific dependence of the amplitude on the RR moduli,
the leading contribution is provided by the D-instantons in that topological class
and is unambiguously defined.

The rest of the paper is summarized as follows. We begin in \S\ref{sback} with a review of
the background material that we use in the rest of the paper. This includes conventions for the
world-sheet theory underlying ten-dimensional type IIA string theory as well as some details
of compactification on CY 3-folds and its complex structure moduli space.
In \S\ref{sduality} we describe the result for the
instanton corrected metric on the hypermultiplet moduli space in the particular weak
coupling limit where we keep only the leading term in each topological sector.
The result is shown in \refb{square}, whereas
the actual computation that leads to it is presented in appendix \ref{ap-metric}.

Sections
\ref{sfour}, \ref{sstrategy} and \ref{sec-maincomput} describe direct world-sheet computation
of the same terms.
In \S\ref{sfour} we describe the computation of the overall normalization by appropriately
`regulating' the exponential of the annulus amplitude, both for single instantons and
multiple coincident instantons. \S\ref{sstrategy} describes the strategy for computing
the correction to the hypermultiplet metric. In general, while extracting corrections to the effective
action from string amplitudes, we need to compare the on-shell S-matrix elements. Since
for massless scalars the first such amplitude is the four-point function, which receives
contributions from the Riemann tensor of the moduli space metric, one needs to compare
the four-point functions in string theory and the effective field theory. However, we show that
due to the presence of the exponential factor $e^{-\TT_\gamma}$ in the correction to the metric, the
computation of the leading term in a given topological sector can be reduced to the
computation of a disk amplitude with one closed string insertion and a pair of open string
fermion zero mode insertions. In section \ref{sec-maincomput} we compute all the
required disk amplitudes and use them to make the final prediction of the instanton
correction to the metric. The result is shown in \refb{e686}, and agrees with
\refb{square} up to field redefinition. Some of the technical details
required for the analysis of this section, {\it e.g.} determination of certain phases and signs,
the construction of the RR vertex operator in different pictures and some identities involving covariantly constant spinors
in CY 3-folds are relegated to appendices.

\section{Background} \label{sback}

In this section we shall review the necessary background material that will be needed
for our analysis.

\subsection{World-sheet conventions in ten dimensions}

In this section we shall describe the conventions for the world-sheet theory of type IIA string
theory following \cite{Sen:2021tpp}.

Since we shall be working in the large volume approximation, locally the metric will be
nearly flat and we can use the results for flat ten-dimensional target space.
We shall denote by
$x^M$ with $0\le M\le 9$ a set of 10 free scalar fields describing the target space-time
coordinates and by  $\bar\psi^M$, $\psi^M$ their superpartner left and right-moving fermions.
The world-sheet theory also contains
Grassmann odd ghost fields $b$, $c$, $\bar b$, $\bar c$ and Grassmann even
ghost fields $\beta,\gamma,\bar\beta,\bar\gamma$. We introduce
scalar fields $\phi,\bar\phi$, and fermionic fields $\xi,\eta,\bar\xi,\bar\eta$ by bosonizing
the $\beta$,$\gamma$ system \cite{Friedan:1985ge}:
\be\label{ebosonize}
\beta=\p\xi\, e^{-\phi},
\qquad
\gamma=\eta\, e^\phi,
\qquad
\bar\beta=\bar\p\bar\xi\, e^{-\bar\phi},
\qquad
\bar\gamma=\bar\eta\, e^{\bar\phi}\, .
\ee

In the $\alpha'=1$ unit that we shall use,
the operator product expansions (OPE) between various  free fields take the form:
\bea
\label{ematterope}
\p X^M(z) \p X^N(w) &=& -{\eta^{MN}\over 2 (z-w)^2}+\cdots,
\qquad
\psi^M (z) \psi^N(w) = -{\eta^{MN}\over 2(z-w)}+\cdots\, ,
\non\\
 c(z) b(w) &=&(z-w)^{-1}+\cdots,
\qquad\qquad\ \
\xi(z)\eta(w) = (z-w)^{-1}+\cdots,
\vphantom{1 \over 2}
\\
\p \phi (z)\,  \p\phi(w) &=& -{1\over (z-w)^2} +\cdots \, ,
\qquad \ \
e^{q_1\phi(z)} e^{q_2\phi(w)} = (z-w)^{-q_1q_2} e^{(q_1+q_2)\phi(w)}+ \cdots \, ,
\non
\eea
where $\cdots$ denote less singular terms.
The mostly $+$ signature
Minkowski metric $\eta^{MN}$ will be replaced by $\delta_{MN}$
in the Euclidean computation.
There are similar OPE involving anti-holomorphic
fields which we shall not write down explicitly. We assign ghost and picture numbers to various fields
as follows:
\be
\begin{split}
\hbox{ghost number}\ &:\ 1 \ \hbox{for} \ c,\bar c,\gamma,\bar\gamma,\eta,\bar\eta, \quad
-1\ \hbox{for} \
b,\bar b,\beta,\bar\beta,\xi,\bar\xi , \quad 0 \ \hbox{for others},
\\
\hbox{picture number}\ &:\ q \ \hbox{for}\  e^{q\phi}, e^{q\bar\phi},\quad 1 \ \hbox{for} \
\xi,\bar\xi, \quad
-1\  \hbox{for}\ \eta,\bar\eta, \quad\hspace{.35cm}  0 \ \hbox{for others}\, .
\end{split}
\ee
The $SL(2,\IC)$ invariant vacuum is defined to carry zero ghost number and picture number.

The stress tensor $T_m(z)$ and its fermionic partner
$T_F(z)$ for the matter sector take the form:
\be
\label{emattertdef}
T_m(z) = - \partial X_M \partial X^M +
\psi_M \p \psi^M ,
\qquad
T_F(z) = -\psi_M \p X^M \, ,
\ee
with similar expressions for their anti-holomorphic counterparts.
The OPE involving $T_m$ and $T_F$ take the form:
\be\label{etope}
\begin{split}
& T_m(z) T_m(w) = {15\over 2} {1\over (z-w)^4}+{2\over (z-w)^2}\, T_m(w)
+{1\over z-w}\,\p T_m(w)+\cdots,
\\
& T_F(z) T_F(w)
=  {5\over 2} \, {1\over (z-w)^3}+ {1\over 2}\, {1\over z-w}\, T_m(w) + \cdots,
\\
&
T_m(z) T_F(w) = {3\over 2} {1\over (z-w)^2}\, T_F(w) + {1\over z-w}\, \p T_F(w)
+\cdots\, .
\end{split}
\ee
The stress tensors of the ghost fields, the BRST current $j_B$ and the BRST charge $Q_B$ are
given by:
\be
T_{b,c}=-2 \, b\, \p\, c + c\, \p\, b,
\qquad
T_{\beta,\gamma} = {3\over 2}\, \beta\p\gamma + {1\over 2}\, \gamma\p\beta
= T_\phi + T_{\eta,\xi}\, ,
\ee
\be \label{e2.4}
T_{\eta,\xi} = -\eta \p \xi\, ,
\qquad
T_\phi = -{1\over 2}\, \p\phi \p\phi - \p^2 \phi\, ,
\ee
\be \label{ebrstcurrent}
\jmath_B =c \, \bigl(T_{m} + T_{\beta,\gamma} \bigr)+ \gamma \, T_F
+ b\, c\, \p c
-{1\over 4}\, \gamma^2 \, b\, ,
\ee
\be \label{ebrs1}
Q_B = \ointop dz \jmath_B(z) \, .
\ee
$\ointop$ is normalized to include the $1/(2\pi i)$ factor so that
$\ointop dz/z=1$.

We shall follow the normalization conventions of \cite{Sen:2021tpp} for the
picture changing operator (PCO) $\XX(z)$ \cite{Friedan:1985ge,Verlinde:1987sd}
and the inverse picture changing operator $\YY(z)$:
\be \label{epicture}
\XX(z) = 2\, \{Q_B, \xi(z)\} =2\,  c \, \partial \xi +
2\, e^\phi T_F - {1\over 2} \p \eta \, e^{2\phi} \, b
- {1\over 2} \p\left(\eta \, e^{2\phi} \, b\right), \qquad \YY = 2\, c\, \p\xi \, e^{-2\phi} \, .
\ee
With this definition we have:
\be
[Q_B, \XX(z)]=0,
\qquad
[Q_B,\YY(z)]=0,
\qquad
\YY(z) \XX(w) = 1 + \OO(z-w)\, .
\ee
We also define:
\be
\XX_0 = \ointop dz \, \XX(z),
\qquad
\YY_0 = \ointop dz \, \YY(z)\, .
\ee
As already mentioned before, we have an analogous  set of operators and OPEs
in the anti-holomorphic sector. In the following we shall use the words right (left)-moving
and holomorphic (anti-holomorphic) fields interchangeably. The labelling of picture
numbers will be done accordingly, with picture number $(m,n)$ implying picture number
$m$ from the left-moving fields and $n$ from the right-moving fields.
The BRST charge for open strings can be represented by $Q_B$, whereas
the BRST charge in the closed string sector will be given by the sum of $Q_B$ and $\bar Q_B$.

Next we turn to the spin fields. We denote by  $S^\alpha$ and $S_\alpha$
the 16-component
spin fields in the matter sector, carrying opposite
chirality. We shall use the convention that in the holomorphic sector
$e^{-\phi/2}S_\alpha$ and $e^{-3\phi/2}S^\beta$ are
GSO even operators while in the anti-holomorphic sector
$e^{-\bar\phi/2} \bar S^\alpha$ and $e^{-3\bar\phi/2}\bar S_\beta$ are
GSO even operators.
This assignment is a consequence of the fact that in type IIA string theory the left and the
right-moving Ramond sectors carry opposite chirality.
The relevant OPE involving the spin fields are:\footnote{Even though we deal with Euclidean
D-branes, we use Lorentzian gamma matrix algebra, regarding time and energy as taking
imaginary values. For most of our analysis it makes no difference whether we work in
Euclidean or
Lorentzian signature. One place where the continuation from one signature to the other introduces
an ambiguity is in the sign of the contribution from integration over the fermion zero modes.
As discussed at the end of \S\ref{sfourone} and
appendix \ref{sdd}, this sign can be fixed using the cluster
property of the multi-instanton amplitudes.
}
\be\label{espinope}
\begin{split}
\psi^M(z) \ e^{-\phi/2} S_\alpha(w) =&\, {\I\over 2}\, (z-w)^{-1/2}\,
(\Gamma^M)_{\alpha\beta} e^{-\phi/2}\, S^\beta(w)
+\cdots ,
\\
\psi^M(z) \ e^{-\phi/2} S^\alpha(w) =&\, {\I\over 2}\, (z-w)^{-1/2}\,
(\Gamma^M)^{\alpha\beta} e^{-\phi/2}\, S_\beta(w)
+\cdots \, ,
\\
e^{-3\phi/2} S^\alpha(z)\ e^{-\phi/2} S_\beta(w)   =&\, (z-w)^{-2} \, \delta^\alpha_\beta \,
e^{-2\phi}(w)+\cdots ,
\\
e^{-\phi/2} S_\alpha(z)  \ e^{-\phi/2} S_\beta(w) =&\, \I (z-w)^{-1} \,
(\Gamma^M)_{\alpha\beta} \, e^{-\phi} \, \psi_M(w) +\cdots \, ,
\end{split}
\ee
where the $16\times 16$ matrices $\Gamma^M_{\alpha\beta}$ satisfy the identities:
\be
\{\Gamma^M, \Gamma^N\} = 2\, \eta^{MN}\, I_{16}\, , \qquad (\Gamma^M)_{\alpha\beta}
=(\Gamma^M)_{\beta\alpha}\, , \qquad (\Gamma^M)^{\alpha\beta}
=(\Gamma^M)^{\beta\alpha}\, ,
\ee
with the understanding that when we take product of the $\Gamma^M$'s, the successive
$\Gamma^M$'s will have their indices alternating between upper and lower indices.
Here $I_{16}$ denotes the $16\times 16$ identity matrix.
The OPE in the anti-holomorphic sector are obtained by replacing all the
holomorphic fields in \refb{espinope} by anti-holomorphic fields, replacing the upper spin indices
by lower ones and the lower spin indices by upper ones.
We also have,
\be
(\Gamma^M)^{\alpha\beta}=(\Gamma^M)_{\alpha\beta} \quad \hbox{for $M\ne 0$},
\qquad (\Gamma^0)^{\alpha\beta}=\delta_{\alpha\beta},
\qquad
(\Gamma^0)_{\alpha\beta}=-
\delta_{\alpha\beta}\, .
\ee

As in \cite{Sen:2021tpp}, for the closed string vacuum carrying momentum $k$, we
choose the normalization:
\be\label{eclosednorm}
\langle k| c_{-1}\bar c_{-1} c_0\bar c_0 c_1 \bar c_1\, e^{-2\phi}(0) e^{-2\bar\phi}(0)|k'\rangle=
- (2\pi)^{10}\delta^{(10)}(k+k')\,,
\ee
while for the open string vacuum on a $p$-brane we choose:
\be\label{eopennorm}
\langle k| c_{-1} c_0 c_1 \, e^{-2\phi}(0)|k'\rangle=(2\pi)^{p+1} \delta^{(p+1)}(k+k')\, .
\ee

We shall follow the convention of \cite{Sen:2021tpp} for computing disk amplitudes in the
presence of a (Euclidean) D$p$-brane.
Since this will be used extensively in our analysis, we shall briefly review the convention
for the amplitudes relevant for us.
Let $V_c$ be the unintegrated vertex operator of a closed string. Then the disk
one-point function of this closed string is given by
\be\label{edisk1}
\{ V_c\}= {1\over 2} \, \kappa \, T_p \, \langle c_0^- V_c \rangle_D\, ,
\ee
where we define
\be
c_0^\pm\equiv (c_0\pm \bar c_0)/2,
\qquad
b_0^\pm=b_0\pm \bar b_0,
\qquad
L_0^\pm=L_0\pm \bar L_0\,.
\ee
$\langle~\rangle$ denotes correlation function on the disk,
$\kappa=\sqrt{8\pi G}$
is the gravitational constant and $\ts_p$ is the tension of the
D$p$-brane, related to the string coupling $g_s$ via the relations
\be \label{ekappags}
\kappa = 2^3 \, \pi^{7/2}\, g_s,
\qquad
\ts_p ={1\over (2\pi)^p g_s}\, .
\ee
We also need to insert appropriate number of PCOs in the disk correlation
function in \refb{edisk1}
to make sure that total picture number of all the operators adds up to $-2$.
The factor of $\kappa$ was not present in eq.(3.26) of \cite{Sen:2021tpp} but has been
included since we shall be working with canonically normalized fields. On the other hand,
if we have a disk amplitude with one closed string and $n$ open strings inserted
on the disk, the amplitude takes the form (see eq.(3.28) of \cite{Sen:2021tpp}):
\be\label{edisk2}
\left\{V_c  \prod_{k=1}^n V_o^{(k)} \right\} =
\ve\, \pi\, \kappa\, T_p \,  \int \left\langle V_c  \prod_{k=1}^n V_o^{(k)} \right\rangle_D,
\ee
where $\ve$ is some power of $\I$,
$V_o^{(k)}$ are the open string vertex operators, with one unintegrated
and $(n-1)$ integrated vertex operators, and the integral runs over the locations of the
integrated vertex operators along the real axis. The factor of $\ve$ was not determined
in \cite{Sen:2021tpp} (see footnote 5 of \cite{Sen:2021tpp})
since there one needed four powers of the disk amplitude. However,
in our analysis we need two powers of the disk amplitude and for this we need the
actual value of $\ve$. It is shown in appendix \ref{se} that we have
\be \label{edisk3}
\ve=\I\, .
\ee
As in \refb{edisk1}, we need to insert appropriate number of
PCOs in \refb{edisk2} to ensure that the total picture number adds up to $-2$. Using the
factorization analysis of \cite{Sen:2021tpp}, it is easy to check that $\ve$ is independent of
the number $n$ of open string insertions, but we also check this in appendix \ref{se}.

When we carry out the analysis for D-branes wrapped on Calabi-Yau cycles, we shall use the large volume approximation
to regard the brane as locally flat.
Then to compute the amplitudes, we can use the ten-dimensional formul\ae\, where
the vertex operators do not carry any momentum along the brane.
In this situation, according to \refb{eopennorm}, we get a factor of $(2\pi)^{p+1}\delta^{(p+1)}(0)$ which
can be viewed as the volume of the brane.
We should interpret this as the statement that the result without this
factor needs to be integrated along the brane.

\subsection{Compactification} \label{scompact}

Let us now consider compactification of the theory on a CY 3-fold $\CY$.
We shall label the four-dimensional coordinates by
Greek indices $\mu,\nu\cdots$
and the six-dimensional indices by lower case bold-faced indices ${\bf i,j,k}\cdots$.
The gamma matrix conventions will be as follows.
Let us denote by $\wt\Gamma^{\bf i}$ the six-dimensional
$\gamma$-matrices and define
\be \label{edefgamma7}
\wt\Gamma = \I\, \wt \Gamma^4\cdots \wt\Gamma^9\, ,
\ee
so that
\be
\wt\Gamma^2=I_4, \qquad \{\wt\Gamma,\wt\Gamma^{\bf i}\}=0\, .
\ee
We also
denote by $\gamma^\mu$ the four-dimensional $\gamma$ matrices that commute with the
$\wt\Gamma^{\bf i}$'s. Then the ten-dimensional $\gamma$-matrices will be chosen as:
\be\label{etenfourgamma}
\Gamma^\mu = \wt\Gamma\otimes \gamma^\mu, \qquad \Gamma^{\bf i} = \wt\Gamma^{\bf i}
\otimes I_4\, .
\ee
We shall denote the six-dimensional spinor indices by $\alpha^{(6)}$ and the four-dimensional
spinor indices by $\alpha^{(4)}$ so that the ten-dimensional index $\alpha$ can be
regarded as the pair $(\alpha^{(6)},\alpha^{(4)})$. We shall  use the dotted and
undotted spinor index $\alpha,\dot\alpha$ for the four-dimensional spinor. In this convention
the components of
$\gamma^\mu$ acting on the undotted and dotted spinors will be labelled as
$(\gamma^\mu)_{\dalpha}^{~\beta}$ and $(\gamma^\mu)_{\alpha}^{~\dbeta}$ respectively.
They satisfy the relations:
\be
\{\gamma^\mu, \gamma^\nu\}=2\, \eta^{\mu\nu}  I_2,
\quad
(\gamma^\mu)_{\alpha\dbeta}=(\gamma^
\mu)_{\alpha}^{~\dgamma} \, \eps_{\dgamma\dbeta},
\quad
(\gamma^\mu)_{\dalpha\beta}=(\gamma^
\mu)_{\dalpha}^{~\gamma} \, \eps_{\gamma\beta},
\quad
(\gamma^\mu)_{\dalpha\beta}
=(\gamma^\mu)_{\alpha\dbeta}\, ,
\ee
with the understanding that in the product of $\gamma$-matrices, an (un)dotted index on
the top is contracted with an (un)dotted index at the bottom.\footnote{A similar convention
must also be followed for the product of six-dimensional gamma matrices.}
$I_2$ and $\eps$ represent respectively the $2\times 2$ identity matrix
and the $2\times 2$ antisymmetric matrix with $\eps_{12}= \eps_{\dot 1\dot 2}= 1$.
Although we shall use the indices $\alpha,\beta,\cdots$ to label both the ten-dimensional
spinor indices and four-dimensional undotted spinor indices, it should be clear
from the context whether a given spinor index $\alpha$ corresponds
to ten- or four-dimensional index.
Since the ten-dimensional spinors carry definite chirality, the chiralities of the four- and
the six-dimensional spinors should be correlated. Therefore, the dotted and undotted four-dimensional
spinor indices must be accompanied by six-dimensional spinors of opposite chiralities.
Furthermore this correlation should be opposite in the left and the right-moving sectors, since
in type IIA string theory the left and right-moving spinors in ten dimensions have opposite
chiralities.

CY 3-folds have a complex conjugate pair of covariantly constant spinors which we
shall denote by $\eta$ and $\bar\eta$. Covariant constancy ensures that $\bar\eta \eta$ is
a constant. We shall choose the normalization and chirality of $\eta$ such that
\be\label{eetanorm}
\bar\eta\, \eta = {1}, \qquad \wt\Gamma\eta=\eta\, .
\ee
The unbroken supersymmetry transformations in the compactified theory are related to the
supersymmetry transformations in the ten-dimensional theory coming from the
holomorphic sector by taking the ten-dimensional
supersymmetry transformation parameters to be the product of $\eta$ and an arbitrary undotted
spinor or $\bar\eta$ and an arbitrary dotted spinor. For supersymmetry transformations
coming from the anti-holomorphic sector, $\eta$ will be accompanied by a dotted spinor and
$\bar\eta$ will be accompanied by the undotted spinor.

The D-instanton contribution we shall analyze will be one or several Euclidean D2-branes
wrapped on some 3-cycle $\CG$ of $\CY$.
In order to preserve half of the supersymmetries, the specific 3-dimensional subspace that
the D2-brane wraps must be a special Lagrangian submanifold, but there may be more than one special
Lagrangian submanifold in a given homology class labelled by $\gamma$. The
D-instanton boundary
conditions break half of the space-time supersymmetry by relating the supersymmetry
transformation parameters associated with the holomorphic sector to that associated with
the anti-holomorphic sector. Associated with the broken supersymmetry there will be
Goldstino zero modes on the instanton which we shall label by two component
spinors $\tilde\chi^\alpha$ and
$\tilde\chi^\dalpha$. Since the zero modes are in one to one correspondence with
broken supersymmetry generators, which we shall identify with the supersymmetry generators in
the holomorphic sector, we can also represent them as ten-dimensional
spinors $\XX^\alpha$ and $\wh\XX^\alpha$, carrying opposite four-dimensional chiralities.
Using the dictionary given earlier, we can take
\be \label{exxdef}
\XX = \eta\otimes \tilde\chi^\alpha,
\qquad
\wh\XX=\bar\eta\otimes \tilde\chi^\dalpha\, .
\ee
Note that for each four-dimensional spinor $\tilde\chi^\alpha$ and
$\tilde\chi^\dalpha$ we have a zero mode, but once these spinors are given,
the components of the ten-dimensional spinors $\XX$ and
$\wh\XX$ are fixed by \refb{exxdef}. In particular, we have
$\wh\XX^\alpha (\Gamma^\mu)_{\alpha\beta} \XX^\beta =
\bigl(\bar\eta \wt\Gamma\eta \bigr)\( \tilde\chi^\dalpha (\gamma^\mu)_{\dalpha\beta}\tilde\chi^\beta\)$.

\subsection{Complex structure moduli space of Calabi-Yau threefold}
\label{subsec-MC}

We shall now review some useful properties of the moduli space describing complex structure deformations of
a CY threefold and evaluate a few integrals that will be needed for our analysis.

For a given CY threefold $\CY$,
let $\{A^\Lambda,B_\Lambda\}\in H_3(\CY)$ with $\Lambda=0,\dots,h^{2,1}(\CY)$ be a symplectic basis of 3-cycles such that
the only non-vanishing intersection numbers are
$A^\Lambda\cap B_\Sigma=\delta^\Lambda_\Sigma$.
We also introduce a dual basis of 3-forms $\{\alpha_\Lambda,\beta^\Lambda\}\in H^3(\IZ,\CY)$
such that
\be
\int_{A^\Lambda}\alpha_\Sigma=\delta^\Lambda_\Sigma,
\qquad
\int_{B_\Lambda}\beta^\Sigma=-\delta_\Lambda^\Sigma,
\qquad
\int_{B_\Lambda}\alpha_\Sigma=\int_{A^\Lambda}\beta^\Sigma=0.
\ee
The complex structure on $\CY$ is encoded into a covariantly constant holomorphic 3-form $\Omega$.
Expanding it in the basis introduced above
\be
\Omega=z^\Lambda\alpha_\Lambda-F_\Lambda \beta^\Lambda,
\label{expOm}
\ee
we get coefficients $z^\Lambda$ and $F_\Lambda$ given by
\be
z^\Lambda= \int_{A^\Lambda}\Omega,
\qquad
F_\Lambda =\int_{B_\Lambda}\Omega,
\ee
which describe deformations of the complex structure and parametrize a moduli space $\cM_C$.
As it is easy to show (see, e.g., \cite{Strominger:1990pd}), they are not independent.
In fact, the moduli space $\cM_C$ carries the structure of a projective special K\"ahler manifold which means
that (locally) there exists a homogeneous of degree two holomorphic function $F(z)$ such that $F_\Lambda=\p_{z^\Lambda}F$
and the metric on $\cM_C$ is captured by the following K\"ahler potential
\be
\begin{split}
&\,\cK=-\log K,
\qquad
K= \I \int_{\CY}\bar\Omega\wedge \Omega=z^\Lambda N_{\Lambda\Sigma} \bz^\Sigma,
\\
&\quad N_{\Lambda\Sigma}\equiv  -2\Im F_{\Lambda\Sigma},
\qquad
F_{\Lambda\Sigma}\equiv \p_{z^\Lambda}\p_{z^\Sigma} F\, ,
\label{def-Kahlerpot}
\end{split}
\ee
where we have used the Riemann bilinear identity
\be
\int_{\CY}\chi\wedge \theta=\sum_\Lambda\(\int_{A^\Lambda}\chi\int_{B_\Lambda}\theta-\int_{B_\Lambda}\chi\int_{A^\Lambda}\theta\)
\label{Riemident}
\ee
to evaluate the integral over $\CY$.
Note that under a holomorphic rescaling of the coordinates $z^\Lambda\to f(z) z^\Lambda$,
the K\"ahler potential transforms by a K\"ahler transformation $\cK\to \cK-\log f-\log \bar f$ which does not affect the metric.
This originates from the fact that $\Omega$ is defined only up to such a rescaling. As a result,
the complex structures are parametrized by only $h^{2,1}(\CY)$ coordinates which can be taken to be $z^a/z^0$ with
$a=1,\dots,h^{2,1}$. We will work in the gauge $z^0=1$ and use $z^a$ as independent coordinates on $\cM_C$.

It is useful to note that there is also an alternative basis in the space of 3-forms on $\CY$
adapted to the Hodge decomposition $H^3=H^{3,0}\oplus H^{2,1}\oplus H^{1,2}\oplus H^{0,3}$.
It is given by $\{\Omega,\chi_a,\bchi_a,\bar\Omega\}$ where
\be
\chi_a(z)=\p_{z^a}\Omega+\Omega\, \cK_a, \qquad \cK_a\equiv \p_{z^a}\cK.
\label{defchia}
\ee
The relative coefficient between the two terms is fixed by requiring $\int\bar\Omega\wedge
\chi_a=0$.

Let us now consider an arbitrary 3-form $C$. It can be expanded in either of the two bases introduced above.
The two decompositions give rise to two sets of coefficients:
\be
C=\zeta^\Lambda \alpha_\Lambda-\tzeta_\Lambda\beta^\Lambda
\label{decC3}
\ee
and
\be
C=\rho \,\Omega+\vrh^a\chi_a+\bar\vrh^{a}\bchi_{a}+\bar\rho\,\bar\Omega.
\label{decC-alt}
\ee
A relation between them can be found by substituting \eqref{expOm} and \eqref{defchia} into \eqref{decC-alt}, which gives
\be
\begin{split}
\zeta^0=&\,2\Re(\rho+\vrh^a \cK_a),
\\
\zeta^a=&\,2\Re \bigl((\rho +\vrh^b \cK_b)z^a+\vrh^a\bigr),
\\
\tzeta_a=&\, 2\Re \bigl((\rho +\vrh^b \cK_b) F_a+F_{ab} \vrh^b\bigr),
\\
\tzeta_0=&\, 2\Re \bigl((\rho+\vrh^a \cK_a) F_0+\vrh^a F_{a0}\bigr).
\end{split}
\label{rel-zeta-rho}
\ee
Setting $\vrh^0\equiv 0$, these relations can be written in a compact form as
\be
\begin{split}
\zeta^\Lambda=&\, 2\Re(\xi^\Lambda),
\qquad
\xi^\Lambda=\vrh^\Lambda+(\rho+\vrh^a\cK_a)z^\Lambda,
\\
\tzeta_\Lambda=&\, 2\Re(F_{\Lambda\Sigma}\xi^\Sigma).
\end{split}
\label{rel-zeta-rho2}
\ee
In particular, one finds
\be
z^\Lambda\tzeta_\Lambda-F_\Lambda\zeta^\Lambda=\I K\bar\rho.
\label{nicecomb}
\ee

Finally, let us choose a 3-cycle $\CG=q_\Lambda A^\Lambda -p^\Lambda B_\Lambda$ and its dual 3-form
$\omega_\gamma =p^\Lambda\alpha_\Lambda-q_\Lambda \beta^\Lambda$, parametrized by a vector of integers $\gamma=(p^\Lambda,q_\Lambda)$.
We are interested in evaluating three integrals over this cycle
\be
\Theta_\gamma\equiv \int_{\CG} C,
\qquad
\Theta^*_\gamma\equiv \int_{\CG} \star\, C,
\qquad
\Theta^J_\gamma\equiv \int_{\CG} J(C),
\label{treeints}
\ee
where $J$ is the complex structure and
\be \label{edefjc}
J(C)_{\bf ijk}={1\over 3} \( J_{{\bf i}}^{\bf m}C_{\bf mjk} + J_{{\bf j}}^{\bf m}C_{\bf mki}
+ J_{{\bf k}}^{\bf m}C_{\bf mij}\) .
\ee
The first integral is trivially evaluated using \eqref{decC3},
\be
\Theta_\gamma=q_\Lambda\zeta^\Lambda-p^\Lambda\tzeta_\Lambda.
\label{intCC}
\ee

The second integral can be found rewriting it as an integral over the whole CY 3-fold:
\be
\Theta^*_\gamma\equiv -\int_{\CY} C\wedge \star\omega_\gamma
\ee
and using the identities \cite{Suzuki:1995rt,Grimm:2004ua}
\be
\begin{split}
\int_{\CY}\alpha_\Lambda\wedge \star \alpha_\Sigma=&\, -\Bigl(\Im \cM+\Re \cM(\Im \cM)^{-1}\Re\cM\Bigr)_{\Lambda\Sigma},
\\
\int_{\CY}\beta^\Lambda\wedge \star \beta^\Sigma=&\, -{(\Im \cM)^{-1}}^{\Lambda\Sigma},
\\
\int_{\CY}\alpha_\Lambda\wedge \star \beta^\Sigma=&\, -\Bigl(\Re \cM(\Im \cM)^{-1}\Bigr)_\Lambda^{~\Sigma},
\end{split}
\ee
where
\be
\cM_{\Lambda\Sigma}=\bF_{\Lambda\Sigma}-\I\, \frac{(Nz)_\Lambda (Nz)_\Sigma}{(zN\bz)}\, .
\label{def-cM}
\ee
It is straightforward to find
\be
\hf {(\Im \cM)^{-1}}^{\Lambda\Sigma}=N^{\Lambda\Sigma}-\frac{1}{K} \(z^\Lambda\bz^\Sigma+\bz^\Lambda z^\Sigma\),
\ee
where $N^{\Lambda\Sigma}$ is the inverse of $N_{\Lambda\Sigma}$, so that
\bea
\hf\, \Bigl(\Re \cM(\Im \cM)^{-1}\Bigr)_\Lambda^{~\Sigma}&=&
\Re F_{\Lambda \Theta} N^{\Theta\Sigma}-\frac{1}{K} \(F_\Lambda\bz^\Sigma+\bF_\Lambda z^\Sigma\),
\\
\Bigl(\Im \cM+\Re \cM(\Im \cM)^{-1}\Re\cM\Bigr)_{\Lambda\Sigma}&=& \hf \, N_{\Lambda\Sigma}
+2\(\Re F\, N^{-1} \,\Re F\)_{\Lambda\Sigma}-\frac{2}{K}\(F_\Lambda\bF_\Sigma+\bF_\Lambda F_\Sigma\).
\non
\eea
These results can be used to obtain
\bea
\Theta^*_\gamma
&=&2 \mathscr{C}_\gamma
-\frac{4}{K}\, \Re\[\bZ_\gamma(z^\Lambda\tzeta_\Lambda-F_\Lambda\zeta^\Lambda)\],
\label{resThetastar}
\eea
where we introduced
\be
Z_\gamma=\int_{\CG}\Omega=q_\Lambda z^\Lambda-p^\Lambda F_\Lambda
\label{defZg}
\ee
and
\be
\mathscr{C}_\gamma = N^{\Lambda\Sigma}\(q_\Lambda-\Re F_{\Lambda\Xi}p^\Xi\)\(\tzeta_\Sigma-\Re F_{\Sigma\Theta}\zeta^\Theta\)
+\frac14\, N_{\Lambda\Sigma}\,p^\Lambda\,\zeta^\Sigma.
\label{def-niceC}
\ee
Using \refb{rel-zeta-rho2}, $\mathscr{C}_\gamma$ may also be expressed as
\be
\begin{split}
\mathscr{C}_\gamma
=&\,  \Im\((q_\Lambda-F_{\Lambda\Sigma}p^\Sigma)\xi^\Lambda\)
\\
=&\,  \Im\(\vrh^a\p_{z^a} Z_\gamma+(\rho+\vrh^a\cK_a)Z_\gamma\).
\end{split}
\label{nicecomb2}
\ee

To find the last integral in \eqref{treeints}, we have to evaluate the action of the complex structure
on a 3-form. This is where the basis adapted to the Hodge decomposition becomes very convenient.
Indeed, using \eqref{decC-alt}, one finds
\be
J(C)=\I\(\rho \,\Omega-\bar\rho\,\bar\Omega\)+\frac{\I}{3}\(\vrh^a\chi_a-\bar\vrh^a\bchi_a\).
\label{decCJ-alt}
\ee
Taking into account \eqref{defZg}, this gives
\be
\Theta^J_\gamma= -2\Im(\rho Z_\gamma)-\frac{2}{3}\,\Im\Bigr[\vrh^a(\p_{z^a}Z_\gamma+\cK_a Z_\gamma)\Bigl].
\ee
It remains to express this result in terms of $\zeta^\Lambda,\tzeta_\Lambda$.
Using \refb{nicecomb} and \refb{nicecomb2},
one immediately obtains
\be
\Theta^J_\gamma
= -\frac{4}{3K}\, \Re\[\bZ_\gamma(z^\Lambda\tzeta_\Lambda-F_\Lambda \zeta^\Lambda)\]
-\frac{2}{3}\,\mathscr{C}_\gamma.
\label{resThetaJ}
\ee

\section{D-instanton corrected hypermultiplet metric from twistors}
\label{sduality}

In this section we present a prediction for the instanton corrected hypermultiplet metric
following from a combination of various dualities, wall crossing and mirror symmetry and obtained
using a twistorial description of quaternionic manifolds.

First of all, let us recall that after compactification on a CY threefold $\CY$,
the low energy effective theory contains $n=h^{2,1}(\CY) +1$ hypermultiplets. Each hypermultiplet comprises 4 real scalars which
parametrize a $4n$-dimensional quaternion-K\"ahler moduli space $\cM_H$. In type IIA theory
we have the following hypermultiplet scalar fields:
\begin{itemize}
\item
the fields $z^a$ ($a=1,\dots,h^{2,1}$) describing deformations of the complex structure of $\CY$ which were reviewed
in \S\ref{subsec-MC};

\item
the RR fields $\zeta^\Lambda,\tzeta_\Lambda$ ($\Lambda=0,\dots,h^{2,1}$)
arising as period integrals of the RR 3-form of type IIA string theory over a symplectic basis of cycles in $H_3(\CY,\IZ)$
(see \eqref{decC3});

\item
the four-dimensional dilaton $r=e^\phi$;

\item
the NS-axion $\sigma$ which is dual to the $B$-field in four dimensions.
\end{itemize}

At tree level the metric on $\cM_H$ can be obtained by Kaluza-Klein reduction from the ten-dimensional type IIA supergravity action
or via the c-map \cite{Cecotti:1989qn,Ferrara:1989ik}.
The resulting metric is completely determined by the holomorphic prepotential $F(z)$ on the space of complex structure deformations
and is given by
\be
\begin{split}
\de s_{\rm tree}^2=& \, \frac{1}{r^2}\,\de r^2
-\frac{1 }{2r}\, \[(\Im\cM)^{-1}\]^{\Lambda\Sigma}\(\de\tzeta_\Lambda -\cM_{\Lambda\Lambda'}\de\zeta^{\Lambda'}\)
\(\de\tzeta_\Sigma -\bar\cM_{\Sigma\Sigma'}\de\zeta^{\Sigma'}\)
\\ &
+ \frac{1}{16 r^2} \(\de\sigma+\tzeta_\Lambda\de\zeta^\Lambda-\zeta^\Lambda\de\tzeta_\Lambda\)^2
+4\cK_{a\bar b}\de z^a \de \bz^{\bar b}  ,
\end{split}
\label{hypmettree}
\ee
where $\cK_{a\bar b}$ are the second derivatives of the K\"ahler potential $\cK$ given in
\eqref{def-Kahlerpot} and
the matrix $\cM_{\Lambda\Sigma}$ is defined in \eqref{def-cM}.  We have used conventions
where the Einstein-Hilbert Lagrangian density
and the kinetic term for the hypermultiplet scalars, collectively denoted by $\vp^i$ with $i=1,\dots,4n$, read as
\be
L_{\rm kin}=- R(g)-\hf\, G_{ij}(\vp) \, \p_\mu \vp^i \p^\mu \vp^j.
\label{kinL}
\ee
Note that $\cM_H$ carries an action of the symplectic group. In particular, $(z^\Lambda,F_\Lambda)$ and $(\zeta^\Lambda,\tzeta_\Lambda)$
transform in the vector representation, whereas $r$ and $\sigma$ are symplectic invariant.\footnote{Recall that we work in the gauge $z^0=1$
which is not invariant under a generic symplectic transformation. Besides, the prepotential $F$ is not invariant either.}

The metric \eqref{hypmettree} receives a one-loop correction and instanton corrections from Euclidean D2-branes
and NS5-branes wrapping 3-cycles and the whole CY, respectively.
In this paper we restrict our attention only to the corrections due to the first type of instantons.
These instantons are characterized by integer charges $\gamma=(p^\Lambda,q_\Lambda)$ labelling homology classes in $H_3(\CY)$.
The charge lattice carries an integer pairing
\be
\langle\gamma,\gamma'\rangle=q_\Lambda p'^\Lambda-q'_\Lambda p^\Lambda.
\label{DSZ}
\ee
In the small string coupling limit each instanton of charge $\gamma$ produces a factor
$\Omega_\gamma\, e^{-\cT_\gamma}$ where $\Omega_\gamma$ is the integer valued generalized Donaldson-Thomas (DT) invariant of $\CY$,
which roughly counts the number of supersymmetric cycles in homology class $\gamma$, and
\be
\cT_\gamma=8\pi\sqrt{\frac{r}{K}}\, |Z_\gamma|+2\pi\I \Theta_\gamma.
\label{actD2}
\ee
Here $Z_\gamma$, defined in \refb{defZg},
is the central charge of the supersymmetry subalgebra
left unbroken by the instanton,
and $\Theta_\gamma$ is a linear combination of RR axion fields given by \eqref{intCC}.
The full D-instanton corrected metric includes in addition a $g_s$-perturbative expansion around instantons as well as
multi-instanton contributions.

A formal\footnote{It is formal because it involves various expansions which are at best asymptotic.}
expression for the metric incorporating contributions of all D2-instantons with all perturbative expansions around them has been found
in \cite[Eq.(2.31)]{Alexandrov:2017mgi}. It was derived from a twistorial construction \cite{Alexandrov:2008gh,Alexandrov:2009zh}
that encodes the D-instantons into the so called holomorphic contact structure on the twistor space, a $\CP$ bundle over $\cM_H$
\cite{MR1327157,Alexandrov:2008nk}. Although this construction allows to capture all D-instantons by
a single holomorphic function (dilogarithm), to extract the metric
from it requires solving an integral equation of Thermodynamic Bethe Ansatz (TBA) type.
Generically, this can be done only by a perturbative expansion in
powers of DT invariants
and leads to a highly complicated result.
We will argue now that for our purposes it is actually not needed and this complication can be avoided.

The point is that contributions non-linear in DT invariants, in particular,
those generated by an iterative solution of the TBA equation mentioned above,
involve products of the form $\prod_i \Om{\gamma_i}\, e^{-k_i\cT_{\gamma_i}}$
with $k_i\in\ZZZ^+$.
They are to be compared with the contribution of a single instanton of charge
$\gamma=\sum_i k_i \gamma_i$.
Since $\Theta_\gamma$ depends linearly on charge, the two contributions
have the same axionic coupling and thus belong to the same topological sector.
However, $\sum_i  k_i |Z_{\gamma_i}|\geq |Z_\gamma|$ with equality possible only if all central charges are aligned.
This can happen only at very special loci in the moduli space unless all the $\gamma_i$'s are aligned.
If two charges $\gamma,\gamma'$ are mutually non-local, which means that they satisfy $\langle\gamma ,\gamma'\rangle\ne 0$,
the loci with $\arg Z_\gamma=\arg Z_{\gamma'}$ are known as walls of marginal stability and coincide with codimension
1 hypersurfaces in $\cM_C$ where DT invariants jump, so that they are not even well-defined at these walls.
The metric however is smooth across the walls and we can safely assume that we stay always away from any such wall.
As a result, contributions from mutually non-local multi-instantons turn out to be
exponentially suppressed compared to one-instantons of the same total charge.
Since we are interested here only in the leading $g_s$ corrections in a given topological sector,
such multi-instantons can be ignored.

This leads to crucial simplifications as we do not need anymore to solve any integral equations ---
the first iteration gives rise to an exact solution.
An explicit expression for the metric in this approximation has been obtained in
\cite{Alexandrov:2014sya} (see \cite{Cortes:2021vdm} for a recent construction of the same metric by mathematicians)
and we present it in appendix \ref{ap-metric}.
In the same appendix we argue that multi-instanton contributions non-linear in DT invariants
for mutually local charges are also suppressed, this time by a power of
the string coupling\footnote{In fact, the argument in the previous paragraph shows that multi-instantons
with mutually local charges, but central charges non-aligned, are actually suppressed exponentially.},
and extract instanton corrections to the tree level metric \eqref{hypmettree}
that are linear in DT invariants.
The result can be conveniently represented in the following form
\be
\begin{split}
\de s_{\rm inst}^2=&\, \sum_\gamma \frac{\Omega_\gamma\, \Sigma_\gamma}{16 r\sqrt{2\pi \cTR_\gamma}}\Biggl[
\frac{|Z_\gamma|^2}{r K}
\(\de\sigma+\tzeta_\Lambda\de\zeta^\Lambda-\zeta^\Lambda\de\tzeta_\Lambda-\frac{4\I \sqrt{rK}}{|Z_\gamma|}\,\cC_\gamma
+8 r\Im \p\log\frac{Z_\gamma}{K} \)^2
\\
&
-\frac{1}{\pi^2}\(\de\cT_\gamma-8\pi|Z_\gamma|\sqrt{\frac{r}{K}}\, \de\log\frac{|Z_\gamma|^2}{K}
-\frac{16\pi\I}{K}\,\Im\[\bZ_\gamma(z^\Lambda\de\tzeta_\Lambda-F_\Lambda\de\zeta^\Lambda)\] \)\de \cT_\gamma\Biggr],
\end{split}
\label{square}
\ee
where the sum goes over all charge lattice, $\p=\de z^a\p_{z^a}$ is the Dolbeault holomorphic differential on $\cM_C$,
\be
\cC_\gamma= N^{\Lambda\Sigma}\(q_\Lambda-\Re F_{\Lambda\Xi}p^\Xi\)\(\de\tzeta_\Sigma-\Re F_{\Sigma\Theta}\de\zeta^\Theta\)
+\frac14\, N_{\Lambda\Sigma}\,p^\Lambda\,\de\zeta^\Sigma
\label{connC}
\ee
is a combination of differentials of the RR fields (\textit{cf.} equation~\eqref{def-niceC}),
$\cTR_\gamma=\Re\cT_\gamma$, and the function $\Sigma_\gamma$ is defined as:
\be
\Sigma_\gamma=\sum_{k=1}^\infty \frac{\sigma_{k\gamma}}{\sqrt{k}}\,  e^{-k\cT_\gamma}.
\label{defSumk}
\ee
Here $\sigma_\gamma$ is a sign factor, known as quadratic refinement, which satisfies the defining relation
$\sigma_{\gamma_1}\sigma_{\gamma_2}=(-1)^{\langle\gamma_1,\gamma_2\rangle}\sigma_{\gamma_1+\gamma_2}$.
A few remarks are in  order:

\begin{itemize}
\item
It is easy to see that the terms real in the square brackets in \eqref{square} are even with respect to $\gamma\to-\gamma$,
whereas the pure imaginary terms are odd. Therefore, since $\bar\TT_\gamma=\TT_{-\gamma}$,
the sum over charges produces a real expression, as it should be.

\item
Note that in contrast to multi-instanton contributions non-linear in DT invariants, the
multi-instanton contributions linear in DT invariants, captured by the $k>1$ terms in \eqref{defSumk},
are not suppressed by powers of string coupling in the small string coupling limit.
We shall see that these contributions come from multiple branes fused together
in a sense that will be made precise at the end of \S\ref{snonmmulti}.

\item
Due to the summation over $k$ in the definition of $\Sigma_\gamma$, the term with a given
$\gamma$ in \refb{square} contributes to multiple topological sectors labelled by integer
multiples of $\gamma$. We can rectify this by rewriting the formula where we replace
$\Sigma_\gamma$ by the $k=1$ term in the sum and replace $\Omega_\gamma$
by the rational invariant $\bOm{\gamma}$ defined as:
\be
\bOm{\gamma}=\sum_{d|\gamma}d^{-2}\Omega_{\gamma/d}.
\label{def-bOm}
\ee

\item
An important feature of the result \eqref{square} is that the contribution of each charge is a full square
up to terms $O(\de \cT_\gamma)$. As we shall see, the latter
terms can be removed by field redefinition at the leading order
and therefore are not relevant for our purposes.
On the other hand, the full square structure is precisely what follows from the analysis of these amplitudes
in \S\ref{sec-maincomput}.

\item
Another useful remark is that a simple rescaling of fields makes it clear that both metrics \eqref{hypmettree}
and \eqref{square} have a uniform scaling in the small string coupling limit. Indeed, let us redefine fields as
\be\label{rescalefields}
r= r'/g_{(4)}^{2},
\qquad
\zeta^\Lambda=\zeta'^\Lambda /g_{(4)} ,
\qquad
\tzeta_\Lambda= \tzeta'_\Lambda/g_{(4)},
\qquad
\sigma=\sigma'/g_{(4)}^2, \qquad z^a=z^{\prime a},
\ee
where $g_{(4)}$ is the four-dimensional string coupling which can be thought as the
vacuum expectation value of $r^{-1/2}$. If we rewrite the metric in terms of the primed
variables, then
it is easy to see that $g_{(4)}$ drops out from the perturbative metric \eqref{hypmettree},
while in the instanton contribution \eqref{square} we get
two overall factors: $g_{(4)}^{1/2}$ in front
and $g_{(4)}^{-1}$ in the instanton action.
Thus, keeping the primed fields fixed in the weak coupling limit,
the metrics have an overall scaling $g_{(4)}^{0}$ and $g_{(4)}^{1/2}\, e^{-g_{(4)}^{-1}\cT'_\gamma}$, respectively.

\end{itemize}

To finish this section, let us establish a precise relation which holds at leading order between the four-dimensional dilaton $r$
appearing as one of the coordinates on $\cM_H$ and the closed and open string couplings,
$g_s$ and $g_o$, that will appear in the amplitude computations.
To this end, let us note that both couplings are related to the real part of the instanton action.
On the one hand, it has the canonical expression
as a product of the D2-brane tension and the volume
$V_\gamma$ of the 3-cycle $\CG$ wrapped by the brane
\be
\cTR_\gamma=\ts_2 V_\gamma=\frac{V_\gamma}{4\pi^2 g_s}\, ,
\label{Realact}
\ee
where we used \eqref{ekappags} for $p=2$. On the other hand, it can be written
as \cite{Polchinski:1998rq,Polchinski:1998rr,Sen:1999xm,Sen:2021tpp}
\be \label{e35xy}
\cTR_\gamma ={1\over 2\pi^2 g_o^2}\, .
\ee
In the following we shall regard $g_s$ and $g_{(4)}$ as constants, related to the vacuum
expectation
values of the fields. Therefore, \refb{Realact} and \refb{e35xy}, as well as \refb{rel-rgs} and
\refb{expr-g0} below, should be regarded as relations involving vacuum expectation
values. However, in general, {\it e.g.} in
\refb{square},  we shall regard $\TT_\gamma$ and $Z_\gamma$ as functions of fields.

As shown in \cite{Becker:1995kb}, the BPS condition requires the cycle $\CG$
to be special Lagrangian,
i.e. the pullback of the K\"ahler form $\omega$ on $\CY$ to $\CG$ must vanish, and the pullback of the holomorphic 3-form $\Omega$ has
to be proportional to the volume form of $\CG$ which we denote by $v_\gamma$:
\be
\omega|_{\CG}=0,
\qquad
\Omega|_{\CG}=f \,v_\gamma,
\label{BPScycle}
\ee
where $f=e^{\I\theta} e^{\hf(\cK'-\cK)}$, $\theta$ is a real constant and $\cK$ and $\cK'$
are K\"ahler potentials on the space of complex and K\"ahler structure
deformations of CY, respectively. Integrating the second relation over $\CG$ and using equation~\eqref{defZg},
one finds
\be
f\, V_\gamma=Z_\gamma.
\label{rel-fVZ}
\ee
Then taking into account the explicit form of $f$
and that in the large volume approximation
\be
\cK'=-\log\(\frac43\int_\CY \omega\wedge\omega\wedge \omega\)=-\log(8V),
\ee
where $V$ is the volume of $\CY$, the absolute value of the relation \eqref{rel-fVZ} becomes
\be
V_\gamma=\sqrt{\frac{8V}{K}}\, |Z_\gamma|.
\ee
Substituting this into \eqref{Realact} and comparing with \eqref{actD2}, one finally obtains
\be
r=\frac{V}{2(2\pi)^6 g_s^2}=\frac{\pi V}{2\kappa^2}\, ,
\label{rel-rgs}
\ee
while \eqref{e35xy} implies
\be
g_o^2=\frac{2g_s}{V_\gamma}=\frac{g_s}{\sqrt{2}|Z_\gamma|}\, \sqrt{\frac{K}{V}}=\frac{\sqrt{K}}{16\pi^3|Z_\gamma| \sqrt{r}}\, .
\label{expr-g0}
\ee

\section{Normalization of the instanton amplitudes} \label{sfour}

The leading order contribution to an instanton amplitude involves the product of the
usual factor of $\I$ accompanying all amplitudes,
exponential of the annulus amplitude and a set of disk amplitudes. In this section we
shall focus on the computation of the exponential of the annulus amplitude that fixes
the overall normalization of the amplitude.
First, we consider the case of a single instanton, and then turn to the case of $k$ identical instantons.

\subsection{Single instanton} \label{sfourone}

Since our analysis in this section will follow closely to that in \cite{Sen:2021tpp}, we shall
begin by briefly reviewing the steps followed there.
\begin{enumerate}
\item Formally the  annulus
amplitude vanishes due to supersymmetry \cite{Blumenhagen:2009qh}, but the zero
mode contributions from the NS and R sector suffer from infrared divergences. We regulate
the infrared divergences by putting slightly shifted boundary condition on the two boundaries
of the annulus and express the exponential of the annulus amplitude  as a `path integral'
over open string modes. The regulator produces a non-zero $L_0$ eigenvalue
$h$ for all the modes which otherwise would have had zero $L_0$ eigenvalue.
\label{p1}
\item The variables involved in the path integral include the usual bosonic zero modes
associated with the breaking of space-time translation invariance by the
instanton and the fermionic zero modes
associated with broken supersymmetry. However there are two additional Grassmann odd
modes which can be identified as the Faddeev-Popov ghosts arising from Siegel gauge
fixing. In the limit when we remove the regulator by taking the $h\to 0$ limit, the
ghost action vanishes, indicating a breakdown of the gauge fixing procedure. We
remedy this by working with the original gauge invariant form of the path integral
whose gauge fixed version produces the exponential of the annulus amplitude. This
gauge invariant version does not have the integral over the ghost modes, but has an
integration over an extra Grassmann even variable that was set to zero in the Siegel
gauge and also contains a division by an integral over the gauge parameter that was
removed by the gauge fixing procedure. \label{p2}
\item We can explicitly perform the integration over the
extra Grassmann even mode in the numerator.
The remaining bosonic zero modes are related to the position of the D-instanton in
space-time. The precise relation is found by comparing the dependence of an
amplitude on these zero
modes with the expected dependence of the
amplitude on the position of the D-instanton. Using this result we express the integral
over these bosonic zero modes in terms of the integration over the instanton position.
The latter integration eventually
produces the usual momentum conserving delta function. Similarly the
integral over the gauge transformation parameter in the denominator is performed
by first finding the relation between the gauge transformation parameter and the
parameter $\wt\theta$
labelling the rigid $U(1)$ symmetry on the instanton under which an open
string with one end on the instanton picks up a phase $e^{i\wt\theta}$.
Once this relation is found, the
factor in the denominator can be expressed as an integral over the rigid U(1) transformation
parameter $\wt\theta$. The latter integral gives $2\pi$ since $\wt\theta$ has period $2\pi$.
\item Finally the integration over the fermion zero modes associated with the
supersymmetries broken by the instanton tells us that in order to get a non-vanishing
amplitude, we must sprinkle the vertex operators associated with these zero modes
in the rest of the factors in the amplitude. Once this is done, the integration over the
zero modes can be performed using the standard rules for Grassmann integration.
\label{p4}
\end{enumerate}

We shall now give the results obtained during different steps.
The D-instanton we shall analyze is a Euclidean D2-brane wrapped on a special Lagrangian 3-cycle.
We shall assume the 3-cycle to be rigid so that the only bosonic
zero modes on the instanton are those associated with translation invariance of the instanton
along the non-compact directions. The vanishing of the annulus amplitude now fixes the number
of fermionic zero modes. After taking into account the pair of ghost zero modes that arise
in fixing the Siegel gauge, one can show that the only fermion zero modes left are the
four zero modes associated with broken supersymmetry. With this information,
at step \ref{p1} we arrive at the following expression for the overall normalization:
\be\label{estrange}
\NN=\I \int \left\{\prod_{\mu=0}^3 {d\xi^\mu\over \sqrt{2\pi}}\right\}\,  dp\,  dq\,
\exp\left[-{1\over 2} h \sum_{\mu=0}^3 \xi_\mu \xi^\mu - h \, p\, q\right]
\int \prod_{\delta,\ddelta=1}^{2}
d\chi^\delta \, d\chi^\ddelta\, \exp\left[m_{\alpha\dbeta} \chi^\alpha \chi^\dbeta
\right] .
\ee
Here $\xi^\mu$ are the Grassmann even variables
associated with broken translation symmetry,
$\chi^\alpha$ and $\chi^\dalpha$ are the Grassmann odd variables
associated with broken supersymmetry, and
$p,q$ are the Grassmann odd Faddeev-Popov ghosts. $m_{\alpha\dbeta}$ is a
matrix with $|\det m|=h$.
\refb{estrange}
differs from eq.(4.7) of \cite{Sen:2021tpp} in that here $\mu$ ranges from 0 to 3 instead of
from 0 to 9,
reflecting the fact that the instanton can move only along the four non-compact space-time
directions and that $\alpha,\dalpha$ together range
over only 4 values instead of 16, reflecting the fact that
the instanton breaks only 4 out of 8 supersymmetries. For $h>0$, the integrals appearing in
\refb{estrange} can be carried out explicitly, and give the result $\I$, reflecting the vanishing
of the annulus amplitude.

At the end of step \ref{p2} we arrive at the following integral in the $h\to 0$ limit:
 \be
 \NN=\I \int_{-\infty}^\infty d\phi \, e^{-\phi^2/4}\,
 \int \prod_{\mu=0}^3 {d\xi^\mu\over \sqrt{2\pi}}\, \int \prod_{\delta,\ddelta=1}^{2}
d\chi^\delta \, d\chi^\ddelta\,
\Bigg/ \int d\theta\, ,
\ee
where $\phi$ is the extra mode that appears when we `unfix' Siegel gauge and $\theta$
represents the gauge transformation parameter. This part of the analysis is identical to that in
\cite{Sen:2021tpp}.
After carrying out integration over $\phi$ we get the
analog of eq.(4.32) of \cite{Sen:2021tpp}:
\be\label{enewernn}
 \NN=\I \, (2\pi)^{-2}\, 2\sqrt \pi\,
 \int \prod_{\mu=0}^3 d\xi^\mu \,\int \prod_{\delta,\ddelta=1}^{2}
d\chi^\delta \, d\chi^\ddelta
\Bigg/ \int d\theta\, .
\ee

The procedure outlined in step 3 gives the analog of eq.(4.38) of \cite{Sen:2021tpp}:
\be \label{eximurel}
\xi^\mu = g_o^{-1} \, \pi^{-1}\, 2^{-1/2}\, \wt \xi^\mu\, ,
\ee
where $\wt\xi^\mu$ is the space-time coordinate of the D-instanton and $g_o$ is
the open string coupling constant.
Also the relation between $\theta$ and $\wt\theta$ takes the form $\theta= 2\wt\theta/g_o$,
leading to the analog of eq. (4.45) of \cite{Sen:2021tpp}:
\be\label{e3.6aa}
\int d\theta = 4\pi / g_o\, .
\ee
Using these results, we get:
\be \label{enewnntwo}
\NN =\I
g_o^{-4} \, \pi^{-4}\, 2^{-2}\, (2\pi)^{-2} \, 2\sqrt \pi\,  {g_o\over 4\pi}\, \int
\prod_\mu d\wt\xi^\mu\, \int \prod_{\delta,\ddelta=1}^{2}
d\chi^\delta \, d\chi^\ddelta\, .
\ee

The fermion zero modes $\chi^\alpha$, $\chi^\dbeta$ appearing in \refb{estrange}
are canonically normalized so that their
vertex operators are accompanied by a factor of the open string coupling
constant $g_o$.
It will be more convenient for our analysis to define new variables $\tilde\chi^\alpha$,
$\tilde\chi^\dbeta$
by absorbing the factor of $g_o$ so that the vertex operators do not carry any such factor.
To this end we define:
\be \label{e1.6}
\tilde\chi^\alpha = g_o \, \chi^\alpha, \qquad \tilde\chi^\dbeta = g_o \, \chi^\dbeta, \qquad
\int \prod_{\delta,\ddelta=1}^{2}
d\chi^\delta \, d\chi^\ddelta\,   = g_o^{4}
\int \prod_{\delta,\ddelta=1}^{2}
d\tilde\chi^\delta \, d\tilde\chi^\ddelta \, ,
\ee
From \refb{enewnntwo} and
\refb{e1.6} we get the analog of eq.(4.49) of \cite{Sen:2021tpp}:
\be
\NN = \I g_o\, 2^{-5} \, \pi^{-13/2} \, \int
\prod_\mu d\wt\xi^\mu\,\int\prod_{\delta,\ddelta=1}^{2}
d\tilde\chi^\delta \, d\tilde\chi^\ddelta\,.
\ee

We need to multiply this by the product of disk amplitudes of external closed strings (one on each disk
to get the lowest order term) and insertion of the vertex operators of the open string
fermion zero modes $\tilde\chi^\alpha$, $\tilde\chi^\dbeta$,
distributed on the disks in all possible ways.
Then we use
\be \label{esignchoice}
\int \left\{ \prod_{\delta,\ddelta=1}^{2}
d\tilde\chi^\delta \, d\tilde\chi^\ddelta\right\}
\tilde\chi^\alpha \tilde\chi^\dalpha \tilde\chi^\beta  \tilde\chi^\dbeta
=\eps^{\alpha\beta}\eps^{\dalpha\dbeta}\, .
\ee
Therefore, if we denote  by
$A_{\alpha\dalpha\beta\dbeta}\, \prod_i e^{\I p_i.\wt\xi}$
the product of disk amplitudes with closed string insertions carrying momenta
$p_1,p_2,\cdots$ and open string zero mode $\tilde\chi^\alpha,
\tilde\chi^\dalpha, \tilde\chi^\beta,
\tilde\chi^\dbeta$ insertions,
then the final result for the amplitude takes the form:
\be\label{e1.9}
\I\,  e^{-\TT_\gamma} \NN^{(0)}_1
(2\pi)^4 \delta^{(4)}\left(\sum_i p_i\right) \AAA_1,
\qquad
\NN^{(0)}_1=g_o\, 2^{-5}  \pi^{-13/2} ,
\qquad \AAA_1= \eps^{\alpha\beta} \eps^{\dalpha\dbeta}
A_{\alpha\dalpha\beta
\dbeta}\, .
\ee
The momentum conserving delta function comes from integration over the $\wt\xi^\mu$'s and the
$\eps^{\alpha\beta}
\eps^{\dalpha\dbeta}$ comes from integration over the $\tilde\chi^\delta$'s and
$\tilde\chi^\ddelta$'s.
This is the analog of eq.(6.5) of \cite{Sen:2021tpp}.
The notations $\NN^{(0)}_1$ and $\AAA_1$ were not introduced in \cite{Sen:2021tpp}, but are convenient for
comparison with the multi-instanton amplitude.

Note that in carrying out the integration over the fermion zero modes using
\refb{esignchoice}, there is a phase
ambiguity associated with the integration measure over the fermions \cite{Sen:2021tpp}.
This reflects the ambiguity associated with the sign factors $\sigma_\gamma$ in \refb{defSumk}.
However, we show in appendix \ref{sdd} that this choice is correlated with the sign of the
multi-instanton contribution analyzed in \S\ref{snonmmulti}. In particular, \refb{esignchoice} is
compatible with \refb{e21}. If on the other hand we have a sign $\sigma_\gamma$ on the
right hand side of \refb{esignchoice}, then \refb{e21} will get an extra
factor of $\sigma_\gamma^{k-1}$
and $\AAA_1$ given in \refb{e1.9} (and its generalization for multi-instanton amplitudes)
will get a factor of $\sigma_\gamma$. Therefore
the $k$ instanton contribution will get an extra
factor of $(\sigma_\gamma)^{k}=\sigma_{k\gamma}$.
For a given $\gamma$, this phase can be absorbed
into the shift in the RR scalar field appearing
in the axionic coupling $e^{-2\pi \I k \Theta_\gamma}$ accompanying the
$k$ instanton amplitude.
Our choice of sign, encoded in \refb{esignchoice} and \refb{e21} will turn out to
agree with \refb{square}, \refb{defSumk} for the choice
$\sigma_{k\gamma}=1$ for a given $\gamma$
and all $k$.
However, the compatibility condition described below \refb{defSumk} shows
that this choice cannot be made  for all $\gamma$ simultaneously.
It should be possible to see the origin of this compatibility condition
by carefully analyzing the contribution from multiple instantons carrying
mutually non-local charges. But this has not been done so far and we shall not discuss the $\sigma_\gamma$
factors any further.

\subsection{Multiple instantons} \label{snonmmulti}

Let us now describe the computation of the overall normalization of
$k$ instanton amplitude which is expected to capture the $k$-th term in the sum in
\refb{defSumk}.
The normalization factor is given by the exponential of the annulus amplitude for open
strings living on the configuration of $k$ identical instantons. To compute it, we shall follow the
procedure described in \cite{Sen:2021jbr}.

At the initial stage, we
can analyze this system by regarding the open string spectrum as $k^2$ copies of
that on a single instanton. This leads to the analog of \refb{enewernn}:
\be\label{enewernn2}
 \NN_\NP=\I \, (2\pi)^{-2\NP^2} (2\sqrt \pi)^{\NP^2}
 \int \prod_{a=0}^{\NP^2-1} \left\{\prod_{\mu=0}^3 d\xi^\mu_a \right\}  \left\{\prod_{\delta,
 \ddelta=1}^2 d\chi^\delta_a d\chi^\ddelta_a\right\}
e^S \Bigg/ \int\prod_{b=0}^{k^2-1} D\theta_b\, .
\ee
The main distinction from \refb{enewernn} is the inclusion of the $e^S$ factor. At quadratic
order the action vanishes, but in order to integrate over the non-abelian modes we need
to include higher order terms in the action.

Next we change variables from $\theta_b$ to the parameters $\wt\theta_b$ of the rigid $U(k)$
symmetry group living on the D-instanton. This relation takes the form
$\theta_b=2\wt\theta_b/g_o$, and we get, as in \refb{e3.6aa},
\be
\int \prod_{b=0}^{k^2-1} D\theta_b = 2^{k^2} (g_o)^{-k^2}\int \prod_{b=0}^{k^2-1} D\wt\theta_b.
\ee
In this integral, $\wt\theta_0$ represents the $U(1)$ generator. Since the correctly normalized generator
of $U(1)$ is $I_k/\sqrt k$, $\wt\theta_0$ has period $2\pi\sqrt k$ and integration over
$\wt\theta_0$ produces the factor of $2\pi\sqrt k$. The rest of the $\wt\theta_a$ parametrize the
group $SU(k)$, but since $U(k)=(SU(k)\times U(1))/\ZZZ_k$, integration over the
$\wt\theta_a$'s for $1\le a\le (k^2-1)$ produces the volume of the group $SU(k)/\ZZZ_k$. This gives
\be\label{enewonenn3pre}
\NN_\NP= \I (2\pi)^{-2\NP^2}
(2\sqrt \pi)^{\NP^2}\,
2^{-\NP^2} (g_o)^{\NP^2}\, {1\over 2\pi\sqrt \NP}\, {1\over V_{SU(\NP)/\ZZZ_{\NP}}}\,
\int \prod_{a=0}^{\NP^2-1} \left\{\prod_{\mu=0}^3
d\xi^\mu_a \right\}  \left\{\prod_{\delta,
 \ddelta=1}^2 d\chi^\delta_a d\chi^\ddelta_a\right\}
e^S\, .
\ee
Since the action does not depend on the center of mass coordinates $\xi^\mu_0$ and
$\chi^\alpha_0$, we can factor out these integrals and relate them
to the location $\wt\xi^\mu$ of the `center of mass' of the D-instanton system and the
fermion zero modes $\tilde\chi^\alpha$ normalized so that their vertex operators do not
carry any factor of the open string coupling constant.
These relations take the form,
\be \label{eximurelnew}
\xi^\mu = \sqrt k\,
g_o^{-1} \, \pi^{-1}\, 2^{-1/2}\, \wt \xi^\mu\, , \qquad \tilde\chi^\alpha = g_o \, \chi^\alpha\, ,
\ee
generalizing \refb{eximurel} and \refb{e1.6}. The extra factor of $\sqrt k$ in the relation
between $\xi^\mu$ and $\wt\xi^\mu$ can be traced to the fact that the Chan-Paton
factor accompanying the correctly normalized
vertex operator for $\xi^\mu$ is given by $1/\sqrt k$ times the identity matrix \cite{Sen:2021jbr}.
Therefore \refb{enewonenn3pre} can be rewritten as:
\be\label{enewonenn3}
\NN_\NP= \I\, \NN^{(0)}_\NP
\int\prod_{\mu=0}^3 d\wt\xi^\mu \prod_{\delta,\ddelta=1}^{2}
d\tilde\chi^\delta \, d\tilde\chi^\ddelta\, ,
\ee
where,
\be\label{enewonenn30}
\begin{split}
\NN_\NP^{(0)}=&\, (2\pi)^{-2\NP^2}
(2\sqrt \pi)^{\NP^2}\left({\sqrt \NP\over g_o\, \pi\sqrt  2}\right)^{4}
2^{-\NP^2}\, (g_o)^{\NP^2}\, {1\over 2\pi\sqrt \NP}\, {1\over V_{SU(\NP)/\ZZZ_{\NP}}}\, g_o^{4}
\\
&\times \int \prod_{a=1}^{\NP^2-1} \left\{\prod_{\mu=0}^3 d\xi^\mu_a \right\}
\left\{\prod_{\delta,
 \ddelta=1}^2 d\chi^\delta_a d\chi^\ddelta_a\right\}
e^S\, .
\end{split}
\ee

The action $S$ appearing in \refb{enewonenn30} is given by the dimensional reduction
of $\NN=1$ supersymmetric Yang-Mills theory in four dimensions, with canonically
normalized kinetic terms, to zero dimensions. Therefore, it has a quartic interaction involving
$\xi^\mu_a$ with coefficient $g_{YM}^2/4$ and a $\xi$-$\chi$-$\chi$ Yukawa
coupling with coefficient $g_{YM}$ where
$g_o=\sqrt 2 \, g_{YM}$ \cite{Polchinski:1998rq,Polchinski:1998rr}.
We can remove the dependence
of $S$ on $g_o$ by making a
change of integration variables
\be \label{e13}
x^\mu_a= g_o^{1/2} \, \xi^\mu_a\, ,
\qquad
y_a^\alpha = g_o^{1/4} \, \chi_a^\alpha, \qquad y_a^\dalpha = g_o^{1/4} \, \chi_a^\dalpha,
\ee
and define
\be
X^\mu = x^\mu_a \, T^a ,
\qquad
Y^\alpha = y_a^\alpha\, T^a , \qquad Y^\dalpha = y_a^\dalpha\, T^a \, ,
\label{Tvar}
\ee
where the generators $T^a$ are normalized as $\Tr(T^a T^b)=\delta^{ab}$.
In terms of these variables, we have:
\be \label{e14}
S = {1\over 8}\, \Tr\bigl( [X_\mu, X_\nu] [X^\mu,X^\nu]\bigr) + {1\over \sqrt 2}\, \gamma^\mu_{\alpha\dbeta}
\Tr\bigl(Y^\alpha[X_\mu, Y^\dbeta]\bigr)
\ee
and
\be\label{eratio2}
{\NN_k^{(0)} \over \NN_1^{(0)}} =(2\pi)^{-2 (\NP^2-1)}(\sqrt \pi)^{\NP^2-1}
\,{\NP^{3/2}\over V_{SU(\NP)/\ZZZ_\NP}}\,
\, \int \prod_{a=1}^{\NP^2-1} \left\{\prod_{\mu=0}^3 dx^\mu_a \right\}
\left\{\prod_{\delta,
 \ddelta=1}^2 dy^\delta_a dy^\ddelta_a\right\}
e^S\, .
\ee
The integral appearing in this expression
can be read out from the conjecture of \cite{Krauth:1998xh}
(derived in \cite{Moore:1998et}) (see eq.(26), (27) of \cite{Krauth:1998xh}  for $D=4$ and $N$ replaced by $\NP$). We get
\be \label{e21}
\int \prod_{a=1}^{\NP^2-1} \left\{\prod_{\mu=0}^3 {dx^\mu_a\over \sqrt{2\pi}} \right\}
\left\{\prod_{\delta,  \ddelta=1}^2 dy^\delta_a dy^\ddelta_a\right\}
e^S={2^{\NP(\NP+1)/2} \pi^{(\NP-1)/2} \over 2\sqrt{\NP} \prod_{j=1}^{\NP-1} j!}\,
 {1\over \NP^2}\, .
\ee
We also have \cite{Marinov:1980jn,Marinov:1981} (see \cite{Sen:2021jbr} for
translation of these results to
the normalization convention used here):
\be\label{egrvol}
V_{SU(\NP)/\ZZZ_\NP}
=  2^{(\NP^2-1)/2}\,
 {2^{(\NP-1)/2} \pi^{(\NP-1)(\NP+2)/2} \over \sqrt{\NP} \prod_{j=1}^{\NP-1} j!}\, .
\ee
Substituting \refb{e21} and \refb{egrvol} into \refb{eratio2} and using \refb{e1.9} we get
\be\label{e2.11not}
{\NN^{(0)}_k\, \over \NN^{(0)}_1} = \NP^{-1/2} \, ,
\qquad
\NN^{(0)}_k=g_o\, 2^{-5}  \pi^{-13/2} \NP^{-1/2} \, .
\ee

We shall end this section with a few observations:
\begin{enumerate}
\item
The analysis carried out above has been done for the system where all $k$
D2-branes wrap a particular special Lagrangian
submanifold in a given homology class $\gamma$. Typically there are multiple
special Lagrangian submanifolds in a given homology class whose number is given by the
DT invariant $\Omega_\gamma$. For amplitudes that are protected by supersymmetry, all
the special Lagrangian submanifolds give the same contribution so that the amplitude is multiplied
by $\Omega_\gamma$. We can also have configurations in which different D2-branes wrap
different special Lagrangian submanifolds --- these will be discussed below.

\item
During our analysis we have assumed that the special Lagrangian submanifolds
along which the D2-branes are wrapped
are rigid, i.e.\ they cannot be deformed  inside the CY 3-fold. It is known that this requires
the first homology group of the special Lagrangian submanifold to be
trivial \cite{mclean}. If we relax this assumption, then the instanton will have additional
degrees of freedom associated with the motion in the moduli space of this
submanifold and we need to integrate over this moduli space. For a single instanton this
integral yields the Euler number of the moduli space up to a sign \cite{AlvarezGaume:1983at},
which coincides with the definition of DT invariant when it does not reduce to a simple counting.
However, for multiple instantons the effect will be generically non-trivial since it will add new degrees of
freedom to the integral appearing in \refb{eratio2}.

\item
Since the action \refb{e14} has no dependence on string coupling, the integral over the
$x^\mu_a$ in \refb{eratio2} gets contribution from a range of order unity. \refb{e13} now
shows that $\xi^\mu_a$ integrals get contributions from a range of order $g_o^{-1/2}$.
On the other hand, the set of $\xi^\mu_a$'s in the Cartan subalgebra of $SU(k)$ can be interpreted as
the mutual distance between the instantons. Therefore, after taking into account the scaling given
in \refb{eximurelnew} that relates the $\xi^\mu$ variables to physical positions $\wt\xi^\mu$,
we see that the integral in \refb{eratio2} receives contribution from the range when the
physical separation between the instantons is of order $g_o^{1/2}$. Since it vanishes
in the limit of weak string coupling,  the instantons fuse into each other in this
limit.
\end{enumerate}
The last observation provides an explanation for why the ratio $\NN_k^{(0)}/\NN_1^{(0)}$
is independent of $g_o$. In general the normalization constant for multiple instantons
should contain a factor that is the product of the normalization constant for each instanton,
and since $\NN_1^{(0)}$ is proportional to $g_o$, we would expect $\NN_k^{(0)}$ to be
proportional to $g_o^k$. However due to
appearance of massless open string states on coincident branes, there is extra enhancement
from the region of small separation between the instantons, and this is what cancels the
extra factors of $g_o$, leading to the result \refb{e2.11not} for $\NN_k^{(0)}/\NN_1^{(0)}$
that is independent of $g_o$.

This situation can be contrasted with the case when we have multi-instanton system with
different instantons wrapped on different special Lagrangian submanifolds. In this case even in the
limit of weak string coupling, the separation between the branes remain fixed and there is
no extra enhancement. The only exception are the points where the
submanifolds intersect. However, as long as the intersection is transverse,
there are no tachyonic or massless
open string modes at the intersection points and we do not expect any enhancement.
Therefore, contributions from such configurations will always be suppressed by powers
of the string coupling. Furthermore, if the wrapped cycles belong to different
homology classes $\gamma_1$ and $\gamma_2$,
then $|Z_{\gamma_1}|+|Z_{\gamma_2}|$ will typically be larger than
$|Z_{\gamma_1+\gamma_2}|$. Thus, the contribution from such configurations,
being proportional to $e^{-\TT_{\gamma_1}-\TT_{\gamma_2}}$,
will be exponentially suppressed compared to the contribution
proportional to $e^{-\TT_{\gamma_1+\gamma_2}}$ from an instanton
wrapping a cycle in the homology class
$\gamma_1+\gamma_2$.\footnote{As discussed already in
\S\ref{sduality}, the full D-instanton corrected metric and, in particular,
the one given in \eqref{mett1} does have such exponentially suppressed contributions. It is natural to expect that
these contributions also appear in the components of the Riemann tensor which are expected to be directly related to
four-point amplitudes in string theory \cite{Becker:1995kb}.
However, the action of such multi-instantons does not saturate the BPS bound.
While this raises a puzzle why such terms are present, one can note that similar terms do appear
even in ten-dimensional type IIB string theory \cite{Green:2005ba,Green:2014yxa,Chester:2020vyz}.}

\section{Strategy for computing D-instanton corrections to the metric}  \label{sstrategy}

In this section we shall describe the general strategy for computing the D-instanton
contribution to the moduli space metric. Since for massless scalars the two and three-point
amplitudes vanish, one needs to compute the four and higher
point amplitudes for finding
corrections to the effective action, including the moduli space metric. However, we shall
show that by exploiting the momentum non-conservation on individual disk amplitudes,
we can compute the leading correction to the metric by examining the two-point function
of a pair of scalars. The latter can be non-zero before imposing momentum conservation
and can be further decomposed into a product of two disk amplitudes, each with one scalar
and a pair of fermion zero modes.
A similar analysis can be carried out for scalars obtained by dualizing the two-form fields.

\subsection{Scalar amplitudes} \label{sscalar}

Let us suppose that the action contains a term of the form:
\be\label{escalar}
-{1\over 2} \, \int d^4 x\, G_{ij}(\vec\vp) \, \p_\mu \vp^i \p^\mu \vp^j \, ,
\ee
where $\vp^i$ are real scalars and $\vec\vp$ denotes the collection of all the
$\vp^i$'s. We use the Greek indices $\mu,\nu,\cdots$  for labelling
the four-dimensional
space-time indices and the roman indices other than
$k$ for labelling the fields. $k$ will label the instanton number.
In the following discussion we shall assume that the fields $\vp^i$ are
normalized as the primed fields introduced in \refb{rescalefields} so that the
dependence of various terms on the string coupling $g_{(4)}$ is simple. In these
variables $G_{ij}(\vec\vp)$ is expected to have an expansion of the from:
\be\label{ematricexpand}
G_{ij}(\vec\vp) = G^{(0)}_{ij} +  \sum_\gamma \, e^{-\TT_\gamma} H^{(\gamma)}_{ij}(\vec\vp)
\, ,
\ee
where $G^{(0)}_{ij}$ is the perturbative metric with an expansion in powers of $g_{(4)}$,
with the leading tree level term being independent of $g_{(4)}$.
The term proportional to $e^{-\TT_\gamma}$ is the contribution from the topological sector
with charge $\gamma$.  $\TT_\gamma$
is proportional to the inverse power of the string coupling $g_{(4)}$
and $H^{(\gamma)}_{ij} (\vec\phi)$ has an expansion in powers of $g_{(4)}$ with the leading
term proportional to $(g_{(4)})^{1/2}$. It may even have exponentially suppressed
contributions, but importantly $H^{(\gamma)}_{ij}(\vec\vp)\, \p_\mu \vp^i \p^\mu \vp^j$
is invariant under the shift symmetry of the RR moduli.
The sole breaking of the shift symmetry comes from the factor $e^{-\TT_\gamma}$, which thus uniquely distinguishes
the instanton sector of charge $\gamma$.
We denote by $h^{(\gamma)}_{ij}(\vec\phi)$ the leading term in the expansion of
$H^{(\gamma)}_{ij} (\vec\phi)$

If we denote by $\phi^i$ the expectation
value of $\vp^i$ and  by $\lambda^i=\vp^i-\phi^i$ the fluctuations of $\vp^i$,
then by expanding the metric to quadratic order in $\lambda^i$ we get the scalar
field kinetic term:
\ben
&& -{1\over 2} \, \int d^4 x\, \left[G_{ij}(\vec\phi) +
\lambda^m \, \p_m G_{ij}(\vec\phi)
+ {1\over 2} \lambda^m \lambda^n \, \p_m\p_n G_{ij}(\vec\phi)
\right]\p_\mu \lambda^i \p^\mu \lambda^j + \cdots\, .
\een
Since
$\TT_\gamma$ has inverse power of $g_{(4)}$, the leading term proportional to $e^{-\TT_\gamma}$ in the
expression for $\p_m\p_n G_{ij}(\vec\phi)$ will come when both derivatives act on
$\TT_\gamma$ in the exponent in \refb{ematricexpand}.
Therefore, the leading four-$\lambda^i$ term in the action proportional to
$e^{-\TT_\gamma}$ will be of the form:
\be \label{eleadscalar}
-{1\over 4} \, \int d^4 x\,  e^{-\TT_\gamma}\, \p_m \TT_\gamma \, \p_n\TT_\gamma \,
 h^{(\gamma)}_{ij}(\vec\phi)\,  \lambda^m \,
\lambda^n \, \p_\mu\lambda^i
\p^\mu\lambda^j \, .
\ee
Such term in the action induces a contribution to the four-scalar amplitude.
Let the $\alpha$-th external state represent the scalar $\eps^{(\alpha)}_m\lambda^m$ and
carry momentum $p^{(\alpha)}$. Then the contribution to the amplitude induced from the action
\refb{eleadscalar} is given by:\footnote{In order to avoid writing factors of $\I$, in the analysis
of this and later sections we shall
refer to amplitudes as the ones computed in Euclidean signature, so that
the amplitude generated by a term in the action is directly given by that term
written in momentum space, without any extra factor.}
\be \label{eampin}
\begin{split}
\AAA =&\,   (2\pi)^4 \delta^{(4)}\left(\sum_\alpha p^{(\alpha)}\right)
\bigg[\eps^{(1)m} \, \eps^{(2)n}  \, \eps^{(3)i}\,  \eps^{(4)j}  \,
\p_m \TT_\gamma\,  \p_n\TT_\gamma \,  h^{(\gamma)}_{ij}(\vec\phi)\, e^{-\TT_\gamma}  p^{(3)}.p^{(4)}
\\
& \hskip .5in +\, \hbox{inequivalent perm. of
(1),(2),(3),(4)}
\bigg]\, ,
\end{split}
\ee
where the field indices $i,j$ etc. are raised and lowered by the lowest order metric $G^{(0)}_{ij}$
and its inverse.

We shall now compare this contribution with the explicit computation of the instanton amplitude.
We shall classify terms by the homology classes $\gamma$ that determine
the imaginary part of $\TT_\gamma$, i.e. the dependence on the RR moduli.
For a given $\gamma$,  the amplitude could receive contribution from
either single instanton or multiple instantons. In the latter case
it follows from the discussion at the end
of \S\ref{sfour} that the leading contribution comes from the configurations where all the
instantons wrap the same special Lagrangian submanifold. Therefore, we can have
$k$ instantons of charge $\gamma/k$ wrapping the same cycle, where $k$ is any integer that divides $\gamma$.
This contribution will be proportional to $\NN^{(0)}_k$ and therefore to $g_o \sim
(g_{(4)})^{1/2}$,
in agreement with the scaling found in \S\ref{sduality} (see below \eqref{rescalefields}).

The leading contribution to the four-point amplitude includes,
besides the overall normalization constant
$\NN^{(0)}_k$,  the product of four disk amplitudes
each with one closed string vertex operator corresponding to one of the $\eps^{(\alpha)}_m\lambda^m$ combinations,
and four open string fermion zero modes distributed among the
four disks. Since each disk must carry even number of fermion zero modes, we can either
have all four fermion zero modes inserted on one disk, or have two of them on one disk
and two of them on another disk. The contribution from the disks without fermion zero mode
insertion, and the insertion of the vertex operator corresponding to $\lambda^m$
is given simply by
the derivative of the instanton action with respect to the field inserted, i.e.\
$-k\, \p_m \TT_{\gamma/k} =-\p_m\TT_\gamma$.
The factor of $k$ comes from the trace over the Chan-Paton factors
on the boundary of the disk. Therefore, the contribution from the amplitude where all four fermion
zero modes are inserted on a single disk will have the momentum dependence of the amplitude
coming entirely from the fourth disk. However, by Lorentz invariance this must be a function of
$p^2$ where $p$ is the momentum carried by the vertex operator on the fourth disk. Since
$p^2=0$, such an amplitude does not have any momentum dependence and would give rise
to a potential term. Since we do not expect instanton corrections to generate any potential, we
expect this amplitude to vanish.

Therefore, the only surviving contribution is the one where we have two of the zero modes on
one disk, two on another disk, and two disks without zero modes. It will have the structure:
\be\label{e67xy}
\begin{split}
& (2\pi)^4 \delta^{(4)}\left(\sum_\alpha p^{(\alpha)}\right) \,
\bigg[\eps^{(1)m} \, \eps^{(2)n}  \, \eps^{(3)i}\,  \eps^{(4)j}  \,
\p_m \TT_{\gamma}\,  \p_n\TT_{\gamma} \, \NN^{(0)}_{k,\gamma/k} \, e^{-\TT_\gamma}\,
\AAA^{(k,\gamma/k)}_{ij}(\vec\phi)
\\
& \hskip .5in +\, \hbox{inequivalent perm. of
(1),(2),(3),(4)}
\bigg]\, ,
\end{split}
\ee
where $\AAA^{(k,\gamma/k)}_{ij}$ is the product of two disk amplitudes,
each with a pair of fermion zero mode
insertions and insertion of vertex operators of $\lambda_i$ or $\lambda_j$,
contracted with $\eps^{\alpha\beta}\eps^{\dalpha\dbeta}$ as in \refb{e1.9}.
The superscript $(k,\gamma/k)$ on $\AAA$
represents that the instanton configuration responsible
for the
amplitude consists of $k$ instantons each of charge $\gamma/k$. The $\gamma/k$ added to
the subscript of $\NN^{(0)}_k$ is a reflection of the fact that it has implicit dependence on
the instanton charge $\gamma/k$ due to the dependence of $\NN^{(0)}_k$ on $g_o$ via \refb{e2.11not}
and the charge dependence of $g_o$ via \eqref{expr-g0}.
If $\Omega_\gamma$ denotes the number of special Lagrangian 3-cycles in homology
class $\gamma$, then the contribution \refb{e67xy} is multiplied by $\Omega_{\gamma/k}$.
Comparing this with \refb{eampin}  we get:\footnote{Note that a term proportional to $e^{-k\TT_\gamma}$
may get contribution from different sources, {\it e.g.}
terms proportional to $\Omega_{\gamma}$ and
$\Omega_{k\gamma}$. They are distinct contributions and all of them are included
in the sum in \refb{e56}.
Using that $\NN^{(0)}_{k,\gamma/k} \sim k^0$ and, as will be established below, $ \AAA^{(k,\gamma/k)}_{ij}\sim k^{-2}$,
one can reorganize the sum in \refb{e56} so that
we only have the $k=1$ term in the sum and $\Omega_\gamma$ is replaced by the
rational index $\bar\Omega_\gamma$ defined in \refb{def-bOm}.
}
\be \label{e56}
\begin{split}
& \sum_{k|\gamma} \, \Omega_{\gamma/k}\,
\NN^{(0)}_{k,\gamma/k} \,   \AAA^{(k,\gamma/k)}_{ij}(\vec\phi) =
h^{(\gamma)}_{ij}  \, p.p'
\\
\Rightarrow \qquad &
p.p'\, \sum_\gamma \, h^{(\gamma)}_{ij} \, e^{-\TT_\gamma}
= \sum_\gamma \Omega_{\gamma} \sum_{k=1}^\infty
\NN^{(0)}_{k,\gamma} \,   \AAA^{(k,\gamma)}_{ij}(\vec\phi) \, e^{-k\, \TT_\gamma}\,  \, ,
\end{split}
\ee
where $p$ and $p'$ are the momenta carried by $\lambda_i$ and $\lambda_j$. Therefore the
knowledge of $\AAA^{(k,\gamma)}_{ij}(\vec\phi)$
can be used to determine the instanton corrected metric,
keeping the leading term in each topological sector.
Note however that \refb{e56}
vanishes for $p'=-p$, and so we implicitly need to keep in mind the presence
of the extra pair of $\lambda$'s that allow momentum to be not conserved.
In practice we can just analyze the $\lambda^i$-$\lambda^j$ amplitude without using
momentum conservation.

There is one important limitation of this approach arising due to possible ambiguities in
rewriting the action \refb{eleadscalar} using integration by parts, or equivalently, rewriting the
amplitude \refb{eampin} using momentum conservation. To examine this, it will be convenient to
choose a basis for the field variables so that the instanton action depends on only one of the
fields. For example we could take the instanton action
itself as a (possibly complex) field variable, and denote
by $\xi$ its fluctuation. Then the action \refb{eleadscalar} takes the form:
\be \label{eleadscalarnew}
-{1\over 4} \int d^4 x \, e^{-\TT_\gamma}\,
 h^{(\gamma)}_{ij}(\vec\phi)\,  \xi^2 \, \p_\mu\lambda^i\,
\p^\mu\lambda^j \, .
\ee
Now suppose that $\lambda^i$ is also $\xi$. In that case \refb{eleadscalarnew} reduces to
\be
-{1\over 4}  \int d^4 x \, e^{-\TT_\gamma}\,
 h^{(\gamma)}_{\xi j}(\vec\phi)\,  \xi^2 \, \p_\mu\xi\,\p^\mu\lambda^j
= -{1\over 12}  \int d^4 x\, e^{-\TT_\gamma}\,
 h^{(\gamma)}_{\xi j}(\vec\phi)\,   \p_\mu(\xi^3)\,
\p^\mu\lambda^j \, .
\ee
We can now integrate by parts to write the integrand as proportional to $\xi^3 \p_\mu\p^\mu\lambda^j$
which vanishes once we use the on-shell condition for the field $\lambda^j$.
This shows that we cannot use the two-point function to compute the leading correction
to the metric proportional to $ h^{(\gamma)}_{i j}$ if either $i$ or $j$ represents the fluctuation $\xi$.
In a general basis of field variables this translates to the statement that we cannot compute
components of the metric along the fluctuation directions $\p_m\TT_\gamma \, \lambda^m$.

Physically, one can understand the origin of this limitation as follows. Since the S-matrix of string
theory determines the action only up to field redefinition, we can consider a field redefinition
of the form $\vp^m\to \vp^m+ e^{-\TT_\gamma} g^m(\vec\vp)$ for any set of functions $g^m(\vec\vp)$ without
changing the S-matrix. Here we should regard $\TT_\gamma$ as a function of the field variables $\vec\vp$
and not just of the background $\vec\phi$. Under this transformation:
\be
d\vp^m\to d\vp^m -  e^{-\TT_\gamma} g^m(\vec\vp)\,\p_n\TT_\gamma\, d\vp^n +\cdots\, ,
\ee
where $\cdots$ are subleading terms in the power series expansion in $g_{(4)}$.
Therefore, the leading
effect of such transformations on the perturbative terms will be to generate terms in the
metric proportional to $e^{-\TT_\gamma}\,  \p_n\TT_\gamma\, d\vp^n$. This can be used to change
the leading terms in the metric proportional to $\de \cT_\gamma$ from instantons of charge $\gamma$.
For this reason, terms proportional to $\de \cT_\gamma$ in the instanton induced metric cannot be
computed unambiguously.

\subsection{Two-form field  and its dualization}

In our analysis we shall also encounter a dual version of this problem where instead of scalar
fields $\phi^i$ we have 2-form fields $B_{i,\mu\nu}$ with action
\be\label{etensor1}
-{1\over 6} \int d^4 x\, G^{ij}(\vec\vp) H_{i,\mu\nu\rho} H_{j}^{\mu\nu\rho},
\qquad H_{i,\mu\nu\rho}=\p_\mu B_{i,\nu\rho}+ \p_\nu B_{i,\rho\mu} + \p_\rho B_{i,\mu\nu}\, .
\ee
When we dualize the 2-form fields into scalar fields, we get back an action of the
form given in \refb{escalar} with
$G_{ij}$ given by the inverse of the metric $G^{ij}$. Now $G^{ij}$ has an expansion similar to the
one given in \refb{ematricexpand}:
\be\label{ematricinvexpand}
G^{ij}(\vec\phi) = G^{(0)ij}(\vec\phi) + \sum_{\gamma} e^{-\TT_\gamma} F^{(\gamma)ij}(\vec\phi)
+\cdots\, .
\ee
If we denote by $f^{(\gamma)ij}$ the leading term in the expansion of $F^{(\gamma)ij}(\vec\phi)$ in powers
of $g_{(4)}$, we get the analog of \refb{eleadscalar}  for the
leading term in the action proportional to $e^{-\TT_\gamma}\lambda^m \lambda^n   H_{i,\mu\nu\rho} H_{j}^{\mu\nu\rho}$:
\be \label{eleadtensor}
-{1\over 12} \, \int d^4 x\,  \p_m\TT_\gamma\, \p_n\TT_\gamma\,   f^{(\gamma)ij}\, e^{-\TT_\gamma}  \, \lambda^m\,
\lambda^n \, H_{i,\mu\nu\rho} H_{j}^{\mu\nu\rho}\, .
\ee
This leads to a $\lambda^m$-$\lambda^n$-$B_{i,\mu\nu}$-$B_j^{\mu\nu}$ amplitude of the form:
\be \label{eamptensor}
\AAA={ 1\over 3} \,  \p_m\TT_\gamma\, \p_n\TT_\gamma\,
f^{(\gamma)ij}\, e^{-\TT_\gamma}  \,
(p^{\mu} b^{\nu\rho}+p^{\nu} b^{\rho\mu}+ p^{\rho} b^{\mu\nu})
(p'_\mu b'_{\nu\rho}+p'_\nu b'_{\rho\nu}+ p'_\rho b'_{\mu\nu}) \,,
\ee
where $p,p'$ are the momenta and $b_{\mu\nu}$ and $b'_{\mu\nu}$ are the
polarizations of $B_{i,\mu\nu}$ and $B_{j,\mu\nu}$. In this formula we have not explicitly written
down the momentum conserving $\delta$-function, but it is certainly present.

This needs to be compared with the product of four disk amplitudes, two with insertions of
$\lambda^m$ and $\lambda^n$ and two with insertions of $B_{i,\mu\nu}$ and $B_{j,\mu\nu}$, and
and four fermion zero mode insertions spread among the four disks.
In this case using Lorentz invariance one can see that the disk amplitude with just $B_{i,\mu\nu}$
insertion vanishes unless it also has fermion zero mode insertions. Therefore, we must insert
two fermion zero modes on the disk containing $B_{i,\mu\nu}$ and two fermion zero modes
on the disk containing $B_{j,\mu\nu}$.
Following the same logic as in the case
of scalar amplitudes, we can now conclude that if for $k$ instantons of charge
$\gamma/k$, we denote the product of the two disk
amplitudes containing $B_{i,\mu\nu}$ and $B_{j,\mu\nu}$,  by
\be\label{etensoramp}
{ 1\over 3} \,
\BBB^{(k,\gamma/k)ij}\,
(p^{\mu} b^{\nu\rho}+p^{\nu} b^{\rho\mu}+ p^{\rho} b^{\mu\nu})
(p'_\mu b'_{\nu\rho}+p'_\nu b'_{\rho\nu}+ p'_\rho b'_{\mu\nu}),
\ee
then we have
\be\label{etensequal}
f^{(\gamma)ij}=  \sum_{k|\gamma} \Omega_{\gamma/k}\,
\NN^{(0)}_{k,\gamma/k}\,  \BBB^{(k,\gamma/k)ij}\, .
\ee
Thus, the disk amplitudes induce the following term in the effective action
\be\label{etensorfin}
-{1\over 6} \, \sum_\gamma\, \Omega_\gamma  \, \sum_{k=1}^\infty \,
\int d^4 x \, e^{-k \TT_\gamma}\, \NN^{(0)}_{k,\gamma}\,  \BBB^{(k,\gamma)ij}
\, H_{i,\mu\nu\rho} H_{j}^{\mu\nu\rho}
\, .
\ee

Since in four space-time dimensions a 2-form field $B_{\mu\nu}$ is dual to a scalar, we can
express the instanton correction \refb{etensorfin} to the action as a correction to the metric
of the dual scalar field. Let us recall the rules for the duality transformation and derive their consequences
for the instanton corrections.
Suppose that we have an action of the form:
\be \label{eoriginal}
\int d^4 x\( -{f \over 12} \, H_{\mu\nu\rho} H^{\mu\nu\rho} +{1\over 6}\,
\eps^{\mu\nu\rho\tau}  H_{\mu\nu\rho} A_\tau\) ,
\ee
where $f$ and the one-form $A$
could be  functions of other fields. Then we introduce the dual scalar field
$\tilde\sigma$ via the relation
\be \label{ehprelation}
H^{\mu\nu\rho}=f^{-1}\, \eps^{\mu\nu\rho\tau} \, (\p_\tau\tilde\sigma+A_\tau) \, ,
\ee
and replace the action \refb{eoriginal} by:
\be\label{edual}
-{1\over 2} \int d^4 x\, f^{-1}\, (\p_\mu\tilde\sigma+A_\mu)\, (\p^\mu\tilde\sigma+A^\mu)\, .
\ee
In this case the equation of motion of $B_{\mu\nu}$ derived from the action \refb{eoriginal}
is automatically satisfied by \refb{ehprelation}, whereas the Bianchi identity $dH=0$ becomes the
equation of motion of $\tilde \sigma$ derived from the action \refb{edual}.

Note that the actions \refb{eoriginal} and \refb{edual} are not equal.
If however we deform the action \refb{eoriginal} by deforming the function $f$ by $\delta f$,
then, to
first order in $\delta f$, the deformation of the scalar action can be computed
simply by expressing the deformation of \refb{eoriginal} using the relation \refb{ehprelation}.
To see this, suppose $f$ is deformed to $f+\delta f$. Then according to \refb{edual} the
deformed scalar action will be
\be \label{e621}
\begin{split}
& -{1\over 2} \int d^4 x\, (f+\delta f)^{-1}\, (\p_\mu\tilde\sigma+A_\mu)\, (\p^\mu\tilde\sigma
+A^\mu) \\  =&
-{1\over 2} \int d^4 x\, f^{-1}\, (\p_\mu\tilde\sigma+A_\mu)\,
(\p^\mu\tilde\sigma
+A^\mu)
+ {1\over 2} \int d^4 x\, f^{-2}\, \delta f \, (\p_\mu\tilde\sigma+A_\mu)\,
(\p^\mu\tilde\sigma
+A^\mu)
\, .
\end{split}
\ee
On the other hand, evaluating the deformation of \refb{eoriginal} using \refb{ehprelation}, we get
\be \label{ehdeformed}
\begin{split}
 -{1\over 12} \, \int d^4 x\, \delta f \, H_{\mu\nu\rho} H^{\mu\nu\rho}
& = -{1\over 12} \, \int d^4 x\, \delta f \, f^{-2} \, \eps_{\mu\nu\rho\tau} (\p^\tau\tilde\sigma+A^\tau)
\eps^{\mu\nu\rho\tau'}(\p_{\tau'}\tilde\sigma+A_{\tau'}) \\
& = {1\over 2} \, \int d^4 x\,   \delta f \, f^{-2} \, (\p^\mu\tilde\sigma+A^\mu)\,
(\p_\mu\tilde\sigma+A_\mu)\, ,
\end{split}
\ee
which agrees with the deformation \refb{e621}. This observation will be useful for us since this
means that in order to rewrite the correction \refb{etensorfin} in terms of scalar fields, we
simply have to express it in terms of the scalars using the perturbative relation between the
three-form $H$ and the dual scalar field.

\section{Explicit computation of D-instanton corrections}
\label{sec-maincomput}

The goal of this section is to determine the D-instanton correction
to the metric on the hypermultiplet moduli space from the computation of amplitudes in string theory.
We shall first outline the general strategy and then turn to the computation of different
components of the metric.

\subsection{General strategy} \label{egenst}

Our main tool for computing the metric will be \refb{e56}.
For this we need to compute $\AAA^{(k,\gamma)}_{ij}$, the product of
two disk amplitudes, each with one closed string vertex operator and two fermion zero modes.
Furthermore, since we are looking for a contribution where each disk gives one power of
momentum,
one of the zero modes must carry dotted index and the other one should carry undotted index
so that we get a contribution proportional to $p_\mu (\gamma^\mu)_{\alpha\dbeta}$.
To compute such disk amplitudes, we shall use the upper half plane description of the disk, insert the
closed string vertex operator at $\I$,
the vertex operator for the undotted zero mode  at the origin and that of
the dotted zero mode  at a point $z$ on the real axis, and integrate over $z$.
Whether we take the dotted vertex operator to be the integrated one or the undotted vertex
operator to be the integrated one is a matter of choice and the final result does
not depend on this choice.
Also the vertex operators of the zero modes can be
represented either as holomorphic or as anti-holomorphic operators.
We take them to be holomorphic and in the $(0,-1/2)$ picture, and the closed string
vertex operator in the $(-1,0)$ or $(-1/2,-1/2)$ picture
so that the total picture number of all the vertex operators adds up to $-2$.
In order to fix the normalization of the closed string vertex operators so that they represent
the fields appearing in \refb{hypmettree}, \refb{square}, we some time  need to
compute disk one-point function of closed string vertex operators without any zero mode
insertion. For such computation we take the closed string vertex operator to be in the $(-1,-1)$
or $(-1/2,-3/2)$ picture and insert it at the point $\I$ on the upper half plane.

While carrying out the integration over the location $z$ of the dotted zero mode vertex
operator, we may encounter singularities as $z\to 0$, i.e.\ when
the two fermion zero mode operators collide. These can be analyzed carefully using the rules of
open string field theory and can be shown to lead to the principal value prescription, but here we shall
describe a simpler procedure. If $\alpha$ and $\dalpha$ are the spinor indices carried by
the zero modes, then the residue at the pole is proportional to
$\gamma^\mu_{\dalpha\alpha}$ contracted with $e^{-\phi}\psi_\mu$
--- the vertex operator of the bosonic zero mode field associated with
translation along the non-compact directions. Therefore, this will produce the amplitude without
any insertion of the zero mode fields multiplied by a factor of $p_\mu \gamma^\mu_{\dalpha\alpha}$.
On the other hand, the insertion of a closed string vertex operator without any open string zero mode
insertion will produce non-zero  result only if the closed string field is along the direction
$\p_m\TT_\gamma$. Hence, the contribution to the effective action of the fluctuating field $\lambda^m$ will
be proportional to $\p_\mu \lambda^m \p_m \TT_\gamma$. Since we have argued in \S\ref{sstrategy} that
at the leading order  these contributions are ambiguous, we shall ignore them.
This allows us to choose the contour so that it passes on one side of the origin --- we
shall choose it to pass  above the origin. Note that this prescription must be followed for
all the amplitudes since individual closed string fields are typically not along the direction $\p_m\TT_\gamma$,
and only after taking the sum of all the contributions we can justify ignoring the residue at the
origin.

If we denote by $\lambda^m$ the closed string field whose amplitude we compute, and by
$p_\mu$ the momentum carried by it, then the
result of the computation produces a term of the form\footnote{We shall use the convention
that while writing an amplitude we also include the fields whose amplitude we compute as
multiplicative factors, so that we can directly read out the effective action from the amplitude.}
\be \label{eterm}
i\, a_m \, p_\mu\,
\gamma^\mu_{\dalpha\alpha} \, \lambda^m\, \tilde\chi^\alpha\tilde\chi^\dalpha\, ,
\ee
for
some constant $a_m$. Note that this result is the same for all $k$ instanton amplitudes since
for $k$ instantons the trace over the Chan-Paton factor gives a factor of $k$ and the two
open string zero modes carry factors of $1/\sqrt k$ each, so that there is no net $k$-dependence.
The product of a pair of such disk amplitudes, one for the field $\lambda^m$ and another for the field
$\lambda^n$, carrying momenta $p_\mu$ and $p'_\mu$ respectively,
will have the form $-a_m\, a_n \, p_\mu \, p'_\nu \, \gamma^\mu_{\dalpha\alpha} \,
\gamma^\nu_{\dbeta\beta}$. Using the result \refb{esignchoice}
that the integration over the fermion zero modes
produces a factor of $\eps^{\alpha\beta}\eps^{\dalpha\dbeta}$
and one has $\eps^{\alpha\beta}\eps^{\dalpha\dbeta}\,\gamma^\mu_{\dalpha\alpha} \,\gamma^\nu_{\dbeta\beta}
=-\Tr(\gamma^\mu\gamma^\nu)=-2\, \eta^{\mu\nu}$,
we see that the net contribution to \eqref{e56} from this particular
$k$ instanton amplitude is given by
\be \label{e6.2aa}
2 \, \Omega_\gamma\, \NN_{k,\gamma}^{(0)}
\, e^{-k\TT_\gamma} \, a_m \, a_n\, p_\mu \, p^{\prime\mu} \, (2\pi)^4 \, \delta^{(4)}(p+p')
\, \lambda^m(p)\lambda^n(p')\, .\ee
The collection of all such amplitudes may be obtained from a term in the effective action of the from:
\be \label{e74pre}
-\sum_\gamma \Omega_\gamma\sum_{k=1}^\infty
\NN_{k,\gamma}^{(0)}  \, e^{-k\TT_\gamma} \, \sum_{m,n} a_m \, a_n\,  \int \, d^4 x\, \p_\mu\lambda^m \, \p^\mu\lambda^n\, .
\ee

String theory computation naturally gives the action
\refb{e74pre} in the background where the four-dimensional string metric is
set equal to $\eta_{\mu\nu}$.
For this metric the coefficient of the Einstein-Hilbert term is $V/(2\kappa^2)$ where
$\kappa$ has been defined in \refb{ekappags} and $V$ is the volume of $\CY$.
On the other hand, the results of
\S\ref{sduality} have been presented in the convention
where the four-dimensional canonical metric is set equal to $\eta_{\mu\nu}$
because this makes the decoupling between the hypermultiplet and the vector multiplet moduli manifest.
As shown in \refb{kinL}, for this metric the coefficient of the Einstein-Hilbert term is
equal to 1.
Since the metrics in the two frames are related by a factor of $2\kappa^2/V$,
the effect of setting the canonical metric to $\eta_{\mu\nu}$ corresponds to setting the string metric to
$2\kappa^2\eta_{\mu\nu}/V$. In this case
we need to
multiply the kinetic term \refb{e74pre} of the scalar field action
by a factor of $2\kappa^2/V$ and the correction to the effective action takes the form:
\be \label{e74}
-2\,\kappa^2 \, V^{-1} \sum_\gamma \Omega_\gamma\sum_{k=1}^\infty
\NN_{k,\gamma}^{(0)}
\, e^{-k\TT_\gamma} \, \sum_{m,n} a_m \, a_n \int  d^4 x\, \p_\mu\lambda^m \, \p^\mu\lambda^n\, .
\ee
Comparison of \refb{escalar} and \refb{e74} now shows that we have
a correction to the moduli space metric of the form
\be\label{e75}
\delta (ds^2) = \sum_\gamma \Omega_\gamma\sum_{k=1}^\infty
\cN_{k,\gamma}\, e^{-k\TT_\gamma} \, \( \sum_m a_m \, d\lambda^m\)^2\, ,
\qquad
\cN_{k,\gamma}=4\kappa^2 \, V^{-1} \NN_{k,\gamma}^{(0)} .
\ee

Below, we shall calculate the coefficients $a_m$ for various fields.
But before doing this, let us express the coefficient $\cN_{k,\gamma}$ in terms of the variables used in \S\ref{sduality}.
Using \refb{e2.11not},  \eqref{rel-rgs} and
\eqref{expr-g0}, we get
\be \label{efincorrect}
\cN_{k,\gamma}=2^{-6}\pi^{-7}k^{-1/2} r^{-5/4}\, K^{1/4} |Z_\gamma|^{-1/2}.
\ee
We remind the reader that even though the computation of the amplitude is done by regarding
the string coupling constant and other moduli as constant background, once we express this
in the form \refb{e75}, we can regard all the moduli, including those appearing inside
the $a_m$'s and $\NN_{k,\gamma}$,
as fields. As explained in \S\ref{sstrategy}, the ability to do this follows from
the fact that the leading contribution to the four-point
amplitude computed from the metric comes from the dependence of the metric on the moduli
through the $e^{-k\TT_\gamma}$ factor, and the latter dependence is captured by disk one-point
function without any zero mode insertion that appears as common factor in all amplitudes.

In the rest of this section we shall work in the limit in which the CY threefold $\CY$ has
large size. Since the moduli associated with the size of $\CY$ belong to
the vector multiplet in type IIA string theory, they cannot affect the metric on the
hypermultiplet moduli space. Therefore, the results derived in the large volume limit
are exact. We shall also make the further approximation of working to leading order in
the string coupling in each charge sector. Since the modulus labelling the string coupling
belongs to the hypermultiplet, we do expect non-vanishing corrections to our results at
higher order in the string coupling.

\subsection{NSNS axion contribution} \label{s7.1}

We shall now compute the contribution to the component of the metric of the scalar field dual to
the NSNS 2-form field which is a universal field present in all string theories.
For this we need to compute
the instanton induced contribution to the disk amplitude with one insertion of the vertex operator
of the 2-form field and a pair of vertex operators associated with open string zero modes.

We denote by $b_{\mu\nu}$ the polarization of the 2-form field $B_{\mu\nu}$ and
take the vertex operator to be in the $(-1,0)$ picture, inserted at the point $\I$
in the upper half plane. The pair of open string fermion zero mode vertex operators are
taken in the
$(0,-1/2)$ picture, inserted at the points 0 and $z$ on the real axis.
The 2-form vertex operator in the $(-1,0)$ picture
in the normalization convention of \cite{Sen:2021tpp} is given by:
\be\label{ebmunuvertex}
V_B = 2\, b_{\mu\nu} \,
c\, \bar c\, \(\p X^{\mu} + \I\, p_\rho\, \psi^\rho\psi^\mu\)
e^{\I p.X}
e^{-\bar\phi}\bar\psi^\nu(\I) +\cdots\, .
\ee
The $\cdots$ terms have non-zero $\xi$-$\eta$ charge and will not contribute to the correlation function.
We denote the unintegrated
$-1/2$ picture undotted vertex operator of the open string fermionic mode by
\be
\XX^\alpha c\, W_\alpha, \qquad W_\alpha \equiv e^{-\phi/2} S_\alpha \, ,
\ee
where $\XX^\alpha$ is the ten-dimensional representation of the fermion
zero mode introduced in \refb{exxdef}.
The integrated $-1/2$ picture dotted vertex operator takes the form:
\be
\wh\XX^\beta\, W_{\beta}= \wh\XX^\beta\, e^{-\phi/2} S_\beta \, .
\ee
Note that we are using the ten-dimensional description of the zero modes, but eventually we shall
transform them into four-dimensional versions by taking $\XX^\alpha$ and $\wh\XX^\beta$ to be
$\eta\otimes \tilde\chi^\alpha$ and $\bar\eta\otimes \tilde\chi^\dalpha$ respectively.
Using \refb{edisk2}, \refb{edisk3}, the disk one-point function in the
presence of a flat  Euclidean D2-brane is given by:
\be
\I \pi\kappa \, \ts_2 \, 2\, b_{\mu\nu} \, \XX^\alpha\wh\XX^\beta
\int_{-\infty}^\infty dz  \left\langle
c\, \bar c \(\p X^{\mu} + \I\, p_\rho\, \psi^\rho\psi^\mu\)
e^{\I p.X}
e^{-\bar\phi}\bar\psi^\nu(\I) \, c e^{-\phi/2} S_\alpha(0) \, e^{-\phi/2} S_\beta(z)
\right\rangle ,
\ee
where $\kappa$ and $\ts_2$ have been defined in \refb{ekappags}.

The correlation function involving $\p X^\mu$ is proportional to $p^\mu$ and vanishes after
using the physical state condition $p^\mu b_{\mu\nu}=0$ for the polarization tensor
$b_{\mu\nu}$. Due to the same reason, we can also drop the $e^{\I p.X}$ factors from the correlation function,
keeping in mind that when we consider the general location $\wt\xi^\mu$ for the D-instanton,
the $e^{\I p.X}$ factor will generate the
$e^{\I p.\wt\xi}$ factor, which in turn will give $(2\pi)^4\delta^{(4)}(
\sum_i p_i)$ factor after integration over the bosonic zero modes $\wt\xi^\mu$.
Furthermore,
using the doubling trick, we can replace
the anti-holomorphic fields in the closed string vertex operator by holomorphic fields placed at
$-\I$ with an additional factor of $-1$ due to the Dirichlet boundary condition on the $X^\mu$
and $\psi^\mu$ fields.
For example, since $\bar c\, e^{-\bar\phi} \bar\psi^\mu$ is an anti-holomorphic dimension 0
operator, using Dirichlet boundary condition on $\psi^\mu$, we can replace this by
$- c\, e^{-\phi} \psi^\mu$ placed at the complex conjugate point.
This expresses the amplitude as
\be\label{ecal1}
-2 \I \pi\, \kappa\, \ts_2 \,  b_{\mu\nu} \,\XX^\alpha\wh\XX^\beta
\int_{-\infty}^\infty dz \, \Bigl\langle  \I \,
c \, p_\rho\psi^\rho\psi^\mu (\I)\,  c\, e^{-\phi}\psi^\nu(-\I)
\, c\, e^{-\phi/2} S_\alpha(0) \, e^{-\phi/2} S_\beta(z) \Bigr\rangle\, .
\ee
As stated earlier, we shall take the $z$ integration contour to pass above the origin.
We shall now deform the $z$ integration contour to pick up residue from the pole at $\I$ using
\refb{espinope}.  The effect of this is to drop the integration over $z$ and the
$e^{-\phi/2} S_\beta(z)$ factor from inside the correlator and replace the
$\psi^\rho\psi^\mu(\I)$
factor
inside the correlator by $(\I\pi/2) \left(\Gamma^{\rho\mu}
\right)_\beta^{\ \gamma} e^{-\phi/2} S_\gamma (\I)$.
The resulting correlation function can be easily evaluated using \refb{espinope},
\refb{eopennorm} and gives:
\be\label{e610}
{1\over 2}\, \pi^2\kappa \, \ts_2 \,   p_\rho\, b_{\mu\nu} \,  \XX^\alpha\wh\XX^\beta\,
(\Gamma^{\rho\mu}\Gamma^{\nu})_{\beta\alpha}\,  (2\pi)^3 \delta^{(3)}(0)=
{1\over 2}\, \pi^2\kappa \, \ts_2 \,   p_\rho\, b_{\mu\nu} \,  \XX^\alpha\wh\XX^\beta\,
(\Gamma^{\rho\mu\nu})_{\beta\alpha}\,  (2\pi)^3 \delta^{(3)}(0)\, ,
\ee
where
\be
\Gamma^{M_1\cdots M_n} \equiv \Gamma^{[M_1} \cdots \Gamma^{M_n]}\,.
\ee
In going from the left to the right hand side of \refb{e610} we have used the fact that
$\Gamma^{\rho\nu}\Gamma^\mu$ differs from
$\Gamma^{\rho\nu\mu}$ by
terms proportional to $\eta^{\mu\nu}$ and $\eta^{\mu\rho}$.
Both terms vanish after we use the
physical state condition for $b_{\mu\nu}$.
The $(2\pi)^3 \delta^{(3)}(0)$ factor is a reflection of the momentum conserving delta
function in \refb{eopennorm}.

We shall now replace $\XX^\alpha$ and $\wh\XX^\beta$ by
$\eta\otimes \tilde\chi^\alpha$ and $\bar\eta\otimes \chi^\dalpha$, respectively, and
replace the $(2\pi)^3 \delta^{(3)}(0)$ factor by an integral over the
world-volume of the Euclidean D2-brane wrapped on a 3-cycle $\CG$ of
$\CY$.
Using \refb{etenfourgamma} and
\refb{eetanorm} the resulting expression may be written as:
\be
{1\over 2}\, \pi^2\kappa \, \cTR_\gamma \,   p_\rho\, b_{\mu\nu} \,  \tilde\chi^\alpha\tilde\chi^\dalpha\,
(\gamma^{\rho\mu\nu})_{\dalpha\alpha}\, ,
\label{ampl-diskB}
\ee
where we used \eqref{Realact} to express the result in terms  of the real part of the instanton action.

We shall now rewrite this in terms of the dual scalar field.
For this we need to first examine
the tree level kinetic term for the 2-form field $B_{\mu\nu}$. In the normalization
convention of \cite{Sen:2021tpp} that we
have been using, the ten-dimensional action for $B_{\mu\nu}$ in the harmonic gauge
$p^\mu B_{\mu\nu}=0$
is given by:
\be
- {1\over 2} \int {d^{10}p\over (2\pi)^{10}} \, B_{\mu\nu}(-p) \, p^2 \, B^{\mu\nu}(p)\, .
 \ee
The gauge invariant action leading to this has the form
\be \label{e717}
-{1\over 6} \int {d^{10}p\over (2\pi)^{10}} \, H_{\mu\nu\rho}(-p) \, H^{\mu\nu\rho}(p)
= -{1\over 6} \int d^{10}x \, H_{\mu\nu\rho}(x) \, H^{\mu\nu\rho}(x)\, ,
\ee
where
\be
H_{\mu\nu\rho}(p)=\I \bigl( p_\mu\, B_{\nu\rho}(p)  + p_\nu\, B_{\rho\mu}(p)+
p_\rho\, B_{\mu\nu}(p)\bigr)\, .
\ee
Upon compactification on a CY 3-fold $\CY$, the action takes the form:
\be \label{etree}
- \int d^{4}x \, {V\over 6} \, H_{\mu\nu\rho}(x) \, H^{\mu\nu\rho}(x)\, ,
\ee
where $V$ is the volume of $\CY$. For the moment, we shall ignore the presence of the
Chern-Simons type terms shown in \refb{eoriginal} but will include the effect of such
terms later. Using \refb{ehprelation} and \refb{edual},
we now see that, ignoring instanton corrections, the scalar field $\tilde\sigma$
obtained by dualizing $B_{\mu\nu}$ satisfies
\be \label{ehpsi}
H^{\mu\nu\rho}={1\over 2 V}\, \eps^{\mu\nu\rho\tau}\, \p_\tau\tilde\sigma\, ,
\ee
and has action
\be\label{e720pre}
-\int d^4 x \, {1\over 4 V}\,  \p_\mu\tilde\sigma\, \p^\mu\, \tilde\sigma\, .
\ee
This is the result when the string metric is set equal to $\eta_{\mu\nu}$.
As discussed below \refb{e74pre}, when the canonical
metric is set equal to $\eta_{\mu\nu}$ the action has a further multiplicative factor of
$2\kappa^2/V$. This gives the action for $\tilde\sigma$ to be:
\be\label{e720}
-\int d^4 x \, {\kappa^2\over 2 V^2}\,  \p_\mu\tilde\sigma\, \p^\mu\tilde\sigma\, .
\ee

We can now use the discussion below \refb{edual}
to rewrite \refb{ampl-diskB} using the momentum space
version of \refb{ehpsi}. This expresses the disk amplitude for $\tilde\sigma$ as:
\be\label{exyz}
{1\over 2}\,  \pi^2 \kappa \, \cTR_\gamma \,   {1\over 6V}\, \eps_{\rho\mu\nu\tau} \, p^\tau
\,  (\gamma^{\rho\mu\nu})_{\dalpha\alpha}\, \tilde\sigma\, \tilde\chi^\alpha \tilde\chi^\dalpha\, .
\ee
Using $\eps_{\rho\mu\nu\tau} \,  (\gamma^{\rho\mu\nu}) = -6\I \gamma_\tau\gamma^5$
where $\gamma^5=\I\gamma^0\gamma^1\gamma^2\gamma^3$
and that $\gamma^5$ acting on an undotted index gives $1$, we can express \refb{exyz} as:
\be
-{\I\over 2V}\,  \pi^2 \kappa \, \cTR_\gamma \,    p^\rho (\gamma_\rho)_{\alpha\dalpha}
\, \tilde\sigma\, \tilde\chi^\alpha \tilde\chi^\dalpha\, .
\label{contrib-chi}
\ee
Comparing this with \refb{eterm}, we get
\be \label{eachi}
a_{\tilde\sigma} =-{1\over 2\, V}\,  \pi^2 \kappa \, \cTR_\gamma \, .
\ee

Our next task will be to find the relation between $\wt\sigma$ and the field $\sigma$
that appears in the analysis of \S\ref{sduality}. For this we compare the
metric \refb{e720} with the perturbative metric for the field $\sigma$ given in \refb{hypmettree}.
In the absence of RR background, the metric involving $\sigma$
is given by $d\sigma^2/ (16 r^2)$,
leading to an action:
\be\label{esigmaaction}
-\int d^4 x\, {1\over 32\, r^2}\, \p_\mu\sigma \p^\mu\sigma\, .
\ee
This agrees with \refb{e720} after we make the identification:
\be \label{e7.36}
\tilde\sigma = -{V \over 4 r \kappa} \, \sigma = -{\kappa\over 2 \pi} \, \sigma\, ,
\ee
where in the second step we used \refb{rel-rgs}.
Note that \refb{e7.36} is correct up to a sign. The correct
sign can in principle be determined by comparing string amplitudes involving $\sigma$
and $\tilde\sigma$ in the two formalisms, but we have not done this. We shall see that
the choice of sign made in \refb{e7.36} reproduces correctly the result of \S\ref{sduality}.
Then \refb{eachi} and \refb{e7.36} imply
\be\label{easigmapre}
a_\sigma d\sigma ={\pi \kappa^2\over 4 V}  \, \cTR_\gamma\,  d\sigma\, .
\ee

We now note from \refb{hypmettree} that when RR fields are not set to 0, $\p_\mu\sigma$ in
\refb{esigmaaction} is replaced by $(\p_\mu\sigma +\tzeta_\Lambda\p_\mu \zeta^\Lambda-\zeta^\Lambda\p_\mu \tzeta_\Lambda)$.
Comparison with \refb{edual} shows that this
is the effect of Chern-Simons type of coupling of $H$ as in \refb{eoriginal} with
$A_\mu=-{\kappa\over 2 \pi} \,(\tzeta_\Lambda\p_\mu \zeta^\Lambda-\zeta^\Lambda\p_\mu \tzeta_\Lambda)$. We also see from
\refb{ehdeformed} that its effect is to
replace $\p_\mu\sigma$ by $(\p_\mu\sigma +\tzeta_\Lambda\p_\mu \zeta^\Lambda-\zeta^\Lambda\p_\mu \tzeta_\Lambda)$
in the correction terms. Therefore, we conclude that \refb{easigmapre} should be replaced by
\be\label{easigma}
(a_\sigma d\sigma)_{\rm mod}
={\pi \kappa^2\over 4 V} \, \cTR_\gamma
(d\sigma + \tzeta_\Lambda \, d \zeta^\Lambda-\zeta^\Lambda\, d \tzeta_\Lambda)\, .
\ee

\subsection{RR three-form contribution}

In this section we shall compute the disk amplitude with one insertion of the vertex
operator of the RR 3-form field
and two insertions of  open string vertex operators representing
fermion zero modes. For RR vertex operators
we shall follow the conventions of appendix \ref{errkin}.

We start with a flat ten-dimensional Euclidean D2-brane and denote by
\be
v ={1\over 3!}\,v_{IJK} dx^I\wedge dx^J\wedge dx^K
\ee
the volume form on the D2-brane.
According to \refb{edisk2}, \refb{edisk3} the relevant disk amplitude
with one RR closed string field with polarization
$F^\gamma_{~\delta}$, normalized so that the vertex operator in the $(-1/2,-1/2)$ picture
is given as in \refb{ephiR},
and two open string zero mode fields $\XX^\alpha$, $\wh\XX^\beta$, is
\be
\I \pi \kappa\, \ts_2 \, F^\gamma_{~\delta} \, \XX^\alpha \wh\XX^\beta
\int dz\, \Bigl\langle c\bar c e^{-\phi/2} S_\gamma e^{-\bar\phi/2} \bar S^\delta(\I)\, c \,e^{-\phi/2} S_\alpha(0)\,
e^{-\phi/2}S_\beta(z)\Bigr\rangle\, .
\ee
We shall take the $z$ contour to lie above the origin as before.
The boundary condition on the real line takes the form:
\be
\bar S^\delta = {1\over 3!} \, v_{IJK} \, (\Gamma^{IJK})^{\delta\delta'}S_{\delta'} \,.
\ee
Therefore, we can
use the doubling trick to represent the closed string vertex operator as
\be
-{1\over 3!}\, c\, e^{-\phi/2} \, S_\gamma(\I) \, c\, e^{-\phi/2} \, v_{IJK}\,
(\Gamma^{IJK})^{\delta\delta'}S_{\delta'}(-\I).
\ee
We can now calculate the correlation function by first deforming the $z$ contour to pick up residue
at $\I$. Using \refb{espinope}, we see that
this makes the operator at $\I$ to be $2\pi\, c\, e^{-\phi} \psi_M (\Gamma^M)_{\beta\gamma}$.
Thus, we are left with a $\psi$-$S$-$S$ correlator that gives a term proportional to
$(\Gamma_M)_{\delta'\alpha}$. The result is:
\be
- {1\over 3!}
\, \pi^2 \kappa\, \ts_2 \, \, F^\gamma_{~\delta}\, \, \XX^\alpha\, \wh\XX^\beta\, (\Gamma^M)_{\beta\gamma}  \, v_{IJK}\,
(\Gamma^{IJK}\Gamma_M)^\delta_{~\alpha}\, (2\pi)^3 \, \delta^{(3)}(0)\, .
\ee
Now suppose that $\wt C_{PQR}$ is the RR 3-form field normalized as in \refb{ecnorm}.
Then interpreting the $(2\pi)^3 \,\delta^{(3)}(0)$ factor as giving the integration over the
D2-brane world-volume and using \refb{ecnorm}, we see that the
amplitude given above leads to an effective action of the form:
\be
\begin{split}
&- \I\,  {\pi^2\over 36} \, \kappa\, \ts_2 \, \XX^\alpha \wh\XX^\beta\, p_N
\int  v \  v_{IJK}\, \wt C_{PQR} (\Gamma^{NPQR})^{\gamma}_{~\delta} \,
(\Gamma^M)_{\beta\gamma} (\Gamma^{IJK}\Gamma_M)^{\delta}_{~\alpha}
\\
=&-\I \, {\pi^2\over 36}\, \kappa\, \ts_2 \, \XX^\alpha \wh\XX^\beta\, p_N
\int v \  v_{IJK}\, \wt C_{PQR} (\Gamma^M \Gamma^{NPQR}
\Gamma^{IJK}\Gamma_M)_{\beta\alpha}\, ,
\end{split}
\label{dA-RR}
\ee
where the integral runs over the D2-brane world-volume and the first factor of $v$ is to be
regarded as a three form.

Now consider the D2-brane wrapped on a 3-cycle $\CG\subset\CY$.
Since we work in the large volume limit, we can use \refb{dA-RR} if we
interpret it as an integral along the D2-brane world-volume and replace $v$ by the
volume form $v_\gamma$ on $\CG$.
We shall use
notations introduced in \S\ref{scompact} for labelling compact and non-compact coordinates.
Since to compute the metric using \refb{eterm},
we need the contribution proportional to $p_\mu$ and the internal components of the
3-form field,
we are interested in the following contribution
to \eqref{dA-RR}
\be
-\I\, {\pi^2\over 36}\, \kappa\, \ts_2 \, \XX^\alpha \wh\XX^\beta\, p_\mu
 \int_{\CG}  v_\gamma \, \wt C_{\bf pqr}\, v_{\gamma,\bf ijk}\,
 (\Gamma^M \Gamma^{\mu \bf pqr}
\Gamma^{\bf ijk}\Gamma_M)_{\beta\alpha} \, .
\ee
Expressing the zero modes
$\XX^\alpha$ and $\wh\XX^\beta$
as $\eta\otimes \tilde\chi^\alpha$ and $\bar \eta \otimes \tilde\chi^\dbeta$ as before,
reduces this contribution to
\be\label{e744}
-\I\,   {\pi^2\over 36}\, \kappa\, \ts_2 \, \tilde\chi^\alpha \tilde\chi^{\dbeta}\, p_\mu
 \int_{\CG} v_\gamma\,  \wt C_{\bf pqr}\,  v_{\gamma,\bf ijk}\,
 (\bar\eta\Gamma^M \Gamma^{\mu \bf pqr}
\Gamma^{\bf ijk}\Gamma_M\eta)_{\dbeta\alpha}\,.
\ee
Note that the contraction with $\eta,\bar\eta$ removes the 6-dimensional part of the spinor index
but leaves behind the 4-dimensional part.
Once we express the sum over $M$ as separate sums over the non-compact index
$\mu$ and compact index $\bf m$, this expression has a fully covariant expression
under coordinate transformations in $\CY$. Furthermore, we can freely switch back and forth
between flat tangent space indices and space-time indices.

Using \refb{etenfourgamma} and \refb{eetanorm}, we get
\be
(\bar\eta\, \Gamma^M \Gamma^{\mu \bf pqr}
\Gamma^{\bf ijk}\Gamma_M\eta)_{\dbeta\alpha}
= (\gamma^\mu)_{\dbeta\alpha} \(-2 \, \bar\eta \, \wt\Gamma^{\bf pqr}
\wt\Gamma^{\bf ijk} \eta -
\bar\eta \, \wt\Gamma^{\bf m} \wt\Gamma^{\bf pqr}
\wt\Gamma^{\bf ijk}\wt\Gamma_{\bf m}\eta\) .
\ee
We can further simplify the gamma-matrices by assuming the general form:
\be
\begin{split}
& -2 \, \wt\Gamma^{\bf pqr}
\wt\Gamma^{\bf ijk}  -
 \wt\Gamma^{\bf m} \wt\Gamma^{\bf pqr}
\wt\Gamma^{\bf ijk}\wt\Gamma_{\bf m} =c_1\, \delta^{\bf pqr,\bf ijk}+{c_2}\,\wt\Gamma^{\bf pqr\bf ijk}
\\
& + c_3 \(\delta^{\bf pq,ij} \, \wt\Gamma^{\bf rk}
 + \hbox{cyclic perm.}\) + c_4 \(\delta^{\bf p,i} \, \wt\Gamma^{\bf qrjk}
+ \hbox{cyclic perm.}\) ,
\end{split}
\label{compGam}
\ee
where `cyclic perm.' denote cyclic permutations of $\bf pqr$ and independently of $\bf ijk$,
producing nine terms inside each parentheses.
Our convention for the $\delta$-symbols are:
\be
\delta^{\bf i_1\cdots i_n\, ,\,  j_1\cdots j_n}
= g^{\bf i_1 j_1}\cdots g^{\bf i_n j_n} + \hbox{$(-1)^P$ weighted
perm. of {$\bf j_1,\cdots, j_n$}}\, ,
\ee
with the understanding that when we use tangent space indices, $g^{\bf ij}$ will be replaced
by $\delta^{\bf ij}$.
We can
determine the coefficients $c_i$ by comparing both sides for the following inequivalent choices
of $(\bf ijk,\bf pqr)$: (456,789), (456,489), (456,459) and (456,456). This gives
\be
c_1=8, \quad c_2=4, \quad c_3=4, \quad c_4=0\, .
\ee
Using these results, we can rewrite \refb{e744} as
\be\label{e744a}
-\I\, \pi^2\kappa \,\ts_2 \, (\gamma^\mu)_{\dbeta\alpha}  \,
 \tilde\chi^\alpha \tilde\chi^{\dbeta}\, p_\mu \, \cI_\gamma\, ,
\ee
where
\be
\cI_\gamma= \frac{1}{36}\int_{\CG}  v_\gamma\,
\wt C_{\bf pqr}\,  v_{\gamma,\bf ijk} \,
\bar\eta \bigg[8\, \delta^{\bf pqr,\bf ijk}+{4}\,\wt\Gamma^{\bf pqr\bf ijk}
 + 4 \(\delta^{\bf pq,ij} \, \wt\Gamma^{\bf rk}
 + \hbox{cyclic perm.}\)\bigg]
\eta\, .
\label{intRR}
\ee
Using \eqref{eetanorm}, \refb{etaeps} and \refb{ereltwo}, this integral can be reduced to
\be
\begin{split}
\cI_\gamma=&\, \frac{1}{9}\int_{\CG} v_\gamma\, \wt C_{\bf pqr}\,  v_{\gamma,\bf ijk}\,
\(2\,\delta^{\bf pqr,\bf ijk}-\I\,\eps^{\bf pqr\bf ijk}
-3\I\, \delta^{\bf pqs,\bf ijk}J_{\bf s}^{\bf r}\)
\\
=&\,4 \int_{\CG} \Bigl( 2\, \wt C +
\I \star  \wt C- 3\I\, J( \wt C)\Bigr)
\, ,
\end{split}
\label{intIg}
\ee
where $J$ is the complex structure of $\CY$ and $J(C)$ has been defined in \refb{edefjc}.
In arriving at the
last expression we have used \refb{projC}.

We now need to find the relative normalization between the field $\wt C_{MNP}$ and the
field $C_{MNP}$ that defines the RR scalars in \S\ref{sduality}. In principle this could be done by comparing the
string field theory kinetic term written in terms of $\wt C_{MNP}$ with the kinetic term for
the fields $\zeta^\Lambda,\tilde\zeta_\Lambda$ following from the metric \refb{hypmettree}.
However, writing down the kinetic term of the RR fields requires introducing an additional free field and then
decoupling one combination of the two fields \cite{deLacroix:2017lif}.
We shall avoid this by computing the one-point
function of $\wt C_{MNP}$ on the disk, and then comparing the result with the imaginary part
$-2\pi\I\Theta_\gamma$ of
the instanton action. For this we need to use the vertex operator of 3-form field in the $(-1/2,-3/2)$
picture. Using \refb{efg3} with $A_\alpha^{~\beta}=0$, \refb{efg7} and \refb{efg11},
we see that the relevant part of the vertex operator is given by:
\be
\wt V = {\I\over 2\cdot 3!} \, \wt C_{MNP} \[(\Gamma^{MNP})_{\alpha\beta}\,
c\, \bar c\, e^{-3\phi/2} S^\alpha e^{-\bar\phi/2} \bar S^\beta
+ (\Gamma^{MNP})^{\alpha\beta}\, c\, \bar c\, \p\xi \, e^{-5\phi/2}
S_\alpha \,  \bar\eta \, e^{\bar\phi/2} \, \bar S_\beta\] e^{\I p.X}\, ,
\ee
where we have dropped the term in the third line of \refb{efg3} since it does not satisfy the
$\xi$-$\eta$ charge conservation and therefore will have vanishing one-point function on the
disk.
Also for this computation we can drop the $e^{\I p.X}$ factor.
From \refb{edisk1} we now see that the disk one-point function of $\wt C_{MNP}$ takes the form:
\be
\begin{split}
& A\equiv  {\kappa\, \ts_2\over 2} \, {\I\over 2\cdot 3!} \, {1\over 2}\,  \wt C_{MNP}\,
\Bigg[(\Gamma^{MNP})_{\alpha\beta}\, \langle (\p c -\bar\p \bar c) \,
c\, \bar c\, e^{-3\phi/2} S^\alpha e^{-\bar\phi/2} \bar S^\beta (\I)
\\ & \hskip 1.7in + (\Gamma^{MNP})^{\alpha\beta}\, \langle (\p c -\bar\p \bar c) c\, \bar c\, \p\xi \, e^{-5\phi/2}
S_\alpha \,  \bar\eta \, e^{\bar\phi/2} \, \bar S_\beta(\I)\,
\rangle
\Bigg]\, .
\end{split}
\ee
After evaluating the correlation function using the doubling trick, we get:
\be
A={1\over 4}\, \kappa\, \ts_2
\, {\I\over 2\cdot 3!} \, \wt C_{MNP} \times {1\over 3!} \, v^{MNP} \times 2 \times 2 \times (-3!)
\times 16 \times (2\pi)^3 \, \delta^{(3)}(0)\, .
\ee
Here the 16 comes from the trace of the identity operator in the spinor representation, one
factor of 2 comes from the $(\p c -\bar\p \bar c)$ term, the second factor of 2 represents the
contribution from two terms inside the square bracket and the $(-3!)$ comes from contraction
of the gamma matrices.
Interpreting the $(2\pi)^3 \, \delta^{(3)}(0)$ factor as the
integral over the 3-cycle $\CG$, we get
\be
A = -8 \I  \kappa\, \ts_2 \int_{\CG} \wt C\, .
\ee
Equating this with $-2\pi \I \int_{\CG} C$, we obtain
\be \label{ectcrel}
\wt C =  {\pi\over 4\kappa\, \ts_2}\, C\, .
\ee

Let us now return to the evaluation of the integral \eqref{intIg}.  Using the relation \refb{ectcrel}
and noticing that the three terms precisely correspond to the three integrals in \eqref{treeints}, we get
\be
\cI_\gamma={\pi\over \kappa\, \ts_2}\, \bigl(2\, \Theta_\gamma+\I\, \Theta^\star_\gamma
-3\,\I\, \Theta^J_\gamma\bigr).
\ee
The three integrals were evaluated in \eqref{intCC}, \eqref{resThetastar} and \eqref{resThetaJ}, respectively.
Substituting these results, one finds
\be
\cI_\gamma={2 \pi\over \kappa\, \ts_2}\,\Bigl[\Theta_\gamma+2 \I\, \mathscr{C}_{\gamma}\Bigr],
\label{contrib-RR}
\ee
where $\mathscr{C}_\gamma$ has been defined in  \eqref{def-niceC}.
Substituting this into \refb{e744a}, we get
\be
-2 \, \I\, \pi^3(\gamma^\mu)_{\dbeta\alpha}  \, p_\mu \,  \tilde\chi^\alpha  \tilde\chi^{\dbeta}\,
\Bigl[\Theta_\gamma+2\I\,\mathscr{C}_{\gamma}\Bigr].
\ee
Comparison with \refb{eterm} now gives:
\be\label{eazeta}
a_\Lambda d\zeta^\Lambda +a^\Lambda  d\tilde\zeta_\Lambda
= -2\, \pi^3\, \Bigl[d\Theta_\gamma+2 \I\,\cC_{\gamma}\Bigr]\, ,
\ee
where $\cC_{\gamma}$ defined in  \eqref{connC} coincides with $\de \mathscr{C}_\gamma$
where the differential acts only on the RR fields.

\subsection{Complex structure moduli contribution} \label{ssixfour}

We shall now use \refb{edisk2}, \refb{edisk3} to
compute the disk amplitude of the metric fluctuation with
components along $\CY$ and a pair of open string fermion
zero modes $\XX^\alpha$ and $\wh\XX^\beta$  in the presence of a
D2-brane instanton.
At this stage we shall not commit ourselves to whether the closed string corresponds to
metric or 2-form components and denote the polarization by $e_{\bf ij}$. Then the vertex
operator has the form
\be
V_e = 2\, e_{\bf ij} \,
c\, \bar c \(\p X^{\bf i} + i\, p_\rho\, \psi^\rho\psi^{\bf i}\)
e^{\I p.X}
e^{-\bar\phi}\bar\psi^{\bf j}(\I) +\cdots\, ,
\ee
so that the amplitude is given by
\be\label{ens1}
2 \, \I\, \pi\, \kappa\, \ts_2\, e_{\bf ij} \, \XX^\alpha \wh\XX^\beta\,
\int dz \left\langle c\bar c \(\p X^{\bf i} + \I\, p_\rho\, \psi^\rho\psi^{\bf i}\)
e^{\I p.X}
e^{-\bar\phi}\bar\psi^{\bf j}(\I)\,  c e^{-\phi/2} S_\alpha(0)\,
e^{-\phi/2}S_\beta(z)\right\rangle\, .
\ee
We can drop the $\p X^{\bf i}$ and $e^{\I p.X}$ terms since they do not contribute to the
correlation function. Also, using the doubling trick, we can replace the
$\bar c\, e^{-\bar\phi}\bar\psi^{\bf j}(\I)$ term by $(P^{\bf j}_{\bf k} - Q^{\bf j}_{\bf k})\,
 c \, e^{-\phi} \psi^{\bf k}(-\I)$ where $P$ and $Q$ are projection operators to subspaces
 tangent to the
 brane and transverse to the brane respectively.
This allows us to replace \refb{ens1} by
\be\label{ens2}
2\, \I\,  \pi \, \kappa\, \ts_2\, \I\, p_\rho\, \XX^\alpha  \wh\XX^\beta\,
e_{\bf ij}  \,(P^{\bf j}_{\bf k} - Q^{\bf j}_{\bf k}) \,
\int dz \left\langle c \, \psi^\rho\psi^{\bf i} (\I)\,
c\, e^{-\phi}\psi^{\bf k}(-\I)\,  c e^{-\phi/2} S_\alpha(0)\,
e^{-\phi/2}S_\beta(z)\right\rangle\, .
\ee
We can now deform the $z$ contour to pick up residue at $\I$ and then evaluate the
resulting correlator. The result  takes the form:
\be
\begin{split}
& -{1\over 2}\, p_\rho\,
 \XX^\alpha \wh\XX^\beta\, e_{\bf ij}  \,
 \pi^2\, \kappa\, \ts_2\,    (P^{\bf j}_{\bf k} - Q^{\bf j}_{\bf k})\,
 (\Gamma^\rho\Gamma^{\bf i}\Gamma^{\bf k})_{\beta\alpha}  \, (2\pi)^3 \, \delta^{(3)}(0)\\
 =&
-{1\over 2}\, p_\rho\,  \tilde\chi^\alpha \tilde\chi^\dbeta\, e_{\bf ij}  \,
\pi^2\, \kappa\, \ts_2\,    (P^{\bf j}_{\bf k} - Q^{\bf j}_{\bf k})
\, \gamma^\rho_{\dbeta\alpha}\,
\bar\eta\wt\Gamma^{\bf i}\wt\Gamma^{\bf k}\eta   \, (2\pi)^3 \, \delta^{(3)}(0)\, ,
\end{split}
\label{elag10}
\ee
where in the second step we have used four-dimensional notation for the zero modes.

In  appendix \ref{sc}, we prove the following result
\be
e_{\bf ij}\, (P^{\bf j}_{\bf k} - Q^{\bf j}_{\bf k}) \,  \bar\eta\wt\Gamma^{\bf i}\wt\Gamma^{\bf k}\eta=4\, e_{ss'}P^{ss'},
\label{eee-main}
\ee
where $s,s'=1,2,3$ label the holomorphic coordinates on $\CY$. Importantly, the r.h.s.
involves only the symmetric, holomorphic components of $e_{\bf ij}$, showing that
neither the internal components of the NSNS 2-form field related to anti-symmetric
components of $e_{\bf ij}$,
nor the K\"ahler moduli of the metric related to the mixed components $e_{s\bar t}$, contribute  to
the disk amplitude we are interested in. This is consistent with the fact that both the 2-form field along $\CY$ and
the K\"ahler moduli are parts of the vector multiplet moduli space and
should not appear in the hypermultiplet action.

On the other hand, the holomorphic components of $e_{\bf ij}$ appearing in \eqref{eee-main}
are related to the $h^{2,1}$ complex structure deformations $\delta z^a$ of the CY metric.
In the normalization convention of \cite{Sen:2021tpp} that we are using, we
identify $e_{\bf ij}=\delta g_{\bf ij}/(2\kappa)$ and (see, {\it e.g.}, eq.(7) in \cite{Bodner:1990zm})
\be
\delta g_{ss'} =  \delta \bz^a \, {\Omega_{str} g^{t\bar t'} g^{ r \bar r'}\over ||\Omega||^2}
\, (\bar\chi_a)_{s' \bar t'\bar r'} \, .
\label{decom-gmna}
\ee
In appendix \ref{sc}, eq. \refb{eappfin} we also show that this variation of the metric satisfies
\be \label{eappfinrep}
P^{ss'}\delta g_{ss'}
=\delta \bz^a \, \frac{(v_\gamma,\chi_a)}{(v_\gamma,\Omega)}, \qquad
(C,C')\equiv {1\over 3!} \, C^{\bf ijk} \, \bar C'_{\bf ijk}\, .
\ee
Using  \eqref{eee-main}, \eqref{eappfinrep} and the relation
$e_{ss'}=\delta g_{ss'}/(2\kappa)$ in \eqref{elag10}
and interpreting the factor $(2\pi)^3 \, \delta^{(3)}(0)$ as a result of integration
over the D2-brane world-volume, we get
\be \label{delz-contr}
- p_\rho\,
\pi^2 \,  T_2    \, \tilde\chi^\alpha \tilde\chi^\dbeta
\,\gamma^\rho_{\dbeta\alpha}
\delta \bz^a \int_{\CG} v_\gamma \, \frac{(v_\gamma,\chi_a)}{(v_\gamma,\Omega)}\, .
\ee
Using \refb{vgammasquare},
\refb{projC}, \eqref{BPScycle}, \eqref{rel-fVZ}, \refb{Realact} and \refb{defchia},
we can rewrite this expression as
\be \label{delz-contr-final}
- \pi^2 \, p_\rho\,\cTR_\gamma   \, \tilde\chi^\alpha \tilde\chi^\dbeta\,
\gamma^\rho_{\dbeta\alpha}
\frac{\delta \bz^a}{\bZ_\gamma} \int_{\CG}\bar\chi_a
= -\pi^2 \, p_\rho\,\cTR_\gamma   \, \tilde\chi^\alpha \tilde\chi^\dbeta\,
\gamma^\rho_{\dbeta\alpha}\bar\delta\log(\bZ_\gamma/K)\, .
\ee
Comparing this with \refb{eterm}, we obtain
\be \label{eaza}
 a_{z^a} dz^a+ a_{\bar z^a} d\bar z^a =
\I\, \pi^2\,  \cTR_\gamma\,
\bar\p\log(\bZ_\gamma/K) \, .
\ee

As a cross check on the normalization of the closed string vertex operator,
we have also verified that the one-point function of the
vertex operator of $z^a$ on the disk without any fermion zero mode insertion agrees with
$-\p\TT_\gamma/\p z^a$. We shall not present the analysis here.

\subsection{Dilaton contribution}

We shall now compute the coefficient $a_m$ associated with the dilaton field $r$.
We start with the dilaton vertex operator in the $(-1,-1)$ picture which, at non-zero
momentum, is given by
\be \label{ebrstoriginal}
\wt V_{-1,-1}= f_{\mu\nu}\,  c\, \bar c\, e^{-\phi}\, \psi^\mu \, e^{-\bar\phi} \bar\psi^\nu \, e^{\I p.X}
\, ,
\ee
where
\be \label{emunuform}
f_{\mu\nu} \propto \left\{\eta_{\mu\nu} - (n.p)^{-1} \, \(n_\mu p_\nu + n_\nu p_\mu\)\right\} ,
\ee
and $n$ is any four-vector for which $n.p\ne 0$.
We shall determine the constant of
proportionality in \refb{emunuform} shortly. \refb{emunuform}
ensures that $p^\mu f_{\mu\nu}=0$, which, together with the
on-shell condition $p^2=0$, ensures BRST invariance of the vertex operator.
The vertex operators for different choices of $n$ differ by BRST exact states.
One can also construct a fully covariant vertex operator in the same BRST
cohomology class, given by:
\be
\wt V'_{-1,-1}\propto \[
\eta_{\mu\nu}\,  c\, \bar c\, e^{-\phi}\, \psi^\mu \, e^{-\bar\phi} \bar\psi^\nu \, e^{\I p.X}
+ {1\over 2}\, c\, \bar c\,  (\eta\, \bar\p \bar\xi\, e^{-2\bar\phi} - \p\xi\, e^{-2\phi} \, \bar\eta)
\, e^{\I p.X}\] .
\ee
We have checked that the computation with this vertex operator gives the same result as
with \refb{ebrstoriginal}. We shall present our analysis using \refb{ebrstoriginal}, since the
computations involved are simpler.

According to \refb{edisk1}, the disk one-point function of the vertex operator is given by:
\be
{\kappa\, \ts_2\over 2} \, {1\over 2} \,\bigl\langle
\(\p c(\I)-\bar\p\bar c(\I)\) \wt V_{-1,-1}(\I)\bigr\rangle
= -{\kappa\, \cTR_\gamma\over 4}\, f_\mu^{~\mu}\, ,
\ee
where we have replaced the $(2\pi)^3\,\delta^{(3)}(0)$ factor by the volume integral over
$\CG$ and used \refb{Realact}.
We would like to normalize the dilaton so that the deformation of the dilaton can be
identified with $\delta r/r$. From the expression for $\TT_\gamma$ given in
\eqref{actD2} we see that the expected
one-point coupling of $\delta r/r$ is $-\cTR_\gamma/2$. Therefore, we choose
\be \label{edilnorm}
f_\mu^{~\mu} = {2\over \kappa}\, {\delta r\over r}\, .
\ee

We shall now use \refb{edisk2}, \refb{edisk3} to
compute the disk amplitude with one dilaton and a pair of open string
fermion zero modes. For this we need the dilaton vertex operator in the $(-1,0)$ picture.
This is given by:
\be
V_{-1,0} = - f_{\mu\nu} \,
c\, \bar c\, \left\{\p X^{\mu} + i\, p_\rho\, \psi^\rho\psi^\mu\right\}
e^{\I p.X}
e^{-\bar\phi}\bar\psi^\nu(\I) +\cdots\, .
\ee
This has the same form as \refb{ebmunuvertex} with $b_{\mu\nu}$ replaced by $-f_{\mu\nu}/2$.
As a consequence the analysis of the amplitude follows exactly the same route as in
\S\ref{s7.1} and
leads us to the left hand side of \refb{e610} with $b_{\mu\nu}$ replaced by $-f_{\mu\nu}/2$:
\be
-{1\over 4}\, \pi^2\kappa \, \ts_2 \,   p_\sigma\, f_{\mu\nu} \,  \XX^\alpha\wh\XX^\beta\,
(\Gamma^{\sigma\mu}\Gamma^{\nu})_{\beta\alpha}\,  (2\pi)^3 \delta^{(3)}(0)
= -{1\over 4}\, \pi^2\kappa \, \ts_2 \,   p_\sigma\, f_{\mu\nu} \,  \tilde\chi^\alpha\tilde\chi^\dalpha\,
(\gamma^{\sigma\mu}\gamma^{\nu})_{\dalpha\alpha}\,  (2\pi)^3 \delta^{(3)}(0)\, .
\ee
We now replace $(2\pi)^3 \delta^{(3)}(0)$ by $V_\gamma$ as usual. Also expressing
$(\gamma^{\sigma\mu}\gamma^{\nu})$ as $\gamma^{\sigma\mu\nu}+\gamma^\sigma
\eta^{\mu\nu}- \gamma^\mu\eta^{\sigma\nu}$, and using the form of $f_{\mu\nu}$
given in \refb{emunuform}, we see that only the $\gamma^\sigma
\eta^{\mu\nu}$ term contributes. Finally, using \refb{edilnorm},
the resulting expression may be written as:
\be
-{1\over 2}\, \pi^2 \, \cTR_\gamma \,   p_\sigma\,  {\delta r\over r}\,
\tilde\chi^\alpha\tilde\chi^\dalpha\,
(\gamma^{\sigma})_{\dalpha\alpha}\, .
\label{edilfin}
\ee
Comparing \refb{edilfin} with \refb{eterm}, we obtain
\be \label{earfin}
a_r dr = \I\, {\pi^2\over 2} \, \cTR_\gamma\, {dr\over r}\, .
\ee

\subsection{Final result}

Combining \eqref{easigma}, \eqref{eazeta}, \eqref{eaza} and \refb{earfin}, one obtains the
combined contribution
of the NSNS 2-form field, the complex structure moduli, the RR moduli and the dilaton:
\ben\label{edifforiginal}
&& (a_\sigma d\sigma)_{\rm mod}
+ \(a_\Lambda d\zeta^\Lambda + a^\Lambda d\wt\zeta_\Lambda\) +  \bigl( a_{z^a} dz^a+ a_{\bar z^a} d\bar z^a\bigr) + a_r dr
\\
&=& \pi^2 \cTR_\gamma\[
{\kappa^2\over 4\pi V}\,  (d\sigma + \tzeta_\Lambda \, d \zeta^\Lambda-\zeta^\Lambda\, d \tzeta_\Lambda)
+ \I\, \bar\p\log(\bZ_\gamma/K)
+ \I \, {dr\over 2 r} \]
- 2\, \pi^3 \Bigl[d\Theta_\gamma+ 2\, \I\, \cC_{\gamma}\Bigr].
\non
\een
From \refb{actD2} we have:
\be
d\cT_\gamma=
\cTR_\gamma\({1\over 2} {dr\over r}-{1\over 2} {dK\over K}
+ {d|Z_\gamma|\over |Z_\gamma|}\) +2\pi\I \, d\Theta_\gamma.
\label{actD2deriv}
\ee
Since we have argued that we can determine the metric only up to terms proportional to
$d\TT_\gamma$, we can set $d\TT_\gamma=0$ for our analysis. Using this we can
eliminate the $d\Theta_\gamma$ term, and then use \refb{rel-rgs} to express
\refb{edifforiginal} as:
\be
\begin{split}
\sum_m a_m d\lambda^m = &\,
\frac{ \pi^2 \cTR_\gamma}{2}\[
{1\over 4r}\,  (d\sigma + \tzeta_\Lambda \, d \zeta^\Lambda-\zeta^\Lambda\, d \tzeta_\Lambda)
-\I\, \(\p\log(Z_\gamma/K)-\bar\p\log(\bZ_\gamma/K)\)\]
\\
&\,
- 4 \pi^3  \I\, \cC_{\gamma}+O(\de\cT_\gamma)\, .
\end{split}
\ee
Substituting this into \refb{e75} and expressing $\cTR_\gamma$ as the real part of
\refb{actD2},
we get the correction to the metric where we keep only the
leading contribution in each homology class.
The result is
\be\label{e686}
\begin{split}
&\frac{\pi^6}{K r} \sum_\gamma
\Omega_\gamma \,|Z_\gamma|^2
\(\sum_{k=1}^\infty \cN_{k,\gamma}\, e^{-k\TT_\gamma}\)
\biggl[(d\sigma + \tilde\zeta_\Lambda \, d \zeta^\Lambda
-\zeta^\Lambda\, d \tilde\zeta_\Lambda)
-  4 \I \, {\sqrt{K\, r}\over |Z_\gamma|} \ \cC_{\gamma}
\\
&\qquad
- 4 \I  r\,  \Bigl(\p\log(Z_\gamma/K)-\bar\p\log(\bZ_\gamma/K)\Bigr)\biggr]^2\,
+ \OO(d\TT_\gamma)\, ,
\end{split}
\ee
where $\cN_{k,\gamma}$ is given in \eqref{efincorrect}. After setting $\sigma_\gamma=1$,
the prediction
\refb{square} based on supersymmetry and duality symmetries
agrees with  \refb{e686} up to terms proportional to $d\TT_\gamma$, which, according
to our arguments in \S\ref{sstrategy}, are ambiguous due to the possibility of field redefinition.
The role of the $\sigma_\gamma$'s has already been discussed at the end of \S\ref{sfourone}.

\bigskip

\noindent {\bf Acknowledgement:}
We wish to thank Boris Pioline for useful discussions and for comments on an earlier version
of the manuscript.
The work of A.S. was supported by the Infosys chair professorship and the
J. C. Bose fellowship of the Department of Science and Technology, India.
B.S. acknowledges funding support
from an STFC Consolidated Grant ‘Theoretical Particle Physics at City, University of London' ST/T000716/1.

\appendix

\section{The phase of the mixed open-closed string amplitude} \label{se}

In this appendix we shall determine the phase $\ve$ that appears in \refb{edisk2}.
In principle this can be determined by carefully studying factorization in different channels,
but we shall determine this by analyzing a special class of amplitudes.

We consider the open string vertex operator associated to the translation zero mode along
the $M$th direction. We need the zero picture unintegrated and integrated vertex operators
which we denote by $V^M_{\rm un}$ and $V^M_{\rm int}$, respectively. They have the form:
\be\label{eappe1}
V^M_{\rm un} =  \I \sqrt 2\,  c\, \p X^M, \qquad V^M_{\rm int} = \I \sqrt 2\, \p X^M\, .
\ee
The overall normalization of the vertex operators will not be important for us but the
relative normalization will be important.

Now consider the effect of inserting such an integrated vertex operator into a closed string
amplitude on the disk carrying momenta $p_1,p_2,\cdots$ transverse to the D-brane.
This effect can be studied using the OPE following from \refb{ematterope}:
\be\label{eappe2}
V^M_{\rm int}(z) \, e^{\I \, p_s . X}(z_s) \simeq  {1\over \sqrt 2}\, {p_s^M\over (z-z_s)} \,e^{\I \, p_s . X}(z_s)\, .
\ee
This gives
\be\label{eappe3}
\int dz\, V^M_{\rm int}(z) \prod_s e^{\I p_s.X^s} (z_s)
= \pi \I \sqrt 2 \, \sum_s p_s^M \,  \prod_s e^{\I p_s.X^s} (z_s)\, .
\ee
In other words the effect of inserting the open string vertex operator into an amplitude
with closed strings carrying momenta $p_1,p_2,\cdots$
is to multiply the original amplitude by $ \pi \I \sqrt 2 \,  \sum_s p_s^M$. We shall now
see if this holds for an amplitude with a single closed string and a single open string zero
mode insertion.

Let us consider the closed string vertex operator,
\be \label{eappe4}
V_c = -2\, e_{NK} \, c\, \bar c\, e^{-\phi}\psi^N \, e^{-\bar\phi}\bar\psi^K\,
e^{\I p.X} \, , \qquad e_{NK} \, p^N=e_{NK} \, p^K=0, \quad p^2=0\, .
\ee
For definiteness we shall choose $e_{NK}$ to have components tangential to
the brane  and $p$ to be orthogonal
to the brane. We shall compute the disk one-point amplitude of this state using
\refb{edisk1}. Representing the disk as upper half plane and placing the vertex
operator $V_c$ at $\I$, we get:
\be\label{eappe5}
\begin{split}
\{ V_c\} &=  {1\over 2}\, \kappa \, T_p\, \(-2\, e_{NK}\) \, \left\langle
{1\over 2} \, (\p c -\bar\p \bar c) \,c \, e^{-\phi}\, \psi^N \, \bar c \, e^{-\bar\phi}\, \bar\psi^K
\, e^{\I p.X}(\I)\right\rangle \\
&= - {1\over 2} \,
\kappa \, T_p\, e_{NK} \, \bigg[ \Bigl\langle \p c \, c \, e^{-\phi}\psi^N\, e^{\I p.X}(\I )
\,\, c\, e^{-\phi}\, \psi^K \, e^{-\I p.X}(-\I)\Bigr\rangle
\\
& \hskip 1in + \Bigl\langle
c \, e^{-\phi}\, \psi^N \, e^{\I p.X}(\I)\,\, \p c \, c \, e^{-\phi}\,
\psi^K \, e^{-\I p.X}(-\I)\Bigr\rangle
\bigg]
\\
&= - {1\over 2} \, \kappa \, T_p\, e_{N}^{~N} \, ,
\end{split}
\ee
where in the second step we have used the doubling trick to replace the anti-holomorphic
fields by holomorphic fields at complex conjugate points.

Next we shall compute the disk two-point amplitude with the closed string state $V_c$
and the open string zero mode introduced in \refb{eappe1}. For this we need to use
the unintegrated open string vertex operator.
Using \refb{edisk2}, we get
\be\label{eappe6}
\begin{split}
\{V_c V^M_{\rm un}\} &= -2\, \ve\, \pi\kappa \, T_p \, e_{NK}\, \left\langle
c \, e^{-\phi}\, \psi^N \, \bar c \, e^{-\bar\phi}\, \bar\psi^K \, e^{\I p.X} (\I)
\   \I \, \sqrt 2\,  c\, \p X^M(0) \right\rangle
\\
&= -2 \, \ve \, \I \, \sqrt 2\, \pi\kappa T_p \, e_{NK}\, \Bigl\langle c \, e^{-\phi}\,
\psi^N \, e^{\I p.X}(\I) \ c \, e^{-\phi}\, \psi^K\,  e^{-\I p.X} (-\I) \
c\, \p X^M(0) \Bigr\rangle
\\
&= -{\ve\over \sqrt 2} \, \pi\, \kappa \, T_p \, p^M \, e_N^{~N}\, .
\end{split}
\ee
According to \refb{eappe3} the
desired result for this is $ \pi \I \sqrt 2 \, p^M $ times \refb{eappe5}. This gives
\be
- {\ve\over \sqrt 2}\, \pi\, \kappa T_p\, e_{N}^{~N} = - \pi \, \I \, \sqrt 2 \,  {1\over 2}\,
\kappa \, T_p\,
e_{N}^{~N}
\quad \Rightarrow \quad \ve = \I\, .
\ee

Note that if we had more external open strings, the corresponding
vertex operators would be integrated. For these we shall automatically get the desired factor
$ \pi \I\sqrt 2 \, p^M $ by our previous analysis and hence $\ve$ is independent of
the number of open string insertions.

\section{The leading instanton contribution to the hypermultiplet metric}
\label{ap-metric}

In this appendix we shall compute the D-instanton correction to the hypermultiplet metric,
keeping only the leading term in each homology class. Our starting point will be the metric computed in \cite{Alexandrov:2014sya}. This
will be reviewed below and then used to extract the leading contributions.

\subsection{The initial metric}
\label{ap-metric-start}

The hypermultiplet metric including perturbative and D-instanton corrections, but
ignoring contributions of multi-instantons with mutually non-local charges,
has been found in \cite[Eq.(3.6)]{Alexandrov:2014sya}.  It
is given by
\be
\begin{split}
\de s^2=&\,
\frac{2}{r^2} \(1-\frac{2r}{\cR^2\Uin}\)(\de r)^2
+\frac{1}{32r^2\(1-\frac{2r}{\cR^2\Uin}\)}\(\de \sigma +\tzeta_\Lambda \de \zeta^\Lambda-\zeta^\Lambda\de \tzeta_\Lambda+\cV \)^2
\\
&\,
+\frac{\cR^2}{2r^2}\,|z^\Lambda\cY_\Lambda|^2
+\frac{1}{r\Uin}\left|  \cY_\Lambda \Min^{\Lambda\Sigma}\bvl_\Sigma-\frac{\I\cR}{2\pi}\, \sum_\gamma \Om{\gamma}\cW_\gamma\de Z_\gamma\right|^2
\\
&\,
-\frac{1}{r}\,\Min^{\Lambda\Sigma}\(\cY_\Lambda
+\frac{\I\cR}{2\pi}\sum_\gamma \Om{\gamma} V_{\gamma\Lambda}\rIgamp{\gamma}\(\de Z_{\gamma}-\Uin^{-1}Z_\gamma\p K\)\)
\\
&,\qquad
\times\(\bar\cY_\Sigma-\frac{\I\cR}{2\pi}\sum_{\gamma'} \Om{\gamma'} \bV_{\gamma'\Sigma}
\rIgamm{\gamma'}\(\de \bZ_{\gamma'}-\Uin^{-1}\bZ_{\gamma'}\bar\p K\)\)
\\
&\,
+\frac{\cR^2 K}{r}\Biggl( \cK_{a\bar b}\de z^a \de\bz^b
-\frac{1}{(2\pi K\Uin)^2}\left|\sum_\gamma \Omega_\gamma Z_\gamma \cW_\gamma\right|^2 |\p K|^2
\Biggr.
\\
&\,\Biggl.\qquad
+\frac{1}{2\pi K}\sum_\gamma \Omega_\gamma \rIg\left|\de Z_\gamma-\Uin^{-1}Z_\gamma\p K\right|^2
\Biggr).
\end{split}
\label{mett1}
\ee
Here we used the following notations:
\begin{itemize}
\item
$\Ig{}$ with various upper indices denote the Penrose-like integrals which have their origin
in the twistorial formulation of D-instantons
\be
\begin{array}{rclrcl}
\Igg{}& = &\displaystyle
\int_{\ellg{\gamma}}\frac{\de t}{t}\,
\log\(1-\qr e^{-2\pi \I \Xi_\gamma(t)}\),
\ \ &
\rIg&=&
\displaystyle \int_{\ellg{\gamma}}\frac{\de t}{t}\,
\frac{1}{\qr e^{2\pi \I \Xi_\gamma(t)}-1},
\\
\Igpm& = &\displaystyle
\pm\int_{\ellg{\gamma}}\frac{\de t}{t^{1\pm 1}}\,\log\(1-\qr e^{-2\pi \I \Xi_\gamma(t)}\),
\ \ &
\rIgpm&=&
\displaystyle \pm \int_{\ellg{\gamma}}\frac{\de t}{t^{1\pm 1}}\,
\frac{1}{\qr e^{2\pi \I \Xi_\gamma(t)}-1},
\end{array}
\label{newfun-expand}
\ee
where $\ell_\gamma$ is the ``BPS ray"
\be
\ell_{\gamma}= \{ t\in \IP^1:\quad Z_{\gamma}/t\in \I\IR^-\}
\ee
and
\be
\label{eqXigi-ml}
\Xigi(t)=\Thkl + \cR\(t^{-1}\Zg-t\bZg\).
\ee
Here $\Theta_\gamma$ and $Z_\gamma$ are the axionic coupling of the instanton and  the central charge defined in
\eqref{intCC} and \eqref{defZg}, respectively, while $\cR$ is a variable which will be related below to the dilaton.
Finally, $\sigma_\gamma$ is the  quadratic refinement introduced below \eqref{defSumk}.

\item
The prepotential $F$, the K\"ahler potential $\cK=-\log K$ and the matrix $N_{\Lambda\Sigma}$ describe
the special K\"ahler geometry of $\cM_C$ and were introduced in \S\ref{subsec-MC}.

\item
$V_{\gamma\Lambda}$ and $\vl_\Lambda$ are the vectors
\be
V_{\gamma\Lambda}= q_\Lambda -F_{\Lambda\Sigma}p^\Sigma,
\qquad
\vl_\Lambda=\sum_\gamma \Om{\gamma} \vv_\gamma V_{\gamma\Lambda},
\label{defVgam}
\ee
where
\be
\vv_\gamma=\frac{1}{4\pi}\( Z_\gamma\rIgp+\bZ_\gamma\rIgm\).
\ee

\item
$\Min^{\Lambda\Sigma}$ is the inverse of the matrix
\be
\Min_{\Lambda\Sigma}=  N_{\Lambda\Sigma}-\frac{1}{2\pi}\sum_\gamma \Om{\gamma}\rIg \bV_{\gamma\Lambda}V_{\gamma\Sigma}.
\label{defmatM}
\ee

\item
$\Uin$ is the function
\bea
\Uin
&=&K-\frac{1}{2\pi}\sum_\gamma \Omega_\gamma|Z_\gamma|^2 \rIg+ \vl_\Lambda \Min^{\Lambda\Sigma}\bvl_\Sigma,
\label{defUin}
\eea
which can be thought of as an instanton corrected version of the K\"ahler potential.

\item
$\cW_\gamma$ is a function on the charge lattice defined by
\bea
\cW_\gamma
&=&\bZ_\gamma \rIg- \rIgp \vl_\Lambda\Min^{\Lambda\Sigma} \bV_{\gamma\Sigma}.
\label{cWdef}
\eea

\item
$\cY_\Lambda$ is the one-form
\be \label{ea10aa}
\cY_\Lambda =
\de\tzeta_\Lambda-F_{\Lambda\Sigma}\de\zeta^\Sigma
-\frac{1}{8\pi^2}\sum\limits_{\gamma} \Omega_\gamma \(q_{\Lambda}-p^\Sigma F_{\Lambda\Sigma}\)\de \Igg{}.
\ee

\item
$\cV$ is the one-form playing the role of the connection in the circle bundle with fiber parametrized by the NS axion
\be
\begin{split}
\cV=&\,
2\cR^2 K\(1-\frac{4r}{\cR^2\Uin}\) \cA_K
+\frac{8r}{\cR\Uin}\sum_\gamma\Omega_\gamma \( \vv_\gamma+\frac{1}{2\pi}\, \rIg V_{\gamma\Lambda}\Min^{\Lambda\Sigma}\bvl_\Sigma\)\cC_\gamma
\\
&\, +\frac{2r }{\pi\I\Uin}\sum_\gamma \Om{\gamma} \[\(\cW_\gamma+\frac{\cR\Uin}{8\pi\I r}\,\Igp\)\de Z_\gamma
-\(\bar\cW_\gamma+\frac{\cR\Uin}{8\pi\I r}\,\Igm\)\de \bZ_\gamma  \],
\end{split}
\label{conn}
\ee
where $\cC_\gamma$ was defined in \eqref{connC} and
\be
\cA_K =  \frac{\I}{2}\(\cK_a\de z^a-\cK_{\bar a}\de \bz^a\)
\label{kalcon}
\ee
is the K\"ahler connection that $\cV$ reduces to in the perturbative approximation.

\item
Finally, the coordinate $\cR$
should be viewed as a function of other coordinates on the moduli space. This function
is defined only implicitly through the following relation
\be
r = \frac{\cR^2}{4}\, K+\frac{\chi(\CY)}{192\pi}
-\frac{ \I\cR}{32\pi^2}\sum\limits_{\gamma}\Omega_\gamma \(Z_\gamma\Igp+\bZ_\gamma\Igm\) .
\label{dil}
\ee
Here the first term relates $\cR$ to the dilaton at the tree level. The second term provides the one-loop correction
parametrized by the Euler characteristic of CY $\chi(\CY)$ and captures all perturbative contributions to the metric.
The final term is the contribution of D-instantons.\footnote{It is worth pointing out that
the definition of the four-dimensional dilaton, appearing as one of the coordinates on the moduli space,
is ambiguous at the non-perturbative level. For example, in \cite{Cortes:2021vdm}
it was chosen to coincide with the first term in \eqref{dil}. Our definition is suggested by geometry since
the function $r$ defined here coincides with the so-called contact potential of quaternionic geometry
\cite{Alexandrov:2008nk}, and by symmetries since it is symplectic invariant and transforms as a modular form
of weight $(-\hf,-\hf)$ in the mirror type IIB formulation \cite{Alexandrov:2008gh}. This ambiguity does not affect
the leading terms that we shall be analyzing.}

\end{itemize}

\subsection{Evaluation of the leading contribution}

There are two possible views on the metric \eqref{mett1}, or in other words, two approximations where it holds:
\begin{enumerate}
\item
The first possibility is to restrict the sums over charges $\gamma$ to a mutually local subset.
Then \eqref{mett1} is an exact quaternion-K\"ahler metric that includes all D-instantons
to all orders in the instanton and string coupling expansions
from this particular subset.
\item
The second possibility is to sum over all charges, but then only terms linear in DT invariants can be trusted.
\end{enumerate}
Our goal here is to extract from this metric the leading contribution in each instanton sector in the small string coupling limit,
which corresponds to taking the $g_{(4)}\to 0$ limit keeping fixed the primed variables
introduced in \refb{rescalefields}.

This limit is analyzed as follows.
First of all, note that, by expanding the logarithm or the ratio,
the twistorial integrals \eqref{newfun-expand} can all be
evaluated in terms of series of modified Bessel functions $K_m(x)$
\be
\begin{split}
\Igg{} =&\,  -2 \Sg{0,1},
\qquad
\Igpm =  2\I \(\frac{\bZ_\gamma}{|Z_\gamma|}\)^{\pm 1} \Sg{1,1},
\\
\rIg=&\, 2 \Sg{0,0},
\qquad
\rIgpm= -2\I \(\frac{\bZ_\gamma}{|Z_\gamma|}\)^{\pm 1} \Sg{1,0},
\label{newfun-K}
\end{split}
\ee
where
\be
\begin{split}
\Sg{m,n}=&\, \sum_{k=1}^\infty \frac{\sigma_{k\gamma}}{k^n}\,e^{-2\pi \I k \Theta_\gamma}  K_m(4\pi k \cR |Z_\gamma|)
\\
=&\, \frac{1}{2\sqrt{2\cR|Z_\gamma|}}
\sum_{k=1}^\infty \frac{\sigma_{k\gamma}}{k^{n+\hf}}\,  e^{-2\pi \I k \Theta_\gamma-4\pi k \cR|Z_\gamma|}
\[1+\frac{4m^2-1}{32\pi k\cR|Z_\gamma|}+\cdots\].
\end{split}
\label{expSmn}
\ee
Taking into account that due to \eqref{dil} $\cR\sim r^{1/2} = (r')^{1/2} g_{(4)}^{-1}$
(see \eqref{rescalefields}), this result implies that in the leading approximation
all these functions are proportional to $\sqrt{g_{(4)}}\, e^{-k\cT_\gamma}$
in the topological sector $k\gamma$.
At the same time, all terms in the metric \eqref{mett1}
can be represented as series expansions in powers of $\Omega_\gamma\Sg{m,n}$.
As a result, all terms non-linear in DT invariants are always of higher power in
$g_{(4)}$ comparing to the terms linear in $\Omega_\gamma$.
Thus, for our purposes, it is enough to restrict to such linear terms.
Therefore, in the weak coupling limit, the terms that survive in the first approximation outlined above are
a subset of those in the second approximation. For this reason we shall stick to the
second approximation and
no restriction on the charges will be imposed.

It is a straightforward exercise to extract the linear approximation to the metric \eqref{mett1}.
Additional simplifications occur once we restrict ourselves for each class of terms only to the leading
contribution in the small string coupling limit. For example, this means that we can neglect the one-loop
correction captured by the Euler characteristic in the relation \eqref{dil} between $\cR$ and the dilaton,
and inverting this relation one finds
\be
\cR\approx 2 \sqrt{\frac{r}{K}}\(1-\frac{1}{8\pi^2 \sqrt{Kr}}\sum_\gamma \Omega_\gamma |Z_\gamma|\Sg{1,1}\).
\ee
Furthermore, it turns out that the second term inside the parentheses can also be dropped because in the weak coupling limit
the contribution $\Omega_\gamma \Sg{1,1}$ is always multiplied by an extra factor of $g_{(4)}$.
By examining \refb{defVgam} to \refb{dil}, one can see that all other terms proportional to
$\Igg$ and $\Igpm$ have similar powers of $g_{(4)}$ multiplying them. The only exception
is the $d \Igg$ term in \refb{ea10aa} whose leading term is that of
$-2S_\gamma^{(0,0)} \, d\TT_\gamma$.
Thus, in our approximation the leading instanton contribution is captured by the function
\be \label{e979n}
\Sg{0}=\frac{(K/r)^{1/4}}{4\sqrt{|Z_\gamma|}}
\sum_{k=1}^\infty \frac{\sigma_{k\gamma}}{\sqrt{k}}\,  e^{-k\cT_\gamma},
\ee
which is the leading term in the expansion \eqref{expSmn} of $\Sg{m,0}$.
In the following we shall use the symbol $\approx$ to denote that we keep only those
terms which produce terms of
order unity in the perturbative metric and terms of order $(g_{(4)})^{1/2} e^{-\TT_\gamma}$ in
the instanton correction to the metric in the topological sector labelled by $\gamma$.
In this convention one finds the following simplifications for the various quantities entering the metric:
\be
\vv_\gamma\approx\frac{ |Z_\gamma|}{\pi\I}\, \Sg{0},
\ee
\be
\Uin \approx K-\frac{1}{\pi}\sum_\gamma \Omega_\gamma|Z_\gamma|^2 \Sg{0},
\ee
\be
\cW_\gamma \approx 2\bZ_\gamma \Sg{0},
\ee
\ben
\cY_\Lambda &\approx&
\de\tzeta_\Lambda-F_{\Lambda\Sigma}\de\zeta^\Sigma
+\frac{1}{2\pi\I}\sum\limits_{\gamma} \Omega_\gamma \(q_{\Lambda}-p^\Sigma F_{\Lambda\Sigma}\)
\Biggl(\de \Theta_\gamma-\frac{2\I|Z_\gamma|}{\sqrt{K r}}\,\de r
\non\\ &&
-4\I\sqrt{\frac{r}{K}}\(\de |Z_\gamma| -\hf\,|Z_\gamma|\de \log K \)
\Biggr)\Sg{0},
\een
\bea
\cV&\approx&
-\(\frac{\chi(\CY)}{24\pi}+\frac{8r}{\pi K}\sum_\gamma\Omega_\gamma|Z_\gamma|^2
\Sg{0}\)\cA_K
\non\\
&&
+\frac{4 r}{\pi\I K}\sum_\gamma\Omega_\gamma \(|Z_\gamma|\sqrt{\frac{K}{r}}\,\cC_\gamma
+\( \bZ_\gamma\de Z_\gamma -Z_\gamma\de \bZ_\gamma  \)\)\Sg{0},
\label{conn-oneinst}
\eea
\bea
1-\frac{2r}{\cR^2\Uin} &\approx&
\hf-\frac{|Z_\gamma|}{2\pi K}\sum_\gamma \Omega_\gamma |Z_\gamma| \Sg{0}.
\eea
The term proportional to $\chi(\CY)$ in \refb{conn-oneinst} actually does not contribute at
the leading order. To see this, note that once we use the primed variables, $\cV$ appears
in the combination $d\sigma'+ g_{(4)}^2 \cV$ in \refb{mett1}, and in the weak coupling limit
the $g_{(4)}^2\chi(\CY)$ term drops out. In contrast the rest of the terms in the expression
for $\cV$ does contribute since $g_{(4)}^2 r=r'$ does not vanish in the weak coupling limit.

Substituting these results into \eqref{mett1}, expanding to linear order in $\Omega_\gamma$ and
keeping only the leading terms in the weak coupling limit, one obtains
\be
\de s^2\approx \de s_{\rm tree}^2+\de s_{\rm inst}^2,
\ee
where the tree level metric is the one given in \eqref{hypmettree} and the instanton contribution reads as
\bea
\de s_{\rm inst}^2 &= & \sum_\gamma \Omega_\gamma\,\Sg{0}\Biggl\{
-\frac{|Z_\gamma|^2(\de r)^2}{\pi r^2K}
+\frac{|Z_\gamma| \de r}{\pi\I r \sqrt{rK}}\(\de\Theta_\gamma
-\frac{4}{K}\, \Im\[\bZ_\gamma(z^\Lambda\de\tzeta_\Lambda-F_\Lambda\de \zeta^\Lambda)\]\)
\non\\
&&
+\frac{|Z_\gamma|^2}{16\pi r^2 K}
\(\de\sigma+\tzeta_\Lambda\de\zeta^\Lambda-\zeta^\Lambda\de\tzeta_\Lambda\)^2
\non\\
&&
+\frac{|Z_\gamma|}{2\pi\I rK}\(\de\sigma+\tzeta_\Lambda\de\zeta^\Lambda-\zeta^\Lambda\de\tzeta_\Lambda\)
\[\sqrt{\frac{K}{r}}\, \cC_\gamma
+|Z_\gamma|\(\p\log \frac{Z_\gamma}{K}-\bar\p\log \frac{\bZ_\gamma}{K}\)
\]
\non\\
&&
-\frac{1}{\pi r}\(\cC_\gamma^2-\frac14\, (\de\Theta_\gamma)^2
+\frac{2}{K}\, \de\Theta_\gamma\,\Im\[\bZ_\gamma(z^\Lambda\de\tzeta_\Lambda-F_\Lambda\de\zeta^\Lambda)\]\)
\non\\
&&
+\frac{8\I}{\pi \sqrt{r} K^{3/2}}\Im\[\bZ_\gamma(z^\Lambda\de\tzeta_\Lambda-F_\Lambda\de\zeta^\Lambda)\]
\(\de|Z_\gamma|-\frac{|Z_\gamma|}{2}\, \de\log K\)
\non\\
&&
-\frac{2|Z_\gamma|}{\pi \sqrt{rK}}\,\cC_\gamma\(\p\log\frac{Z_\gamma}{K}-\bar\p\log\frac{\bZ_\gamma}{K}\)
+\frac{4}{\pi K}\left|\de Z_\gamma-Z_\gamma\p\log K\right|^2
\Biggl\},
\label{e877}
\eea
where $\cC_\gamma$ has been defined in \refb{connC}.
Finally, noting that $\Sg{0}=\sqrt{\pi}\Sigma_\gamma/\sqrt{2\cTR_\gamma}$
with $\Sigma_\gamma$ defined in \eqref{defSumk},
\refb{e877} can be rewritten in the form given in the main text in \eqref{square}.

\section{Fixing the sign of the integration over the fermion zero modes} \label{sdd}

In this appendix we shall test the compatibility between
the choice of sign of integration over fermion zero modes as
given in \refb{esignchoice} and the multi-instanton contribution described
in \refb{e21}. To this end, let us begin with a general choice:
\be \label{ed1a}
\int \left\{\prod_{\delta,\ddelta=1}^2  d \chi_a^\delta d \chi_a^{\ddelta}\right\}
 \chi_a^\alpha  \chi_a^\dalpha  \chi_a^\beta  \chi_a^\dbeta
=s\, \eps^{\alpha\beta}\eps^{\dalpha\dbeta}
\qquad
\hbox{for} \ 0\le a\le k^2 -1\, ,
\ee
where $s$ can be $\pm 1$. We shall however take $s$ to be independent of $a$. This is needed
for cluster property --- when a pair of instantons are widely separated then the contribution
to the amplitude must factorize. We shall now use this to fix the sign of the integral
appearing in \refb{e21}.

We shall consider the case $k=2$,
but this analysis can be generalized to other values of $k$.
We  carry out the integration over the
fermionic modes in \refb{e21} by expanding $e^S$ in a power series expansion
in the fermionic terms in $S$.
Using the commutation relation $[T^a,T^b]=\I\, \sqrt 2\,
\eps^{abc} T^c$ for the $SU(2)$ generators normalized as below \refb{Tvar},
the Yukawa coupling term in \refb{e14} can be written as $-\I\, g_o\, \eps^{abc} (\gamma_\mu)_{\alpha\dalpha} \xi_a^\mu \,
\chi_b^\dalpha \chi_c^\alpha$.
One particular term in the integral will be:
\be\label{ed2a}
- g_o^6 \int \left\{\prod_{a=1}^3 \prod_{\delta,\ddelta=1}^2  d \chi_a^\delta d \chi_a^{\ddelta}\right\}
\[\Bigl( \chi_1^\alpha \, \xi_3^\mu (\gamma_\mu)_{\alpha\dalpha}
 \chi_2^\dalpha\Bigr)
\( \chi_1^\beta \, \xi_3^\nu (\gamma_\nu)_{\beta\dbeta}  \chi_2^\dbeta\)
\times \hbox{cyclic perm. of 1,2,3}\]\, .
\ee
Let us now cyclically rename the dotted zero mode variables, {\it e.g.}
$  \chi_2^\dalpha$ as $\chi_1^\dalpha$ etc. Under this relabelling the integration measure
remains unchanged since we have to exchange pairs of spinor variables together. Now
we can use \refb{ed1a} to express \refb{ed2a} as
\be
-g_o^6\, \((\xi_3^\mu \xi_3^\nu) \, s \, \eps^{\alpha\beta}\eps^{\dalpha\dbeta}\,
(\gamma^\mu)_{\alpha\dalpha}\, (\gamma_\nu)_{\beta\dbeta}\) \, \times
(3\to 1,2)\, .
\ee
Using manipulations described above \refb{e6.2aa}, this may be written as
\be
g_o^6\, s^3 \prod_{a=1}^3 \(2\eta_{\mu\nu} \, \xi_a^\mu\,  \xi_a^\nu\) \, .
\ee
We see that the sign of the contribution of this term
is $s^3$.
Even though there are other contributions we expect that the sign of \refb{e21}
will coincide with the sign of this term, since this represents the `diagonal component' of
the determinant --- the other terms are proportional to $\xi_a\cdot\xi_b$ for $a\ne b$.
Since the integral in \refb{e21} is taken to be positive, we must have
\be
s=1\, .
\ee
For general $k$, we can repeat the same argument to argue that \refb{e21} will have
sign $s^{k^2-1}=s^{k-1}$. Therefore, the same choice $s=1$ will ensure positivity of \refb{e21}.
On the other hand, if we choose $s$ to be $-1$, then the $k$ instanton amplitude will
acquire an extra factor of $(-1)^k$.

\section{Vertex operator for  the RR fields} \label{errkin}

In this appendix we shall present the construction of the vertex operator of the RR field describing the
3-form potential or equivalently the 4-form field strength in ten-dimensional
type IIA string theory.

In the $(-1/2,-1/2)$ picture, the off-shell RR field at level 0, describing massless fields, is given
by a state $|\phi_R\rangle$ satisfying the conditions
$b_0^-|\phi_R\rangle=0$, $L_0^-|\phi_R\rangle=0$.
The general form of $|\phi_R\rangle$ is:
\be\label{ephiR}
|\phi_R\rangle = \int{d^{10}p\over (2\pi)^{10}}\,
F^\alpha_{~\beta}(p) \, c\, \bar c\, e^{-\phi/2} S_\alpha \,e^{-\bar\phi/2}
\bar S^\beta \, e^{\I p.X}(0)|0\rangle\, ,
\ee
for some set of functions $F^\alpha_{~\beta}(p)$. We have
\be
\begin{split}
& (Q_B+\bar Q_B) |\phi_R\rangle =
\int{d^{10}p\over (2\pi)^{10}}\,
F^\alpha_{~\beta}(p) \, \Bigg[ {p^2\over 4} (\p c+\bar \p\bar c) \,
c\, \bar c\, e^{-\phi/2} S_\alpha \,e^{-\bar\phi/2}
\bar S^\beta \, e^{\I p.X}(0)|0\rangle
\\
& \quad
- {1\over 4} (\sp)_{\alpha\gamma} c\, \bar c\, \eta\, e^{\phi/2} S^\gamma e^{-\bar\phi/2}
\bar S^\beta \, e^{\I p.X}(0)|0\rangle   + {1\over 4} (\sp)^{\beta\delta} \,
c\, \bar c\, e^{-\phi/2} S_\alpha \bar\eta\, e^{\bar\phi/2}
\bar S_\delta \, e^{\I p.X}(0)|0\rangle\Bigg]\, .
\end{split}
\ee
Therefore, BRST invariance of the state will require:
\be\label{errbrst}
p^2\, F^\alpha_{~\beta}(p) =0,
\qquad
F^\alpha_{~\beta}(p) (\sp)_{\alpha\gamma}=0,
\qquad
F^\alpha_{~\beta}(p) (\sp)^{\beta\delta} =0\, ,
\ee
where $\sp\equiv p_M\Gamma^M$.
Note that the first condition follows from the other two.
Since we shall be interested in only the component with 4-form field strength, we
use the ansatz:
\be \label{efabexp}
F_\alpha^{~\beta}(p)={1\over 4!}\, F^{(4)}_{MNPQ}(p)\, \(\Gamma^{MNPQ}\)_\alpha^{~\beta}
\, ,
\ee
where $F^{(4)}$ is a 4-form.
Substituting this into \refb{errbrst} we get:
\be
p^2 F^{(4)}_{MNPQ}(p)=0,
\qquad
F^{(4)}_{MNPQ}(p) \(\sp\, \Gamma^{MNPQ}\)=0,
\qquad
F^{(4)}_{MNPQ}(p)\(\Gamma^{MNPQ}\!\! \sp\)=0\, ,
\ee
leading to
\be
p^2 F^{(4)}_{MNPQ}(p)=0,
\qquad
p^M  F^{(4)}_{MNPQ}=0,
\qquad
p_{[R} F^{(4)}_{MNPQ]}=0\, .
\ee
This is the usual on-shell condition on the 4-form field strength.

For our analysis we shall also need the BRST invariant vertex operator for the same
state in the $(-1/2, -3/2)$ picture since this contains information about the potential $\wt
C_{MNP}$
associated with the field strength $F^{(4)}_{MNPQ}$. For this
we begin with the general form of level zero string field in the $(-3/2,-3/2)$ picture:
\bea \label{efg1}
|\wt\phi_R\rangle &=& \int{d^{10}p\over (2\pi)^{10}}\, \bigg[
A_\alpha^{~\beta}\, c\, \bar c \,e^{-3\phi/2}\, S^\alpha \, e^{-3\bar\phi/2}\, \bar S_\beta (0)
+ B_{\alpha\beta} \, (\p\, c+\bar\p\, \bar c)\, c\, \bar c\, e^{-3\phi/2}\, S^\alpha \, \bar\p \bar\xi \,
e^{-5\bar\phi/2}\bar S^\beta(0)
\non\\
&&
+ D^{\alpha\beta} \, (\p\, c+\bar\p\, \bar c)\,  c\, \bar c\, \p \xi \,
e^{-5\phi/2} S_\alpha \, e^{-3\bar\phi/2}\, \bar S_\beta(0)
\bigg] e^{\I p.X}(0)|0\rangle\, .
\eea
From this we get
\be\label{efg2}
\begin{split}
(Q_B+\bar Q_B) |\wt\phi_R\rangle =&\, \int{d^{10}p\over (2\pi)^{10}} \[ {p^2\over 4} \,
A_\alpha^{~\beta}
- {1\over 4} \, B_{\alpha\gamma}  (\sp)^{\gamma\beta}
+ {1\over 4}\, D^{\gamma\beta} (\sp)_{\gamma\alpha} \]
\\ & \hskip 1in
\,\times(\p c+\bar\p\bar c) \,
c\, \bar c \,e^{-3\phi/2} S^\alpha \, e^{-3\bar\phi/2}\, \bar S_\beta\,
e^{\I p.X}(0)|0\rangle\, , \end{split}
\ee
\be\label{efg3}
\begin{split}
|\wt\phi_{-1/2,-3/2}\rangle =& \bar\XX_0 \, |\wt\phi_R\rangle =
\int{d^{10}p\over (2\pi)^{10}}\, \bigg[{1\over 2}\,
A_\alpha^{~\beta}\, (\sp)_{\beta\gamma} \, c\, \bar c \,e^{-3\phi/2}\, S^\alpha \,
e^{-\bar\phi/2}\, \bar S^\gamma (0) \\ &
- {1\over 2} \, B_{\alpha\beta} \, c\, \bar c\, e^{-3\phi/2} S^\alpha \,  e^{-\bar\phi/2}\,
\bar S^\beta (0) \\ &
+ {1\over 2} \, D^{\alpha\beta} \, (\sp)_{\beta\gamma}
 (\p\, c+\bar\p\, \bar c)\,  c\, \bar c\, \p \xi \,
e^{-5\phi/2} S_\alpha \, e^{-\bar\phi/2}\, \bar S^\gamma(0) \\ &
- {1\over 2} \, D^{\alpha\beta} \, c\, \bar c\, \p \xi \,
e^{-5\phi/2} S_\alpha \, \bar\eta\, e^{\bar\phi/2}\, \bar S_\beta(0)
\bigg] e^{\I p.X}(0)|0\rangle\, ,
\end{split}
\ee
and the $(-1/2,-1/2)$ picture string field
\be \label{efg4}
\begin{split}
|\phi_{R}\rangle = \XX_0 \, |\wt\phi_{-1/2,-3/2}\rangle = &
\int{d^{10}p\over (2\pi)^{10}}\, \[{1\over 4}\,
A_\alpha^{~\beta}\, (\sp)_{\beta\gamma} (\sp)^{\alpha\delta} - {1\over 4} \, B_{\alpha\gamma} \, (\sp)^{\alpha\delta}
+ {1\over 4} \, D^{\delta\beta} \, (\sp)_{\beta\gamma}\]
\\
& \hskip 1in
\,\times c\, \bar c \,e^{-\phi/2}\, S_\delta \,
e^{-\bar\phi/2}\, \bar S^\gamma \, e^{\I p.X}(0)|0\rangle\, .
\end{split}
\ee
Comparison of \refb{ephiR} and \refb{efg4} leads to the identification
\be \label{efg5}
F^\alpha_{~\beta} = {1\over 4}\( \sp \, A \!\sp\,  -\! \sp\, B + D \!\sp\)^\alpha_{~\beta}\, .
\ee
Also \refb{efg2} leads to the on-shell condition
\be
p^2  A - B\! \sp \,+\! \sp \, D =0\, ,
\ee
which implies  \refb{errbrst}.

From \refb{efg5} we see that the description in terms of the matrices $A$, $B$ and $D$
provides a redundant description of the field strength $F^\alpha_{~\beta}$.  Indeed
$F^\alpha_{~\beta}$ is invariant under the transformation:
\be
A\to A+\Lambda,
\qquad
B\to B + \Lambda' \!\sp\, ,
\qquad
D\to D+\!\sp\, (\Lambda' -\Lambda)\, ,
\ee
for arbitrary matrices $\Lambda,\Lambda'$.
These are related
to gauge transformation parameters in the $(-3/2,-3/2)$ picture.
We shall choose a particular gauge in which $A=0$. Therefore we have, on-shell,
\be\label{efg6}
F = {1\over 4} \(D \!\sp\, - \sp\, B\),
\qquad
\sp\, D = B \!\sp\, .
\ee

We now decompose $B_{\alpha\beta}$ and $D^{\alpha\beta}$ as:
\be\label{efg7}
B_{\alpha\beta} = \sum_{k=1\atop k\in2\ZZZ+1}^5 {1\over k!}
B^{(k)}_{M_1\cdots M_k} (\Gamma^{M_1\cdots M_k})_{\alpha\beta}\, ,
\qquad
D^{\alpha\beta}= \sum_{k=1\atop k\in2\ZZZ+1}^5 {1\over k!}
D^{(k)}_{M_1\cdots M_k} (\Gamma^{M_1\cdots M_k})^{\alpha\beta}\, ,
\ee
with $B^{(5)}$ being self-dual and $D^{(5)}$ being anti-self-dual.
Now, from \refb{efabexp}, \refb{efg6} and \refb{efg7} we get,
\bea\label{efg8}
F^{(4)}_{M_1 M_2M_3M_4} &=& {1\over 16}\, \Tr(\Gamma_{M_1 M_2M_3M_4} F) = {1\over 4}
p^M \(D^{(5)}-B^{(5)}\)_{MM_1 M_2M_3M_4}
\\ &&
-{1\over 4} \[
p_{M_1} \(D^{(3)}+B^{(3)}\)_{M_2M_3M_4} +\hbox{cyclic perm. of $M_1,M_2,M_3,M_4$
with sign}\].
\non
\eea
Also, multiplying the second equation in \refb{efg6} by $\Gamma_{MNPQ}$
and taking the trace, we get
\be\label{efg9}
\begin{split}
&\,  p^M \(D^{(5)}-B^{(5)}\)_{MM_1 M_2M_3M_4}
\\
=-&\,  \[
p_{M_1} \(D^{(3)}+B^{(3)}\)_{M_2M_3M_4} +\hbox{cyclic perm. of $M_1,M_2,M_3,M_4$
with sign}\].
\end{split}
\ee
Using this, we can express \refb{efg8} as
\be\label{efg10}
F^{(4)}_{M_1 M_2M_3M_4} = -{1\over 2} \[
p_{M_1} \(D^{(3)}+B^{(3)}\)_{M_2M_3M_4} +\hbox{cyclic perm. of $M_1,M_2,M_3,M_4$
with sign}\].
\ee

Physically $B^{(3)}+D^{(3)}$ may be interpreted as the RR potential for the field strength
$F^{(4)}_{MNPQ}$ and $B^{(5)}-D^{(5)}$, or more
precisely its Hodge dual $B^{(5)}+D^{(5)}$,
may be regarded as the potential for the Hodge dual 6-form field strength. Therefore,
string field theory naturally uses the democratic formalism of \cite{Townsend:1995gp}.
Indeed, this is the reason why the open-closed string field theory constructed in
\cite{FarooghMoosavian:2019yke} can
describe coupling to all D-branes at the same time, including the ones that are
electric-magnetic dual to each other. Equations of motion of course relate the two field
strengths and allows us to express the result in terms of just the 3-form potential as in
\refb{efg10}.

We further note that if we had tried to express the 2-form field strength
component $F^{(2)}_{MN}$
of $F_\alpha^{~\beta}$ using an expansion similar to that in \refb{efabexp}, and
repeated the analysis described above, then $F^{(2)}_{M_1M_2}$ would have contained a
term proportional to $p^M \(D^{(3)}-B^{(3)}\)_{MM_1 M_2}$. Therefore, if we want to generate
purely 4-form field strength, we can set
\be \label{efg11}
D^{(3)}=B^{(3)}=-\I\, \wt C\, .
\ee
Eqs. \refb{efg10} and \refb{efabexp} now reduce to, respectively,
\be \label{efcrelation}
F^{(4)}_{MNPQ} = \(\I\, p_M \wt C_{NPQ} +\hbox{cyclic permut. of
$M,N,P,Q$ $\times$ with sign}\)
\ee
and
\be
F^\alpha_{~\beta} = {\I\over 3!} \, p_M \, \wt C_{NPQ}\, \(\Gamma^{MNPQ}\)^\alpha_{~\beta}\, .
\label{ecnorm}
\ee
\refb{efg3}, \refb{efg7} and \refb{efg11} with $A_\alpha^{~\beta}=0$
can be used to read out the vertex
operator of the 3-form field $\wt C_{MNP}$
in the $(-1/2,-3/2)$ picture, while \refb{ephiR} and \refb{ecnorm} may be
used to read out the same vertex operator in the $(-1/2,-1/2)$ picture.

\section{Some spinor identities in Calabi-Yau threefolds} \label{sc}

In this appendix we shall evaluate the contractions with the covariantly constant spinor $\eta$ appearing
in \refb{intRR} and \refb{elag10}.

First, due to anti-symmetry we have
$
\bar\eta \wt\Gamma^{\bf pqr ijk}\eta =c\, \eps^{\bf pqr ijk}.
$
Choosing $({\bf pqr ijk})=(456789)$ one obtains $c=-\I \bar\eta\wt\Gamma\eta=-\I$ so that
\be
\bar\eta \wt\Gamma^{\bf pqr ijk}\eta =-\I\, \eps^{\bf pqr ijk}.
\label{etaeps}
\ee
This relation is valid both for tangent space and space-time indices,
provided for space-time indices the right hand side is
understood to contain a factor of
$(\det g)^{-1/2}$ to convert it into a tensor.

Next we shall analyze
$
\bar\eta \wt\Gamma^{\bf rk} \eta$. For this
let us work in the local
coordinate system in which the metric and the K\"ahler form on $\CY$ take the form:
\be\label{elag11}
ds^2=\sum_{{\bf m}=4}^9 (dx^{\bf m})^2,
\qquad
\omega=  \sum_{s=1}^3  dx^{2s+2} \wedge dx^{2s+3}\, .
\ee
The associated complex coordinates
are taken to be
\be\label{e788}
\hat w^s=x^{2s+2}+i\, x^{2s+3}, \qquad 1\le s\le 3\, .
\ee
Let us  denote the holomorphic
and anti-holomorphic components of $\wt\Gamma^{\bf m}$ by
\be
\begin{split}
\hGam^1=\wt\Gamma^4+\I\wt\Gamma^5,
\qquad
\hGam^2=\wt\Gamma^6+\I\wt\Gamma^7,
\qquad
\hGam^3=\wt\Gamma^8+\I\wt\Gamma^9,
\\
\hGam^{\bar 1}=\wt\Gamma^4-\I\wt\Gamma^5,
\qquad
\hGam^{\bar 2}=\wt\Gamma^6-\I\wt\Gamma^7,
\qquad
\hGam^{\bar 3}=\wt\Gamma^8-\I\wt\Gamma^9.
\end{split}
\ee
Furthermore, we postulate that $\hGam^{s}\eta=0$, $\bar \eta\hGam^{\bar s}=0$
for $1\le s\le 3$. One can easily verify that these conditions are compatible with the
condition $\wt\Gamma\,\eta=\eta$.
Then using the anti-commutation relations $\{\hGam^s,\hGam^{\bar t}\}=4\, \delta^{st}$,
we get
\be
\bar\eta \wt\Gamma^{s\bar t} \eta=\hf\, \bar\eta \hGam^s\hGam^{\bar t}\eta=2\, \delta^{st}
= \I \, \omega^{s\bar t} ,
\ee
since we have $\omega^{s\bar t}=g^{s\bar u}g^{\bar t v}\omega_{\bar u v}=-2\I\delta^{st}$.
We also  have
$\bar\eta \wt\Gamma^{st}\eta=\omega^{st}=0$ as well as similar relations for their complex conjugates.
Therefore, we find $\bar\eta \wt\Gamma^{\bf ij}\eta = \I \omega^{\bf ij}$
in this coordinate system. However, since both sides are tensors, we can take this to be a
general relation valid in any coordinate system, i.e.
\be\label{ereltwo}
\bar\eta \wt\Gamma^{\bf rk} \eta =\I\, \omega^{\bf rk} = -\I J_{\bf m}^{\bf r}g^{\bf mk}\, ,
\ee
where  in the last term
we used the relation $\omega^{\bf ij}=g^{\bf ik}J_{\bf k}^{\bf j}=-J_{\bf k}^{\bf i}g^{\bf kj}$
following from the definition of the K\"ahler form.

Our next task will be to evaluate
\be
e_{\bf ij}\, (P^{\bf j}_{\bf k} - Q^{\bf j}_{\bf k})\,  \bar\eta\wt\Gamma^{\bf i}\wt\Gamma^{\bf k}\eta\, ,
\label{eee}
\ee
where $P$ and $Q$ are projection operators along the brane and  transverse to the brane, respectively.
They can be expressed as
\be \label{edefPQ}
P_{\bf i}^{\bf j}={1\over 2} \, v_{\gamma,\bf ikl}v_\gamma^{\bf jkl}\, ,
\qquad
Q_{\bf i}^{\bf j}={1\over 2} \, (\star v_\gamma)_{\bf ikl}(\star v_\gamma)^{\bf jkl}\, .
\ee
We define, for any 3-form,
\be
(C,C')=\frac{1}{3!}\, C^{\bf ijk}\bar C'_{\bf ijk},
\qquad
||C||^2=(C,C).
\label{def-norm}
\ee
Note that we have
\be \label{vgammasquare}
||v_\gamma||^2 =1\, .
\ee
We shall use the convention that the $\star$ is an {\it anti-linear} map which,
besides taking the Hodge dual of a form,
complex conjugates the coefficients of the forms written in real coordinates.
On 3-forms we have $\star^2=-1$.

The projectors $P$ and $Q$ satisfy the following properties:
\begin{itemize}
\item
for any 3-form $C$ one has
\be
P^{\bf l}_{\bf i}P^{\bf m}_{\bf j}P^{\bf n}_{\bf k} C_{\bf lmn}
=(C,v_\gamma)\, v_{\gamma,\bf ijk}\,;
\label{projC}
\ee

\item
in the holomorphic coordinates the projectors satisfy
\be
P_{st}=- Q_{st},
\qquad
P_{\bar s\bar t}=-Q_{\bar s\bar t},
\qquad
P_{s\bt}=Q_{s\bt}\, .
\label{holQP}
\ee

\end{itemize}
\refb{projC} is  self-evident. The first two relations in \refb{holQP} follow from the observation that
$P_{\bf ik}+Q_{\bf ik}=g_{\bf ik}$ and $g_{st}=g_{\bar s\bar t}=0$.
The third relation in \refb{holQP} can be proven as follows.
Since the cycle is Lagrangian, we have the identity ({\it cf.} \eqref{BPScycle})
\be
P_{\bf i}^{\bf k}P_{\bf j}^{\bf l}\,\omega_{\bf kl}=0.
\label{Lagr-indent}
\ee
On the other hand, using the relation $\star\omega=\hf\, \omega\wedge \omega$,
we can write
\be
\begin{split}
Q_{\bf i}^{\bf k}Q_{\bf j}^{\bf l}\,\omega_{\bf kl}
=&\, \frac{1}{8}\, Q_{\bf i}^{\bf k}Q_{\bf j}^{\bf l}\,{\eps_{\bf kl}}^{\bf mnpq}\,\omega_{\bf mn}\omega_{\bf pq}
\\
=&\,
\frac{1}{2}\, Q_{\bf i}^{\bf k}Q_{\bf j}^{\bf l}\,{\eps_{\bf kl}}^{\bf mnpq}
\(Q_{\bf m}^{\bf m'}P_{\bf n}^{\bf n'}\omega_{\bf m'n'}\)\(P_{\bf p}^{\bf p'}P_{\bf q}^{\bf q'}\omega_{\bf p'q'}\)=0,
\end{split}
\label{Lagr-indent2}
\ee
where in the second equality we used that $P$ and $Q$ are projectors on 3-dimensional subspaces
and the last equality follows from \eqref{Lagr-indent}.
In the holomorphic basis, the mixed components of the two identities \eqref{Lagr-indent} and \eqref{Lagr-indent2} read as
\be
P_{s u} P^u_{\bar t}=P_{s\bar u}P^{\bar u}_{\bar t},
\qquad
Q_{s u} Q^u_{\bar t}=Q_{s\bar u}Q^{\bar u}_{\bar t}.
\ee
Left hand sides of the two equations are equal due to the  first two relations in \eqref{holQP}.
Therefore, the right hand sides of these equations must be equal as well. This gives
\be
P_{s\bar u} g^{\bar u v}P_{v\bar t}=Q_{s\bar u} g^{\bar u v}Q_{v\bar t}
=(g-P)_{s\bar u} g^{\bar u v}(g-P)_{v\bar t} = P_{s\bar u} g^{\bar u v}P_{v\bar t} + g_{s\bar t} - 2\, P_{s\bar t}.
\ee
This implies $P_{s\bar t}=\frac12\, g_{s\bar t}$ and hence $Q_{s\bar t} = g_{s\bar t}- P_{s\bar t} =  \frac12\, g_{s\bar t}$.
This establishes the last relation in \eqref{holQP}.

Let us now turn to the analysis of \refb{eee}.
Using \refb{ereltwo}, this can be rewritten as
\be
e_{\bf ij}\(P_{\bf k}^{\bf j}-Q_{\bf k}^{\bf j}\)\(g^{\bf ik}-\I J^{\bf i}_{\bf l} g^{\bf lk}\).
\ee
The second factor vanishes when the index $\bf i$ corresponds to
an anti-holomorphic index and is equal to $2 g^{\bf ik}$ when $\bf i$ is a holomorphic
index. This allows us to rewrite the above expression in the holomorphic coordinates, and
we get
\be\label{eestpst}
e_{\bf ij}\, (P^{\bf j}_{\bf k} - Q^{\bf j}_{\bf k})\,
\bar\eta\,\wt\Gamma^{\bf i}\wt\Gamma^{\bf k}\eta =
2\, e_{st} (P^{st}-Q^{st}) + 2\, e_{s\bar t} (P^{s\bar t}-Q^{s\bar t}) = 4\, e_{st}P^{st}\, ,
\ee
where we used \eqref{holQP}.

We now use the relation $e_{\bf ij}=\delta g_{\bf ij}/(2\kappa)$ given above \refb{decom-gmna}
to write the right hand side of \refb{eestpst} as $2\,\delta g_{st} P^{st}/\kappa$. Using
\refb{decom-gmna}
relating $\delta g_{\bf ij}$ to the complex structure deformation,
we can write
\be
P^{st}\delta g_{st}=
\delta \bz^a \, {\Omega_{\bf ijk} g^{\bf jm} g^{\bf kn}\over ||\Omega||^2}
\, (\bar\chi_a)_{\bf lmn}P^{\bf il}
\label{contr-delz}
\ee
because the indices on the projector are automatically holomorphic due to the properties of $\Omega$ and $\chi_a$.
Next, we can compute
\be
\begin{split}
||\Omega||^2=&\, \frac{1}{6}\, \Omega_{\bf ijk}(P^{\bf il}+Q^{\bf il})(P^{\bf jm}+Q^{\bf jm})(P^{\bf kn}+Q^{\bf kn})\bar\Omega_{\bf lmn}
\\
=&\,  \frac{4}{3}\, \Omega_{\bf ijk}P^{\bf il}P^{\bf jm}P^{\bf kn}\bar\Omega_{\bf lmn}
=8\, |(\Omega,v_\gamma)|^2\, ,
\end{split}
\ee
where we used \eqref{holQP} followed by \eqref{projC} and \refb{vgammasquare}.
Similarly, one has
\be
\begin{split}
\Omega_{\bf ijk} g^{\bf jm} g^{\bf kn}(\bar\chi_a)_{\bf lmn}P^{\bf il}
=&\, \Omega_{\bf ijk} (P^{\bf jm}+Q^{\bf jm})(P^{\bf kn}+Q^{\bf kn})(\bar\chi_a)_{\bf lmn}P^{\bf il}
\\
=&\, \frac{4}{3}\,  \Omega_{\bf ijk} P^{\bf il}P^{\bf jm}P^{\bf kn}(\bar\chi_a)_{\bf lmn}
=8\, (\Omega,v_\gamma)(v_\gamma,\chi_a) \, ,
\end{split}
\ee
where in the first two steps we made repeated use of \refb{holQP} and in the last step we used
\refb{projC} and \refb{def-norm}.
Thus, \eqref{contr-delz} becomes
\be \label{eappfin}
P^{st}\delta g_{st}=\delta \bz^a \, \frac{(v_\gamma,\chi_a)}{(v_\gamma,\Omega)}.
\ee

\providecommand{\href}[2]{#2}\begingroup\raggedright\endgroup

%\bibliographystyle{utphys}
%\bibliography{combined}

\end{document}